\documentclass[preprint,11pt,authoryear]{elsarticle}

\usepackage[ruled,vlined,linesnumbered]{algorithm2e}
\usepackage{amssymb}
\usepackage{amsthm}
\usepackage{commath}
\usepackage{comment}
\usepackage{esdiff}
\usepackage{enumitem}
\usepackage{float}
\usepackage{hyperref}
\usepackage{cleveref}
\usepackage{ifthen}
\usepackage{lineno}
\usepackage{listings}
\usepackage{nicefrac}
\usepackage{placeins}
\usepackage{soul}
\usepackage{subfiles}
\usepackage{subcaption}
\usepackage{tikz}
\usepackage[inline]{trackchanges}
\usepackage[dvipsnames]{xcolor}

\graphicspath{{./graphics}}

\usetikzlibrary{
    3d,
    shapes, 
    calc,
    decorations.pathreplacing,
    calligraphy, 
    positioning,
    backgrounds
}

\hypersetup{
    colorlinks=true,
    filecolor=magenta,      
    urlcolor=cyan,
    citecolor=blue,
}
\newfloat{lstfloat}{htb}{lop}
\floatname{lstfloat}{Listing}

\lstdefinestyle{in}{
    basicstyle=\footnotesize\ttfamily,
    commentstyle=\color{gray},
    breaklines=true,
    frame=single,
}

\crefalias{lstfloat}{lstlisting}
\crefname{lstlisting}{listing}{listings}
\crefname{lstlisting}{Listing}{Listings}

\addeditor{Jeremy}
\addeditor{Derek}
\addeditor{Mark}

\soulregister\citep7
\soulregister\citet7
\soulregister\ref7
\soulregister\eqref7
\soulregister\Cref7
\soulregister\cref7

\theoremstyle{definition}
\newtheorem{defn}{Definition}

\theoremstyle{remark}
\newtheorem{remark}{Remark}

\theoremstyle{remark}
\newtheorem{example}{Example}

\renewcommand{\epsilon}{\varepsilon}
\renewcommand{\phi}{\varphi}

\renewcommand{\vec}[1]{\mathbf{#1}}
\newcommand{\mat}[1]{\mathbf{#1}}
\newcommand{\tens}[1]{\mathcal{#1}}
\renewcommand{\tt}[1]{\mathcal{#1}_\text{TT}}
\newcommand{\tto}[1]{\mathcal{#1}_\text{TTO}}

\newcommand{\Rtens}[1]{\mathbb{R}^{{#1}_1 \times \cdots \times {#1}_d}}
\newcommand{\frob}[1]{\norm{#1}_\text{F}}
\newcommand{\ord}[1]{\mathcal{O}(#1)}
\newcommand{\Ord}[1]{\mathcal{O}\left(#1\right)}

\newcommand{\nlat}{n_\text{lat}}
\newcommand{\nlon}{n_\text{lon}}

\DeclareMathOperator{\comp}{comp}
\DeclareMathOperator{\numel}{numel}
\DeclareMathOperator{\round}{round}
\DeclareMathOperator{\vol}{vol}

\journal{Journal of Computational Physics}

\begin{document}


\begin{frontmatter}
    \title{Viability of Tensor Train Methods for Geophysical Fluid Dynamics}

    \author[ccs2,cnls]{Jeremy Lilly\corref{cor}}
    \ead{jlilly@lanl.gov}
    \ead[url]{https://jeremy-lilly.github.io/}
    \cortext[cor]{Corresponding author}
    
    \author[ccs2]{Derek DeSantis}
    \author[ccs2]{Mark R. Petersen}

    \affiliation[ccs2]{
        organization={Computational Physics and Methods Group},
        addressline={Los Alamos National Laboratory}, 
        city={Los Alamos},
        state={NM},
        postcode={87545}, 
        country={USA},
    }
    \affiliation[cnls]{
        organization={Center for Nonlinear Studies},
        addressline={Los Alamos National Laboratory}, 
        city={Los Alamos},
        state={NM},
        postcode={87545}, 
        country={USA},
    }

    \begin{abstract}
        Tensor train (TT) methods have recently gained popularity for accelerating the solving of systems of PDEs.
        Here, we evaluate the performance of TT methods in the context of geophysical fluid dynamics (GFD) using the shallow water equations and a discretization scheme employed by the ocean component of the Energy Exascale Earth System Model (E3SM).
        Through a suite of four test cases of increasing complexity, we evaluate TT methods in terms of how much TT is able to compress the model state, the error incurred by the TT approximation, and the speedup obtained by TT versus an optimal standard non-TT implementation in a representative subproblem.
        We show that though TT is able to effectively compress and speed up simple flows, it struggles to efficiently represent more complex states that are common in realistic GFD applications.
    \end{abstract}

    \begin{keyword}
        tensor train \sep GPU acceleration \sep shallow water \sep TRiSK \sep geophysical fluid dynamics
    \end{keyword}
\end{frontmatter}





\section{Introduction}
\label{sec:introduction}

The central challenge in modeling geophysical fluids is balancing accuracy with computational cost, largely due to the broad range of spatial and temporal scales in play.
As spatial resolution is increased to more accurately resolve flow dynamics at smaller scales, the number of unknowns in the numerical system quickly grows, driving up computational expenses.
The use of variable-resolution, unstructured grids can help somewhat, allowing geophysical fluid dynamics (GFD) modelers to highly resolve areas of interest while saving on the total number of spatial cells.
However, this is often not enough to allow researchers to run the kinds of large ensembles at high resolutions that would accelerate our understanding of vitally important systems like our Earth's climate.

Inspired by new techniques originally developed for quantum physics applications, there has been recent success applying tensor decomposition methods to general computational fluid dynamics (CFD) problems \citep{vonLarcher2019,gourianov2022,kiffner2023,truong2024,danis2025a}.
Given some system of equations, such as that derived from a set of semi-discrete PDEs on a large mesh, one can reformulate the problem in terms of tensor decompositions.
Tensor decompositions can discover a low-dimensional latent space to compress high-dimensional interactions, and perform computations there, making previously intractable simulations feasible (see \Cref{fig:tt_overview}).
By far the most popular tensor decomposition method is the so-called tensor train (TT) decomposition wherein a \( d \)-dimensional tensor is represented by a ``train'' of \( d \) 3-dimensional tensors \citep{oseledets2011}; this means that the cost of storing such a tensor scales linearly with \( d \) rather than exponentially.

A recent study from \citet{danis2025b} applies TT methods to the shallow water equations (SWEs) and shows that these methods are capable of accurately resolving certain dynamics in simple GFD problems while providing speedups up to 124x.
Moving towards more realistic applications and comparing computational efficiency to that of optimal methods, we seek to test TT methods on more complex GFD flows and on more realistic domains to know how this potential 124x speedup generalizes.
Pushing toward the climate scale, we consider configurations as close as possible to those used in real-world applications; the particular configuration that is of most interest is that used by the Energy Exascale Earth System Model (E3SM) \cite{golaz2022}.
The ocean component of E3SM use the TRiSK spatial discretization \cite{ringler2010}.

It has been shown that more complicated flows require larger TTs (i.e. less model compression) to capture all relevant features \citep{vonLarcher2019}; an open question is whether TT methods provide similar benefits under real-world model configurations like those used in E3SM.
For sufficiently complicated flows, the TT may cost more to store in memory than the original ``uncompressed'' model state.
As stated above, the tensor decomposition methods are rooted the quantum physics space where the tensor train is referred to as a matrix product state (MPS).
It can be shown that all realistic quantum states can be efficiently represented as a MPS (i.e. by a tensor train of some high dimensional state tensor) \citep{orus2014}.
For this reason, TT is a vital and efficient tool for such applications.
In the context of GFD however, we have no such guarantee that any flow state can be efficiently represented by a TT (we discuss this further in \Cref{sec:approximating_with_tt}).
Previous work cited above shows that simple flows compress well for a range of CFD applications, but more work is needed to see if TT methods can be generalized to more realistic flows for real-world applications.

In this work, we apply TT methods to the SWEs using an E3SM-like model with the TRiSK spatial discretization.
We investigate TT methods as applied to four test cases of increasing complexity, including canonical SWE test cases from \citet{galewsky2004} and \citet{williamson1992}.
The TT methods are evaluated based on their ability to compress and accurately represent model states, as well as the speedup they supply versus optimized non-TT methods.
We show that performance is strong for simple flows, but that TT methods struggle with the more complex flows from \citet{galewsky2004} and \citet{williamson1992}, which implies that TT methods would also struggle with real-world applications at the climate scale.

The paper is structured as follows.
In \Cref{sec:background}, we provide the necessary background on tensor trains and how they can be used to solve partial differential equations (PDEs).
Next, in \Cref{sec:methods}, we describe relevant aspects of the shallow water code used for timing and experiments, as well as issues related to TT on structured and unstructured grids (further discussion of TT on unstructured grids is given in \Cref{apx:unstructured_tt}).
Additionally, we provide details about how we construct sparse operators in the TT-format.
Finally, in \Cref{sec:experiments}, we describe a collection of four test cases and their results, which illustrate the central issue in this work -- how effective TT is at compressing flows relevant to GFD.



\section{Background}
\label{sec:background}


\subsection{The Shallow Water Equations}
\label{sec:shallow_water}

The shallow water equations (SWEs) are a simplification of the Navier-Stokes equations for general fluid flow.
The SWEs are obtained from Navier-Stokes under two primary simplifying assumptions; first, that the fluid is of a constant density, and second, that the horizontal length scale of the fluid is much larger than the vertical length scale.
The second simplifying assumption has the affect that magnitude of vertical velocities in the fluid become negligible, granting us a two-dimensional flow.
Colloquially, this condition can be stated as, \emph{``The Pacific Ocean is shallow. Your coffee cup is deep.''} \citep{higdon2006}.

Though a relatively simple set of governing equations, the SWEs resolve an important subset of physical processes relevant to general geophysical fluid flow \citep{lilly2023}.
In particular, the SWEs are used to model the so-called barotropic component of full three-dimensional models of the ocean and atmosphere and are therefore often used as a test-bed for new numerical methods targeted at global and climate scale geophysical fluid dynamics.
As such, they serve well for our investigation of tensor train methods for this application.

The SWEs on a rotating sphere are given by
\begin{equation}
\begin{aligned}
    \diffp{\vec{u}}{t} + \left( \nabla \times \vec{u} + f\mathbf{k} \right) \times \vec{u} &= -\nabla\frac{\abs{\vec{u}}^2}{2} - g\nabla (h + z_b) \\
    \diffp{h}{t} + \nabla \cdot \left(h\vec{u}\right) &= 0 \,,
\end{aligned}
    \label{eqn:nonlinear_swe}
\end{equation}
where \( \mathbf{u}(x, y, t) = \left(u(x, y, t),\, v(x, y, t),\, 0\right) \) is the two-dimensional fluid velocity,
\( x \) and \( y \) are the spatial coordinates,
\( t \) is the time coordinate,
\( f \) is the Coriolis parameter,
\( \mathbf{k} \) is the local vertical unit vector normal to the sphere,
\( g \) is the gravitational constant,
\( h \) is the fluid thickness,
and \( z_b \) is the height of the bottom topography.
We will often refer to the equation for the evolution of \( \vec{u} \) as the momentum or velocity equation, and the equation for the evolution of \( h \) as the thickness or mass equation.

In this work, we solve the SWEs using the TRiSK spatial discretization \citep{ringler2010}, which is a finite volume-type spatial discretization made for unstructured, variable-resolution polygonal grids, and is the discretization used in the ocean component of E3SM \citep{golaz2022}.
TRiSK employs C-grid-type discretization \citep{arakawa1977} wherein the mass variable is computed on cell centers and the normal component of velocity is computed on cell edges.
In E3SM, these are Voronoi grids \citep{ju2011, okabe2017} consisting primarily of hexagons as the primal mesh, with a dual mesh consisting of triangles.
However, TRiSK is not limited to grids of this type; in this work we will primarily utilize quadrilateral grids wherein both the primal and dual grids are quadrilaterals (more on this in \Cref{sec:shallow_water_model}).


\subsection{Tensor Decomposition with Tensor Train}
\label{sec:tensor_train}

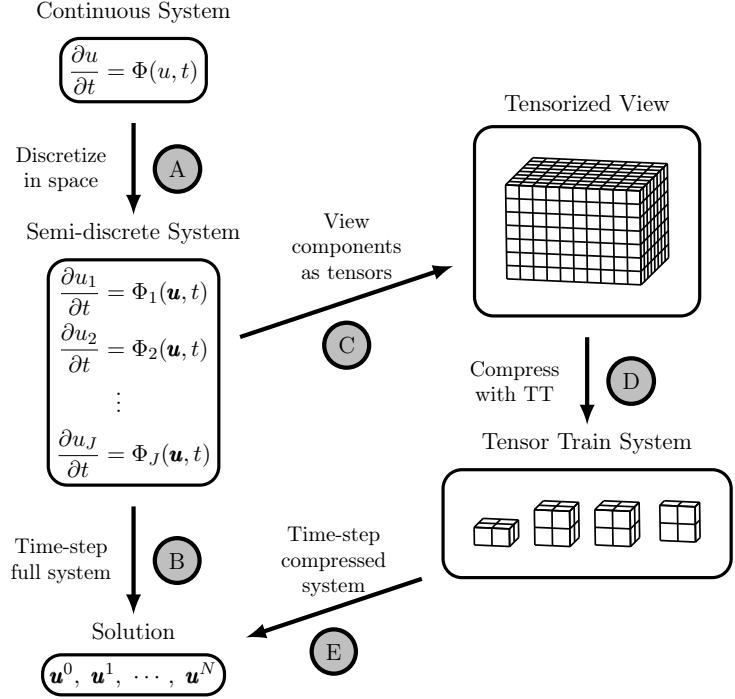
\begin{figure}
    \centering
    \tikzstyle{tn}=[very thick,
                rounded corners=0.25cm,
                outer sep=10pt,
                font=\normalsize,
                anchor=center,
                align=center,
                draw]

\tikzstyle{small_tn}=[font=\small,
                      align=center,
                      text width=2cm,
                      outer sep=2pt]

\tikzstyle{circ_label}=[anchor=center,
                        outer sep=10pt]

\tikzstyle{ar}=[ultra thick, -latex]

\tikzset{pics/cuboid/.style args={#1,#2,#3}{code={
    \begin{scope}[rotate around x=10, rotate around y=-10]
        \foreach \x in {0,...,#1} {
            \draw (\x,0,#3) -- (\x,#2,#3);
            \draw (\x,#2,#3) -- (\x,#2,0);
        }
        \foreach \x in {0,...,#2} {
            \draw (#1,\x,#3) -- (#1,\x,0);
            \draw (0,\x,#3) -- (#1,\x,#3);
        }
        \foreach \x in {0,...,#3} {
            \draw (#1,0,\x) -- (#1,#2,\x );
            \draw (0,#2,\x) -- (#1,#2,\x );
        }
    \end{scope}
}}}

\tikzset{pics/circ/.style={code={
    \node[circle, fill=gray!50!white, draw] {#1};
}}}

\def\gridsize{50mm}
\def\scale{0.8}
\begin{tikzpicture}[black, x=\gridsize, y=\gridsize,
                    scale=\scale,
                    every node/.style={scale=\scale}]
    \node[tn] (pde) at (0,0) {
         \( \begin{aligned}
            \frac{\partial u}{\partial t} = \Phi(u, t)
        \end{aligned} \)%
    };
    \node[above=-5pt of pde] (pde_label) {Continuous System};

    \node[tn] (discrete) at (0,-1) {
        \( \begin{aligned}
            \frac{\partial u_1}{\partial t} &= \Phi_1(\pmb{u}, t) \\
            \frac{\partial u_2}{\partial t} &= \Phi_2(\pmb{u}, t) \\
            &\hspace{7pt}\vdots \\
            \frac{\partial u_J}{\partial t} &= \Phi_J(\pmb{u}, t)
        \end{aligned} \)%
    };
    \node[above=-5pt of discrete] (discrete_label) {Semi-discrete System};

    \node[tn] (solution) at (0,-2) {
        \( \pmb{u}^0,\ \pmb{u}^1,\ \cdots,\ \pmb{u}^N  \)%
    };
    \node[above=-5pt of solution] (solution_label) {Solution};

    \node[tn, inner sep=0.5cm] (tensor) at (1.5,-0.5) {
        \begin{tikzpicture}
            \def\ncubes{7}
            \pic[scale=0.4/\ncubes, thick] {cuboid={10, \ncubes, 8}};
        \end{tikzpicture}%
    };
    \node[above=-5pt of tensor] (tensor_label) {Tensorized View};

    \node[tn, inner sep=0.5cm] (tt) at (1.5,-1.5) {
        \begin{tikzpicture}
            \def\ncubes{2}
            \pic[scale=0.15/\ncubes, thick] at (0,0) {cuboid={\ncubes, 1, \ncubes}};
            \pic[scale=0.15/\ncubes, thick] at (0.2,0) {cuboid={\ncubes, \ncubes, \ncubes}};
            \pic[scale=0.15/\ncubes, thick] at (0.4,0) {cuboid={\ncubes, \ncubes, \ncubes}};
            \pic[scale=0.15/\ncubes, thick] at (0.6,0) {cuboid={\ncubes, \ncubes, 1}};
        \end{tikzpicture}%
    };
    \node[above=-5pt of tt] (tt_label) {Tensor Train System};

    
    \draw[ar] (pde) -- pic[circ_label, midway, right] {circ={A}} node[small_tn, midway, left] {Discretize in space} (discrete_label);

    \draw[ar] (discrete) -- pic[circ_label, midway, right] {circ={B}} node[small_tn, midway, left] {Time-step full system} (solution_label);

    \draw[ar] (discrete) -- pic[circ_label, midway, below] {circ={C}} node[small_tn, midway, above=4pt] {View components as tensors} (tensor) (solution_label);

    \draw[ar] (tensor) -- pic[circ_label, midway, right] {circ={D}} node[small_tn, midway, left] {Compress with TT} (tt_label);

    \draw[ar] (tt) -- pic[circ_label, midway, below] {circ={E}} node[small_tn, midway, above] {Time-step compressed system} (solution);
\end{tikzpicture}
    \caption{
        Tensor train methods find speedups by avoiding time-stepping the full discrete system directly.
        Standard methods involve (A) discretizing the PDE in space, followed by (B) the use of an iterative time-stepping algorithm.
        Tensor train methods can be understood to instead (C) treat the discrete system as a collection of tensors, which are (D) approximated by tensor trains.  
        Speedup is found by (E) time-stepping in the reduced dimensional latent space of the tensor trains, which scales linearly with the size of the problem instead of polynomially as in (B).
    }
    \label{fig:tt_overview}
\end{figure}

A tensor decomposition is a method by which a given tensor can be represented by a set of (usually lower dimensional) tensors.
This allows high-dimensional data with exponential storage costs to be represented in a more memory-efficient way; one can think of this as obtaining a ``compressed'' representation of the original tensor.
At a high level, the goal of the tensor decomposition methods we investigate here (e.g. tensor train) is to obtain an approximate model state that is significantly compressed versus the true model state, then to evolve the compressed model state forward in time.
The idea is that the time evolution of the compressed model state is less computationally expensive at the cost of a controllable amount of error.
\Cref{fig:tt_overview} provides a high-level summary of this strategy.

However, computational savings that one can obtain with such a tensor decomposition method greatly depends on how well the model state can be compressed.
Further, because the model state is changing in time, the compression rate relative to the allowed error will also change in time, potentially greatly affecting the effectiveness a of tensor decomposition method.
\emph{This is the central idea that we will investigate in this work.}

\subsubsection{Groundwork}
\label{sec:groundwork}

To facilitate our discussion of the tensor train (TT) based methods that we use in this work, we begin by carefully defining a number of terms that often appear in different forms across the literature.
\begin{defn}
    \label{defn:tensor}
    A \emph{tensor} is a multi-dimensional array of arbitrary dimension, consisting of real entries, subject to nonlinear point-wise operations and linear transformations.
    That is, for a positive integer \( d \) and a sequence of \emph{mode sizes} \( n_1,\, n_2,\, \cdots,\, n_d \in \mathbb{N} \), a tensor \( \tens{X} \) is an element of \( \mathbb{R}^{n_1 \times n_2 \times \cdots \times n_d} \).
    We say that the tensor \( \tens{X} \) has \emph{shape} \( [n_1,\, n_2,\, \cdots,\, n_d] \).
    
    In this way, a \( n_1 \times n_2 \) matrix is a 2-dimensional tensor in \( \mathbb{R}^{n_1 \times n_2} \) and a length \( n_1 \) vector is a 1-dimensional tensor in \( \mathbb{R}^{n_1} \).
    A \( d \)-dimensional tensor \( \tens{X} \in \mathbb{R}^{n_1 \times n_2 \times \cdots \times n_d} \) is indexed by a multi-index \( \mathbf{i} = (i_1,\, i_2,\, \cdots,\, i_d) \), where \( i_k \in \{1,\, 2,\, \cdots,\, n_k\} \) for all \( k = 1,\, 2,\, \cdots,\, d \); this is written as \( \tens{X}(i_1,\, i_2,\, \cdots,\, i_d) \in \mathbb{R} \).
\end{defn}
\begin{defn}
    \label{defn:tensor_decomposition}
    A \emph{tensor decomposition} is a representation of a given \( d \)-dimensional tensor \( \tens{X} \in \Rtens{n} \) by some type of product of other tensors.  
    There are many types of products (outer, Khatri-Rao, and Kronecker products) and other multilinear operations used (tensor contractions), which formulate different types of tensor decompositions. 
    Typically the component tensors of a tensor decomposition have dimension less than \( d \).

    The typical example of a tensor decomposition is the tensor train decomposition (\Cref{defn:tensor_train}), but others include canonical polyadic decomposition (CPD) \citep{prevost2025}, the Tucker decomposition \citep{bhatt2021}, tensor ring \citep{wu2023}, and Projected Entangles Parallel Sates (PEPS) \citep{orus2014}.
    Some tensor decompositions can be viewed as generalizations of matrix decompositions; for example, the singular value decomposition (SVD), LU factorization, or CUR decomposition can be seen as tensor decompositions of 2-dimensional tensors.
\end{defn}
\begin{defn}
    \label{defn:tensor_train}
    A \emph{tensor train} (TT) is a particular tensor decomposition that represents a given \( d \)-dimensional tensor \( \tt{X} \in \Rtens{n} \) by a sequence of \( d \) 3-dimensional tensors \( \tens{G}_k \) for \( k = 1,\, 2,\, \cdots,\, d \) called \emph{cores}.
    These cores are sometimes also called \emph{carts}, hence the name tensor ``train.''

    Each core \( \tens{G}_k \) has shape \( [r_{k-1},\, n_k,\, r_k] \), where the \( r_k \in \mathbb{N} \) for \( k = 0,\, 1,\, \cdots,\, d \), with \( r_0 = r_d = 1 \), are called the \emph{ranks} of the TT.
    We often refer to the ranks collectively as a vector quantity \( \vec{r} = [r_0,\, r_1,\, \cdots,\, r_d] \) called the \emph{rank}.
    The TT-rank is directly related to how effective a TT representation of a tensor is -- the smaller the TT-ranks, the smaller the TT-cores, and so the fewer elements stored by the TT in total (e.g. see \Cref{rmk:rank_growth} and \Cref{defn:compression}).

    The TT representation of the tensor \( \tt{X} \) can be written as
    \begin{equation}
        \label{eqn:tt_contraction}
    \begin{aligned}
        \tt{X}(i_1,\, \cdots,\, i_d) = \sum_{\alpha_1, \cdots, \alpha_{d-1}}^{r_1, \cdots, r_{d-1}}
                                       &\tens{G}_1(1,\, i_1,\, \alpha_1) \
                                       \tens{G}_2(\alpha_1,\, i_2,\, \alpha_2) \
                                       \cdots \\
                                       &\quad\tens{G}_{d-1}(\alpha_{d-2},\, i_{d-1},\, \alpha_{d-1}) \
                                       \tens{G}_d(\alpha_{d-1},\, i_d,\, 1) \,.
    \end{aligned}
    \end{equation}
    Equivalently, we can also denote the TT-format by the multiple matrix product,
    \begin{equation}
        \label{eqn:tt_matrix_product}
        \tt{X}(i_1,\, i_2,\, \dots,\, i_d) = \mat{G}_1(i_1) \
                                             \mat{G}_2(i_2) \
                                             \cdots \
                                             \mat{G}_d(i_d) \,,
    \end{equation}
    where each term \( \mat{G}_{k}(i_k) \), for \( i_k = 1,\, \cdots,\, n_k \) and \( k = 1,\, \cdots, d \), is a matrix of shape
    \( [r_{k-1},\, r_k] \).

    These statements can be nontrivial to understand at first glance, so we direct to reader's attention to \Cref{fig:tt_contraction}, which provides a visual example of the tensor contraction in \Cref{eqn:tt_contraction,eqn:tt_matrix_product}.
\end{defn}
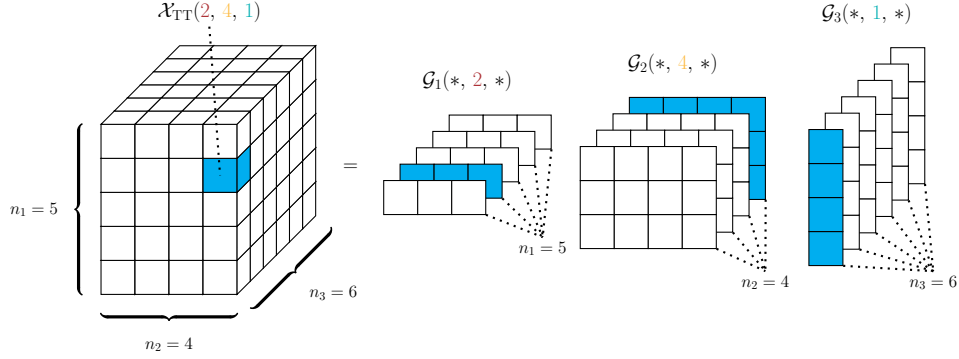
\begin{figure}
    \centering
    \newcommand{\squarearray}[4]{
    \foreach \i in {1,...,#1}{
        \foreach \j in {1,...,#2}{
            \draw[fill=#3] (\j-1,\i-1) rectangle node[name=#4-\i-\j,
                                                      minimum width=1cm,
                                                      minimum height=1cm] {} (\j,\i);
        }
    }
}

\newcommand{\stackarray}[6]{
    \def\sep{1.25}
    \foreach \x in {1,...,#1}{
        \begin{scope}[canvas is xy plane at z={-(#1-\x)*\sep}]
            \ifthenelse{\x=#2}{
                \squarearray{#4}{#5}{#3}{#6-\x}
            }{
                \squarearray{#4}{#5}{white}{#6-\x}
            }
        \end{scope}
    }
}

\newcommand{\cube}[2]{
    \begin{scope}[canvas is yx plane at z=1]
        \draw[fill=#1] (0,0) rectangle node[name=#2-face-1,
                                            minimum width=1cm,
                                            minimum height=1cm] {} (1,1);
    \end{scope}
    \begin{scope}[canvas is xz plane at y=1]
        \draw[fill=#1] (0,0) rectangle node[name=#2-face-2,
                                            minimum width=0cm,
                                            minimum height=0cm] {} (1,1);
    \end{scope}
    \begin{scope}[canvas is yz plane at x=1]
        \draw[fill=#1] (0,0) rectangle node[name=#2-face-3,
                                            minimum width=0.5cm,
                                            minimum height=0.5cm] {} (1,1);
    \end{scope}
}

\newcommand{\cubearray}[8]{
    \foreach \i in {1,...,#1}{
        \foreach \j in {1,...,#2}{
            \foreach \k in {1,...,#3}{
                \begin{scope}[shift={(\j-1,\i-1,\k-1)}]
                    \ifthenelse{\i=#4 \AND \j=#5 \AND \k=#6}{
                        \cube{#7}{#8-cube-\i-\j-\k}
                    }{
                        \cube{white}{#8-cube-\i-\j-\k}
                    }
                \end{scope}
            }
        }
    }
}

\tikzset{dotline/.style={dotted, thick}}
\tikzset{brace/.style={decorate, decoration={calligraphic brace, raise=0.25cm}, very thick}}
\tikzset{bracemirror/.style={decorate, decoration={calligraphic brace, mirror, raise=0.25cm}, very thick}}

\def\none{5}
\def\ntwo{4}
\def\nthree{6}
\def\pone{2}
\def\ptwo{4}
\def\pthree{1}
\def\scale{0.45}
\begin{tikzpicture}[scale=\scale,
                    every node/.style={scale=\scale,font=\Large}, every picture]
    \begin{scope}[local bounding box=middlecore]
        \stackarray{\ntwo}{\inteval{\ntwo+1-\ptwo}}{cyan}{3}{4}{mcore}
    \end{scope}
    \node[font=\LARGE] (G2) at ($(middlecore.north)+(0,1)$) {\( \mathcal{G}_2(*,\, \mathcolor{Dandelion}{\ptwo},\, *) \)};
    \node (n2) at ($(middlecore.south east)+(0,-1)$) {\( n_2 = \ntwo \)};
    \foreach \i in {1,...,\ntwo}{
        \draw[dotline] (mcore-\i-1-4.south east) -- (n2.north);
    }

    \begin{scope}[shift={($(middlecore.south)+(-8.5,1)$)},
                  local bounding box=leftcore]
        \stackarray{\none}{\inteval{\none+1-\pone}}{cyan}{1}{3}{lcore}
    \end{scope}
    \node[font=\LARGE] (G1) at ($(leftcore.north)+(0,1)$) {\( \mathcal{G}_1(*,\, \mathcolor{Maroon}{\pone},\, *) \)};
    \node (n1) at ($(leftcore.south east)+(-0.25,-1)$) {\( n_1 = \none \)};
    \foreach \i in {1,...,\none}{
        \draw[dotline] (lcore-\i-1-3.south east) -- (n1.north);
    }

    \begin{scope}[shift={($(middlecore.south)+(4,-0.5)$)},
                  local bounding box=rightcore]
        \stackarray{\nthree}{\inteval{\nthree+1-\pthree}}{cyan}{4}{1}{rcore}
    \end{scope}
    \node[font=\LARGE] (G3) at ($(rightcore.north)+(0,1)$) {\( \mathcal{G}_3(*,\, \mathcolor{TealBlue}{\pthree},\, *) \)};
    \node (n3) at ($(rightcore.south east)+(0.25,-0.5)$) {\( n_3 = \nthree \)};
    \foreach \i in {1,...,\nthree}{
        \draw[dotline] (rcore-\i-1-1.south east) -- (n3.north);
    }

    \begin{scope}[shift={($(leftcore.west)+(-6,-1.5)$)},
                  local bounding box=fulltensor]
        \cubearray{\none}{\ntwo}{\nthree}{\inteval{\none+1-\pone}}{\ptwo}{\inteval{\nthree+1-\pthree}}{cyan}{tensor}

    \end{scope}
    \node[font=\LARGE] (X) at ($(fulltensor.north)+(0,1)$) {\( \mathcal{X}_\text{TT}(\mathcolor{Maroon}{\pone},\,\mathcolor{Dandelion}{\ptwo},\,\mathcolor{TealBlue}{\pthree}) \)};
    \draw[dotline] (X.south) -- (tensor-cube-\inteval{\none+1-\pone}-\ptwo-\inteval{\nthree+1-\pthree}-face-1.center);
    \draw[bracemirror] (tensor-cube-1-1-\nthree-face-1.south west) -- node[midway, below=0.5cm] {\( n_2 = \ntwo \)} (tensor-cube-1-\ntwo-\nthree-face-1.south east);
    \draw[brace] (tensor-cube-1-1-\nthree-face-1.south west) -- node[midway, left=0.5cm] {\( n_1 = \none \)} (tensor-cube-\none-1-\nthree-face-1.north west);
    \draw[bracemirror] (tensor-cube-1-\ntwo-\nthree-face-1.south east) -- node[midway, below right=0.5cm] {\( n_3 = \nthree \)} (tensor-cube-1-\ntwo-1-face-3.south east);
    \path (fulltensor.east) -- node[midway, font=\LARGE] {\( = \)} (leftcore.west);
\end{tikzpicture}
    \caption{
        Visualization of tensor contraction for a tensor train \( \tt{X} \in \mathbb{R}^{5 \times 4 \times 6} \).
        The entry at multi-index \( \mathbf{i} = (2,\, 4,\, 1) \) is obtained by taking the standard matrix-product of the highlighted slices of TT cores \( \tens{G}_1 \), \( \tens{G}_2 \), and \( \tens{G}_3 \).
        This figure was adapted from a similar figure from \citet{xu2023}. 
    }
    \label{fig:tt_contraction}
\end{figure}

\subsubsection{Approximating Tensors by Tensor Trains}
\label{sec:approximating_with_tt}

In the present context (solving systems of PDEs), it is most common to use tensor train to approximate a given tensor, e.g. (D) in \Cref{fig:tt_overview}. 
Given some tensor \( \tens{X} \in \Rtens{n} \) and some \( \epsilon > 0 \), there exists a tensor train \( \tt{X} \) such that
\begin{equation}
    \label{eqn:tt_approximation}
    \frob{\tens{X} - \tt{X}} < \epsilon \frob{\tens{X}} \,,
\end{equation}
where \( \frob{\, \cdot\, } \) is the Frobenius norm.
In practice, there are a number of ways that \( \tt{X} \) can be obtained from a particular \( \tens{X} \) and \( \epsilon \), but the most common is via the so-called TT-SVD algorithm from \citet[Algorithm 1]{oseledets2011}.
This is a well-known algorithm, so we will not reproduce it in full here.
In summary, TT-SVD produces a TT by taking a total of \( d \) truncated SVDs of certain unfoldings of the tensor \( \tens{X} \).
For each \( k = 1,\, \cdots,\, d \), the \( k \)-th application of the truncated SVD produces the core \( \tens{G}_k \) and rank \( r_k \), where \( r_k \) is exactly the number of singular values left over after the singular value truncation.

Let \( n = \max_k\{n_k\} \) and \( r = \max_k\{r_k\} \), then the cost to store all of tensor \( \tens{X} \) is \( \ord{n^d} \), while the cost to store \( \tt{X} \) is only \( \ord{ndr^2} \) -- linear in both \( n \) and \( d \).
For sufficiently controlled values of \( r \), this means that \( \tt{X} \) is both more efficient to store and more efficient to operate on, e.g. (E) in \Cref{fig:tt_overview}.
The difficulty, which is the primary concern of this paper, is that the TT-rank \( \vec{r} \) is a function of the desired accuracy \( \epsilon \).
When \( \epsilon \) is taken to be small, more singular values are kept during each step of the TT-SVD algorithm, and the ranks \( r_k \) grow appropriately.
Further, the TT-rank \( \vec{r} \) also depends on \( \tens{X} \) -- if the singular values of the unfoldings of \( \tens{X} \) decay slowly, more singular values and vectors need to be kept, again resulting in higher TT-rank.

The TT-format was popularized in the quantum materials space, where it is referred to as the Matrix Product State (MPS).
In this context, a tensor (usually of large dimension) is used to represent some quantum state, and it can be shown that the set of states that are relevant to the study of quantum materials (so-called ``Area-law'' states) can be efficiently represented as MPSs \citep{orus2014}.
In general however, as in our case, there is no guarantee that an arbitrary tensor can be approximated by a TT of low rank.

\begin{remark}
    \label{rmk:rank_growth}
    Given any tensor \( \tens{X} \in \Rtens{n} \), there exists a sequence of TTs \( \left\{ \tt{X}^{(n)} \right\}_{n = 1}^\infty \subset \Rtens{n} \) such that 
    \begin{equation}
        \label{eqn:tt_density}
        \norm{\tens{X} - \tt{X}^{(n)}} \to 0 \quad \text{as} \quad n \to \infty \,.
    \end{equation}
    However, the TT rank of \( \tt{X}^{(n)} \) must grow to cover \( \Rtens{n} \) \citep{orus2014}, so there exists \( \tens{X} \in \Rtens{n} \) such that \Cref{eqn:tt_density} holds while 
    \begin{equation}
        \label{eqn:rank_divergence}
        \max_k \left\{r_k^{(n)}\right\} \to D \quad \text{as} \quad n \to \infty \,,
    \end{equation}
    where \( r_k^{(n)} \) for \( k = 0,\, \cdots ,\, d \) are the ranks of \( \tt{X}^{(n)} \), and \( D \) is the dimension of \( \Rtens{n} \).
    This statement can be generalized to arbitrary Hilbert spaces \( \mathbb{H} \) and \( D = \infty \), see \citet{orus2014}.
    In such a context, it can be said that the set of elements of \( \mathbb{H} \) that can be represented by TTs are \emph{dense} in \( \mathbb{H} \).
\end{remark}

The consequence of \Cref{rmk:rank_growth} in the present context is that while we can approximate any given model state by a TT to arbitrary accuracy, it may be that the TT-rank is unacceptably high, to the point that it is more expensive to store the TT approximation than the original full tensor.
To reiterate, the difference between our application and that of quantum materials is that in the quantum materials context, it can be shown that a realistic given state can be represented efficiently as a TT, whereas here we have no such guarantee. The purpose of this work is to investigate this TT compression efficiency for GFD flows with empirical experiments.

\begin{defn}
    \label{defn:compression}
    The tensor train \emph{compression} rate is simply the ratio between the number of elements stored by a full tensor \( \tens{X} \in \Rtens{n} \) and its TT approximation \( \tt{X} \),
    \begin{equation}
        \label{eqn:compression}
        \comp(\tens{X},\, \tt{X}) = \frac{\numel(\tt{X})}{\numel(\tens{X})} \,,
    \end{equation}
    where \( \numel(\cdot) \) denotes the total number of elements stored.
    From the discussion above, we know that
    \begin{equation}
        \label{eqn:compression_order}
        \comp(\tens{X},\, \tt{X}) \sim \Ord{\frac{dr^2}{n^{d-1}}} \,.
    \end{equation}
    For computational performance, we want \( \comp(\tens{X},\, \tt{X}) \) to be close to zero with an acceptable error \( \frob{\tens{X} - \tt{X}} \).
\end{defn}

\subsubsection{Operating on Tensor Trains}
\label{sec:operating_on_tts}

Tensor trains are, similar to vectors, subject to nonlinear point-wise operations and linear transformations.
That is, we can point-wise multiply and add TTs, and also subject TTs to operations analogous to matrix-vector multiplications via tensor train operators.

\begin{defn}
    \label{defn:tto}
    A \emph{tensor train operator} (TTO) is a linear operator \( L_\text{TTO}: \Rtens{n} \to \Rtens{m} \) that maps tensor trains to tensor trains.
    These can be represented as tensors \( \tto{L} \in \mathbb{R}^{m_1 \times n_1 \times m_2 \times n_2 \times \cdots \times m_d \times n_d} \), which in turn are written in a tensor train format similar to that defined in \Cref{defn:tensor_train}, but with 4-dimensional cores of shape \( [r_{k-1},\, m_k,\, n_k,\, r_k] \) for \( k = 1,\, \cdots,\, d \), with \( r_0 = r_d = 1 \).

    A TTO \( \tto{L} \) acts on a TT, \( \tt{X} \in \Rtens{n} \) on a ``core-by-core'' basis.
    If we denote the cores of \( \tto{L} \) by \( \tens{F}_k \in \mathbb{R}^{r_{k-1} \times m_k \times n_k \times r_k} \) and the cores of \( \tt{X} \) by \( \tens{G}_k \in \mathbb{R}^{\tilde{r}_{k-1} \times n_k \times \tilde{r}_k} \), then for each \( k \) the tensor \( \tens{F}_k \) is contracted with \( \tens{G}_k \) along their axis of shared length \( n_k \) to produce a new core \( \tens{H}_k \in \mathbb{R}^{r_{k-1}\tilde{r}_{k-1} \times m_k \times r_k\tilde{r}_k} \).
    Together, the new cores form a new TT, \( \tt{Y} \in \mathbb{R}^{m_1 \times \cdots \times m_d} \).
    For legibility, we write this operation as \( \tto{L} \tt{X} = \tt{Y} \).
\end{defn}

An important property of the TT-format is that standard numerical operations between TTs return a TT.
This means that (E) from \Cref{fig:tt_overview} can take place entirely in the TT format, without ever forming the full tensor of the model state.
However, the rank of the TT returned by such an operation depends on the rank of the operands as shown in \Cref{tbl:tt_operations}.
To combat rank growth during series of TT operations, a type of re-approximation algorithm needs to be applied.

\begin{table}
    \centering
    \begin{tabular}{cc}
         Operation & Rank of result \\
         \hline
         \( \tt{X} + \tt{Y} \) & \( \vec{r}_x + \vec{r}_y \) \\
         \( \tt{X} * \tt{Y} \) & \( \vec{r}_x * \vec{r}_y \) \\
         \( \tto{L} \tt{X} \) & \( \vec{r}_\ell * \vec{r}_x \)
    \end{tabular}
    \caption{
        The ranks of resultants from TT operations for \( \tt{X},\, \tt{Y} \in \Rtens{n} \) and \( \tto{L} \in \mathbb{R}^{m_1 \times n_1 \times m_2 \times n_2 \times \cdots \times m_d \times n_d} \) with ranks \( \vec{r}_x \), \( \vec{r}_y \) and \( \vec{r}_\ell \) respectively. Here, \( * \) denotes an element-wise product.
    }
    \label{tbl:tt_operations}
\end{table}

\begin{defn}
    \label{defn:tt_rounding}
    A TT \emph{rounding} algorithm (also called a re-truncation algorithm) is a method that approximates a given TT by a new TT of lower rank. For a TT, \( \tt{X} \) with ranks \( \vec{r}_x \) and \( \epsilon > 0 \), this is written as 
    \begin{equation}
        \label{eqn:tt_rounding}
        \tt{Y} = \round(\tt{X},\, \epsilon) \,,
    \end{equation}
    where \( \tt{Y} \) has ranks \( \vec{r}_y \), \( \frob{\tt{X} - \tt{Y}} < \epsilon \), and \( \vec{r}_y \leq \vec{r}_x \).
    The cost of TT-rounding is nontrivial and depends on the rank of the input TT; \( \ord{dnr^6} \) \citep{oseledets2011}.
    
    Often it is possible to find \( \tt{Y} \) such that the rank \( \vec{r}_y \) is much smaller than \( \vec{r}_x \).
    The benefit of a rounding algorithm rather than computing a new TT from scratch is that there is no need to first form the full tensor from \( \tt{X} \).
\end{defn}

The most common TT-rounding algorithm is given in \citet[Algorithm 2]{oseledets2011}, which re-compresses an input TT by first orthogonalizing its cores via a series of QR decompositions, then compresses those cores with a series of truncated SVDs.
During a time-integration procedure (e.g. (E) in \Cref{fig:tt_overview}), it is common practice to perform a TT-round on the state tensor after each time-step, or even after each substage of a given time-stepping method.

\subsubsection{Solving PDEs with Tensor Train}
\label{sec:solving_pdes_with_tt}

Here, we provide a example of using TT to solve a simple PDE.
The example uses the background introduced in this section and serves to illustrate more of the details eluded to by \Cref{fig:tt_overview}.
In this example, we only need to apply linear operations in the TT-format, but in general, more complex nonlinear operations can also be performed in the TT-format via so-called ``cross approximation'' algorithms, see e.g. \citet{savostyanov2011,danis2025b}.

\begin{example}
    \label{ex:advection_with_tt}
    Let \( u \) be some quantity of interest and consider the simple advection of \( u \) in one spatial dimension,
    \begin{equation}
        \label{eqn:ex_advect_continuous}
        \diffp{u}{t} = -\diffp{u}{x} \,.
    \end{equation}
    We discretize this equation in space via some given finite-difference or finite-volume method and write the resulting problem as
    \begin{equation}
        \label{eqn:ex_advect_discrete}
        \diffp{\vec{u}}{t} = -\vec{D} \mat{u} \,,
    \end{equation}
    where \( \vec{u} \) is a vector giving the values of \( u \) at a number of discrete grid points and \( \mat{D} \) is a matrix encoding the given discrete form of \( \nicefrac{\partial}{\partial x} \).
    Moving from \Cref{eqn:ex_advect_continuous} to \Cref{eqn:ex_advect_discrete} corresponds to (A) in \Cref{fig:tt_overview}.

    Then, both \( \mat{D} \) and \( \vec{u} \) can be represented by tensor trains; \( \mat{D} \) will become a TTO that will operate on the TT of \( \vec{u} \).
    But first, \( \vec{u} \) must be written as a tensor of dimension \( d \geq 2 \).
    The choice of how to fold the vector \( \vec{u} \) into a higher dimensional tensor depends on the problem; we discuss this for our application in \Cref{sec:tt_on_grids} and \Cref{apx:unstructured_tt}.
    For now, say that we have chosen to fold \( \vec{u} \) into a 2-dimensional tensor \( \tens{U} \in \mathbb{R}^{n \times m} \).
    Then, the appropriate shape for the TTO of \( \mat{D} \) is \( [n,\, n,\, m,\, m] \). 
    This TTO can be computed using a process analogous to TT-SVD \citep[Algorithm 1]{oseledets2011}, or with \Cref{alg:sparse_to_tto} given below; call the resulting TTO, \( \tto{D} \).
    The process of tensorizing \( \vec{u} \) and \( \vec{D} \) corresponds to (C) in \Cref{fig:tt_overview}, and computing their TT representations corresponds to (D).
    Together, all this allows us to write 
    \begin{equation}
        \label{eqn:ex_advect_tt}
        \diffp{\tt{U}}{t} = -\tto{D} \tt{U} \,.
    \end{equation}
    The linear operation \( \tto{D} \tt{U} \) returns a tensor train and is analogous to the matrix-vector multiplication \( \mat{D}\vec{u} \).
    This means that we can solve \Cref{eqn:ex_advect_tt} without leaving the TT format.
    The time derivative can be treated with any given time-integration method. 
    However, as noted above (see \Cref{tbl:tt_operations}), operations between TTs result in rank growth, so TT-rounding needs to be applied at appropriate intervals. 
    As an example, a TT aware version of the Forward Euler time-integration scheme would be given by
    \begin{equation}
        \label{eqn:ex_fe_tt}
        \begin{aligned}
            \tt{U}^{*} &= \tt{U}^n - \Delta t\, \tto{D}\tt{U}^n \\
            \tt{U}^{n + 1} &= \round\left(\tt{U}^{*},\, \epsilon\right) \,,
        \end{aligned}
    \end{equation}
    where \( \epsilon > 0 \) is some user chosen value for the allowable error. 
\end{example}



\section{Methods}
\label{sec:methods}


\subsection{An E3SM-like Shallow Water Model}
\label{sec:shallow_water_model}

\begin{figure}
    \centering
    \includegraphics[width=0.7\linewidth]{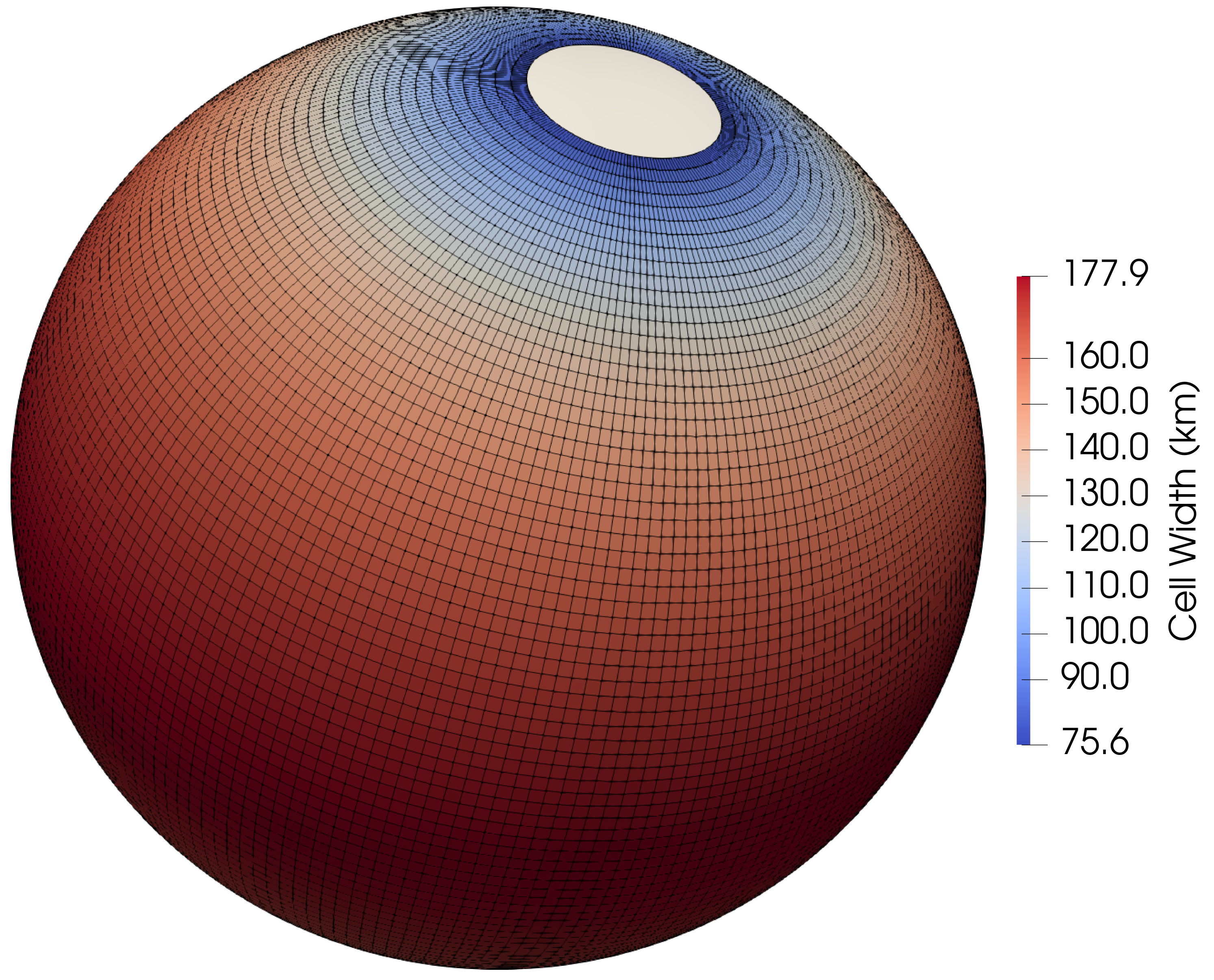}
    \caption{
        A spherical TRiSK domain consisting of quadrilateral cells (QS1.6 from \Cref{tbl:meshes}).
        Cells are colored according to the average width of each cell in km.
    }
    \label{fig:qs1.6}
\end{figure}

The particular shallow water model that we use here, swe-python \citep{swe-python}, is written to use the same TRiSK spatial discretization \citep{ringler2010} and meshes as the ocean component of the US Department of Energy's Energy Exascale Earth System Model (E3SM) \citep{golaz2022}.
This choice was made to evaluate TT methods in a setting as close to that of E3SM as possible.
This model was originally written to run on unstructured, boundary-less, spherical Voronoi grids \citep{ju2011,okabe2017} consisting primarily of hexagons (\Cref{fig:trisk_grid}).
However, the unstructured nature of these types of grids leads to significant difficulties applying tensor train to a tensor representing the state; we will discuss this further in \Cref{sec:tt_on_grids} and  \Cref{apx:unstructured_tt}.
In order to utilize the existing code, and remain in an E3SM-like setting, we have generalized swe-python to accept meshes consisting of quadrilateral cells and mesh boundaries as pictured in \Cref{fig:qs1.6}.
At the boundary, we enforce a no-flow, no-slip boundary condition, which means that the normal velocity \( \vec{u} \), as well as the tangential velocity \( \vec{u}^\perp \) is zero.
This gives us a structured grid on which we can apply our TT methods.

As the performance of TT methods can heavily depend the problem size, we will perform experiments as we vary mesh resolution, using a total of five meshes that we refer to as the prefix QS, meaning ``Quadrilateral Sphere,'' followed by the resolution of the mesh in degrees latitude and longitude.
\Cref{tbl:meshes} give the resolutions and sizes of each of the meshes used.
The resolutions are chosen so that moving from one mesh to the next in resolution increases the problem size by a factor of four.

\begin{table}
    \centering
    \begin{tabular}{c|c|c|c}
        Mesh Name & Number of Cells & Number of Edges & Number of Vertices \\
        \hline
        QS1.6 & 22,500 & 45,225 & 22,725 \\
        QS0.8 & 90,000 & 180,450 & 90,450 \\
        QS0.4 & 360,000 & 720,900 & 360,900 \\
        QS0.2 & 1,440,000 & 2,881,800 & 1,441,800 \\
        QS0.1 & 5,760,000 & 11,523,600 & 5,763,600
    \end{tabular}
    \caption{
        Meshes used in this work.
        The number following QS in each mesh refers to the resolution in degrees of latitude and longitude used by each mesh.
        To avoid a mesh singularity at the north and sough poles, the meshes have boundaries at 80\textdegree\, north and 80\textdegree\, south.
        All domains have a radius of 6371220 m.
    }
    \label{tbl:meshes}
\end{table}

\subsubsection{Discrete Spatial Operators}
\label{sec:discrete_spatial_operators}

An important aspect of the swe-python code is that all discrete spatial operations are implemented as sparse matrix-vector multiplications (SpMV) using SciPy \citep{scipy}, which has two primary benefits.
First, SciPy uses a custom SpMV kernel written in C, which is significantly faster than Just In Time (JIT) compiled loops over spatial operator stencils, leading to improved computational performance comparable to that of HPC codes\footnote{
    By this, we mean that the SciPy SpMV kernel is expected to perform comparably to a similar kernel in a HPC code, not that swe-python's performance, as a single threaded Python application, is itself comparable to that of a HPC code.
}.
Second, as the TRiSK spatial operators are represented as matrices (2-dimensional tensors), they can be directly converted to TTOs of whatever appropriate shape.

As swe-python, and the ocean component of E3SM, implement only explicit time-stepping schemes, this leads to straightforward implementation of spatial operations in the TT format -- the right-hand-side terms of a spatially discretized PDE can be evaluated by a combination of nonlinear element-wise operations and linear transformations.
\Cref{ex:advection_with_tt} provides a brief example of how this can be written.

\subsubsection{Tensor Train on Structured and Unstructured Grids}
\label{sec:tt_on_grids}

As alluded to above, the unstructured, variable resolution Voronoi grids normally used with TRiSK (\Cref{fig:trisk_grid}) are not conducive to favorable TT-compression.
We will discuss this further in \Cref{apx:unstructured_tt}, but will summarize the relevant issues here for completeness of the main text.

The tensor train decomposition of a tensor can be thought of as a separation of variables over the modes of the tensor. 
In the case of structured grids of any given dimension \( d \), this is most commonly realized as a separation of variables over non-temporal dimensions.
For example, a tensor \( \tens{X} \in \mathbb{R}^{n_1 \times n_2 \times n_3} \) might represent a model state in a 3-dimensional domain, with the first, second, and third tensor modes corresponding to spatial coordinates \( x \), \( y \), and \( z \) respectively.
Using the notation of \Cref{defn:tensor_train}, we could then approximate \( \tens{X} \) in the TT-format as
\begin{equation}
    \label{eqn:seperation_of_vars_example}
    \tt{X}(i_1,\, i_2,\, i_3) = \mat{G}_1(i_1) \
                                \mat{G}_2(i_2) \
                                \mat{G}_3(i_3) \,.
\end{equation}
In this context, each core \( \tens{G}_1 \), \( \tens{G}_2 \), and \( \tens{G}_3 \) corresponds to the \( x \), \( y \), and \( z \) spatial dimensions respectively, and the size of rank \( \vec{r} = [1,\, r_1,\, r_2,\, 1] \) encodes how effectively the model state can be represented in this separated state; low rank means that the state is approximately separable, while high rank means it is not, requiring more interaction between the cores to resolve the state.

Applying this strategy to our set of structured 2-dimensional meshes (e.g. \Cref{fig:qs1.6}) is straightforward.
Let \( \nlat \) be the number of grid points in the latitudinal direction and let \( \nlon \) be the number of grid cells in the longitudinal direction.
The thickness variable \( h \) is realized as a 2-dimensional tensor \( \tens{H} \in \mathbb{R}^{\nlat \times \nlon} \), where the data is arranged so that the first mode corresponds to each cell's latitude and the second mode corresponds to each cell's longitude. 
The velocity variable \( \vec{u} = (u,\, v,\, 0) \) is realized as a 2-dimensional tensor \( \tens{U} \in \mathbb{R}^{(2\nlat + 1) \times \nlon} \), where \( u \) and \( v \) are the fluid velocities in the longitudinal and latitudinal directions respectively.
The data for \( \vec{u} \) is arranged similarly to that for \( h \) so that the first mode corresponds to each edges' latitude and the second mode corresponds to each edges' longitude; this has the effect that the rows of \( \tens{U} \) alternate between values for \( v \) and values for \( u \).
Then, when \( \tens{H} \) and \( \tens{U} \) are approximated by TTs, \( \tt{H} \) and \( \tt{U} \). 
These can be viewed as separating variables over latitude and longitude similar to \Cref{eqn:seperation_of_vars_example}.

The problem with applying TT methods to an unstructured Voronoi mesh as in \Cref{fig:trisk_grid} is that the data for both \( h \) and \( \vec{u} \) must be represented by a tensor in a way that some separation of variables can be taken advantage of.
The unstructured, variable resolution grids used in E3SM do not have an obvious mapping to a tensor space that preserves the structure of the data in a way that allows for an advantageous splitting of variables. 
There are several ways that this problem could be dealt with: Cleverly sorting and mapping the grid points, re-gridding the data, or using a different tensor decomposition method that is possibly more amenable to an unstructured mesh.
However, each of these strategies introduce their own set of challenges; we discuss further in \Cref{apx:unstructured_tt}.


\subsection{Working with Sparse Tensor Train Operators}
\label{sec:sparse_ttos}

As described in \Cref{sec:discrete_spatial_operators}, the TRiSK spatial operators are treated as sparse matrices which are then converted to TTOs.
It is possible to achieve this via standard algorithms such as those presented in \citet{oseledets2011}, but most methods and implementations assume that the input tensor is dense.
This is fine for small problem sizes, but as the problem size grows, our TRiSK operators can only be reasonably stored and manipulated in a sparse representation.

To facilitate the rest of our work, we have implemented a collection of TT algorithms from \citet{li2022} meant for working with sparse input tensors.
Our Python implementation uses PyData Sparse \citep{sparse}, which generalizes SciPy sparse arrays to arbitrary dimensions.
Our code is available on GitLab \citep{sparsett}.
We will briefly describe the requisite algorithms from \citet{li2022} before describing our approach to obtain a TTO from a sparse matrix operator in \Cref{alg:sparse_to_tto}.

\begin{algorithm}
    \SetAlgoRefName{Li-4}
    \SetAlgoRefRelativeSize{1} 
    \caption{Compute an exact TT that has sparse cores \citep[Algorithm 4]{li2022}.}
    \label{alg:li4}
    \KwIn{%
        A sparse tensor \( \tens{A} \in \Rtens{n} \) and an integer \( p \) with \( 1 \leq p \leq d \).
    }
    \KwOut{%
        A TT-format tensor \( \tt{A} \) with sparse cores that is equivalent to \( \tens{A} \), possibly with large TT-rank.
        All cores of \( \tt{A} \), except for the \( p \)-th core, consist of only ones and zeros. 
    }
\end{algorithm}

\begin{algorithm}
    \SetAlgoRefName{Li-6}
    \SetAlgoRefRelativeSize{1} 
    \caption{Efficient TT-rounding \citep[Algorithm 6]{li2022}.}
    \label{alg:li6}
    \KwIn{%
        A TT, \( \tt{A} \in \Rtens{n} \), the desired error \( \epsilon \), and an integer \( p \) with \( 1 \leq p \leq d \).
        In the case that \( \tt{A} \) was computed using \Cref{alg:li4}, \( p \) should be chosen to be the same as in that process for the best performance.
    }
    \KwOut{%
        A TT, \( \tt{B} \) approximating \( \tt{A} \) such that \( \frob{\tt{A} - \tt{B}} < \epsilon \frob{\tt{A}} \).
    }
\end{algorithm}

\begin{algorithm}
    \caption{%
        Compute a TTO representation of a sparse matrix operator.
    }
    \label{alg:sparse_to_tto}
    \KwIn{%
        A sparse matrix operator \( \mat{A} \in \mathbb{R}^{m \times n} \) for \( m,\, n \in \mathbb{N} \), the desired accuracy \( \epsilon > 0 \), and \( m_1,\, n_1,\, \cdots,\, m_d,\, n_d \in \mathbb{N} \) giving the desired shape of the TTO \( [m_1,\, n_1,\, \cdots,\, m_d,\, n_d] \) that will act as a map \( \Rtens{n} \to \Rtens{m} \).
    }
    \KwOut{%
        \( \tto{A} \in \mathbb{R}^{m_1 \times n_1 \times \cdots \times m_d \times n_d} \) is a TTO approximating the action of \( \mat{A} \) to accuracy \( \epsilon \).
    }
    Reshape \( \mat{A} \) into a tensor \( \tens{A} \) with shape \( [m_1,\, m_2,\, \cdots,\, m_d,\, n_1,\, n_2,\, \cdots,\, n_d] \).

    Permute the modes of \( \tens{A} \) such that its new shape is \( [m_1,\, n_1,\, m_2,\, n_2,\, \cdots,\, m_d,\, n_d] \).

    Reshape \( \tens{A} \) to have shape \( [m_1 n_1,\, m_2 n_2,\, \cdots,\, m_d n_d] \).
    
    Apply \Cref{alg:li4} to \( \tens{A} \) with \( p = \lfloor\nicefrac{d}{2}\rfloor \) to obtain a TT, \( \tt{A} \) with sparse cores.
    
    Apply \Cref{alg:li6} to \( \tt{A} \) with \( p = \lfloor\nicefrac{d}{2}\rfloor \) and desired accuracy \( \epsilon \) to obtain an approximate TT, \( \tt{B} \) with cores \( \tens{G}_k \) for \( k = 1,\, \cdots,\, d \), each of which has shape \( [r_{k-1},\, m_k n_k,\, r_k] \).
    
    \For{\( k = 1,\, \cdots,\, d \)}{%
        Permute the modes of \( \tens{G}_k \) such that its new shape is \( [m_k n_k,\, r_{k-1},\, r_k] \).

        Reshape \( \tens{G}_k \) to have shape \( [m_k,\, n_k,\, r_{k-1},\, r_k] \).

        Permute the modes of \( \tens{G}_k \) such that its new shape is \( [r_{k-1},\, m_k,\, n_k,\, r_k] \).
    }

    \Return{%
        The cores \( \tens{G}_k \) for \( k = 1,\, \cdots,\, d \), which form a TTO \( \tto{A} \in \mathbb{R}^{m_1 \times n_1 \times \cdots \times m_d \times n_d} \).
    }
\end{algorithm}

\begin{example}
    \label{ex:qs_tt_div}
    Using \Cref{alg:sparse_to_tto}, we can convert an arbitrary sparse matrix operator, and therefore any linear operator, to the TTO format.
    To give a concrete example that is used heavily in \Cref{sec:experiments}, consider an arbitrary QS grid with \( \nlat \) grid points in the latitudinal direction, and \( \nlon \) grid points in the longitudinal direction.
    Let \( \mat{D} \) be the TRiSK divergence operator, which maps the fluid velocity \( \vec{u} \) on grid edges to cell-centered values for the divergence. 
    The vector \( \vec{u} \) has length \( (2\nlat + 1)\nlon \) and \( \mat{D} \) is a sparse matrix of shape \( [\nlat\nlon,\, (2\nlat + 1)\nlon] \).
    As described in \Cref{sec:tt_on_grids}, the fluid velocity can then be represented as a TT, \( \tt{U} \in \mathbb{R}^{(2\nlat + 1) \times \nlon} \).
    We apply \Cref{alg:sparse_to_tto} to the matrix \( \mat{D} \) to obtain a TTO, \( \tto{D} \in \mathbb{R}^{\nlat \times (2\nlat + 1) \times \nlon \times \nlon} \).
    Then, we can write
    \begin{equation}
        \label{eqn:qs_tt_div}
        \tto{D} \tt{U} = \tt{V} \,,
    \end{equation}
    where the result \( \tt{V} \in \mathbb{R}^{\nlat \times \nlon} \) is a TT giving the cell-centered values of the divergence of the fluid velocity.
    We can write an expression similar to that in \Cref{eqn:qs_tt_div} for all TRiSK spatial operators.
\end{example}



\section{Experiments}
\label{sec:experiments}


\subsection{Experiment Design}
\label{sec:experiment_design}

The purpose of the experiments that follow are to evaluate the performance of TT-format operations in context of geophysical fluid dynamics (GFD).
We are primarily concerned with two metrics.
First is the TT-compression (\Cref{defn:compression}) and how this relates to the complexity of the model state in time.
Second is the speedup achieved by the TT-format operations versus standard (non-TT) methods -- specifically as compared to an optimized sparse matrix-vector multiplication kernel.

\subsubsection{Timing Linear Operators}
\label{sec:timing_linear_operators}

For practical reasons that will become clear in \Cref{sec:results}, we have opted to perform our TT performance experiments in an ``offline'' fashion.
By this, we mean that we have opted not to implement a fully TT-enabled version of swe-python that solves the SWEs in the TT-format.
Instead, we solve the SWEs using standard methods while writing the model state to disk at intervals.
Then, we apply our TT-format operations to the model state at each saved time-level.
This allows us to quickly test different TT configurations without the need to run the model from scratch each time.
It has been shown in \citet{danis2025b} that TT-methods can successfully maintain the formal order of accuracy of the methods on which they are based, in both space and time.
\emph{Therefore, our concern is not whether these types of methods can be used to accurately solve the SWEs, but rather if they can achieve a significant speedup over standard methods and therefore be used in real-world application HPC codes.} 

In swe-python, each time-step requires a series of linear transformations on the model state and a small number of element-wise nonlinear operations.
As such, the majority of the computational cost comes from operations of the form of large matrix-vector multiplications.
It is therefore sufficient to test the performance of our TT methods for these types of operations, which is then representative of the model's overall computational performance.

In \Cref{sec:results}, we present timing and error results of taking the divergence of the model velocity field through various methods; sparse matrix-vector multiplication (SpMV), a simple Just In Time compiled Python loop over the operator stencil (JITLoop), and the TT-format linear transformation described in  \Cref{ex:qs_tt_div} (TTMV).
Our choice to use the divergence operator as a representative operation is essentially arbitrary since, again, all linear operations are given in the form of matrix-vector multiplications.
However, we were motivated to choose the divergence operator as it is a map between edge-centered and cell-centered quantities, and so interacts with both grid locations that hold prognostic variable quantities.
The SpMV case is representative of an optimized, high performance implementation, while the JITLoop case is representative of a more na\"{i}ve implementation -- the comparison to both helps to more clearly contextualize the performance of our TT-methods.

As discussed previously, the SpMV kernel is supplied by SciPy \citep{scipy}.
Additionally, we use Numba \citep{numba} for the JIT compilation of JITLoop and torchTT \citep{torchtt} for TT operations, which in turn uses PyTorch \citep{pytorch} as a backend.
All experiments were carried out on a Linux machine with two Intel Xeon Silver 4215R CPUs, two Nvidia Quadro RTX 8000 GPUs, and 64 GB of memory.
This machine was provided by the Center for Nonlinear Studies (CNLS) at Los Alamos National Laboratory (LANL), and has been affectionately dubbed the Computer for Nonlinear Studies. 
For consistency across the various Python libraries used in this work, no attempt at parallelization was made within swe-python or the deployed tensor train libraries; Python was run as default in serial.
Also, we note that results reported below using this machine's GPU were obtained by running on a single GPU.
GPU operations were performed with PyTorch as backend, and timed as shown in \Cref{lst:gpu_timing}.
Finally, we note that when timing all methods, SpMV, JITLoop, and TTMV, the reported timings are the average of 10 total runs, with one additional ``warm-up'' run at the beginning which is not counted in the average.

\begin{lstfloat}
    \begin{lstlisting}[language=Python, style=in]
# send data to GPU in advance of operations
tto_gpu = tto.to('cuda')
state_tt_gpu = state_tt.to('cuda')

# create cuda event trackers
start = torch.cuda.Event(enable_timing=True)
end = torch.cuda.Event(enable_timing=True)

# perform and time TT matrix-vector multiplication
start.record()
result_tt = tto_gpu @ state_tt_gpu
end.record()
torch.cuda.synchronize()

# record time, converting milliseconds to seconds
matvec_time = start.elapsed_time(end) / 1000

# perform and time TT-rounding
start.record()
result_tt = result_tt.round(eps=dynamic_eps)
end.record()
torch.cuda.synchronize()

# record time, converting milliseconds to seconds
round_time = start.elapsed_time(end) / 1000
    \end{lstlisting}
    \caption{
        GPU timing for TT operations using torchTT and PyTorch.
    }
    \label{lst:gpu_timing}
\end{lstfloat}

\subsubsection{Choosing \texorpdfstring{\( \varepsilon \)}{Epsilon}}
\label{sec:choosing_eps}

The computational efficiency of the TTMV as compared to that of the SpMV depends on the rank of the state TT being operated on, which in turn depends on the chosen TT approximation error \( \epsilon \) (e.g. see \Cref{eqn:tt_approximation}).
In our experiments, we choose \( \epsilon \) in a way that would guarantee the preservation of formal order of accuracy (if the TT-methods were being applied in the full model), as described in \citet[Section 3d]{danis2025b};
\begin{equation}
    \label{eqn:dynamic_eps}
    \epsilon = \min\left(10^{-3},\, C_\epsilon \frac{V^{\nicefrac{1}{2}} \Delta x^{\ell - \nicefrac{1}{2}}}{\max_{\vec{q} \in \{\vec{h},\vec{u}\}}{\frob{\vec{q}}}}\right) \,,
\end{equation}
where \( V \) is the volume of the domain, \( \ell \) is the spatial order of accuracy of the spatial discretization (2 in our case), \( \vec{h} \) and \( \vec{u} \) are vectors of the thickness and velocity model state, and \( C_\epsilon \) is a problem dependent variable discussed below.

The presence of \( C_\epsilon \), which must be chosen by the user empirically, creates some difficulty in determining a suitable value for \( \epsilon \).
Ideally, we would choose \( C_\epsilon \) so that the error induced by the TT-approximation of the state was less than the temporal and spatial truncation errors, but these cannot be known \emph{a priori}, so it typically needs to be hand-tuned.
Large \( C_\epsilon \) will allow the TT-approximation to achieve low rank and therefore high computational efficiency, but may not represent the model state with sufficient accuracy.
Small \( C_\epsilon \) can assure high accuracy, but can lead to very high TT-rank.
\citet{danis2025b} suggests choosing \( C_\epsilon \) via a process analogous to a grid-independence study, where \( C_\epsilon \) is taken lower and lower until it no longer has an effect on the model solution accuracy.
As we are not solving the full SWEs in the TT-format, we instead choose a fixed \( C_\epsilon = 1 \) for all test cases, which was observed by \citet{danis2025b} to be sufficient in the majority of cases.
We will share results for other choices of \( C_\epsilon \) in \Cref{apx:additional_results} for completeness.


\subsection{Test Cases}
\label{sec:test_cases}

To investigate the performance of TT-format operations across a range of problems relevant to GFD, we consider a suite four test cases of increasing complexity.
All test cases take place on a rotating sphere with angular velocity \( \Omega = 7.292\times10^{-5} \) s\(^{-1}\) and a radius of 6371.22 km.
The sea-floor topography is uniform unless otherwise specified.
Snapshots of the final states for each test case are given in \Cref{fig:compression_summary}.

\begin{itemize}
    \item[] \textbf{Quasi-Linear Gravity Wave (QLW)}: A simple gravity wave traveling from 0\textdegree\ latitude, to the north/south boundaries, then back again, a process which takes approximately seven days of simulated time.
    This is the case of an external wave mode, also called the Lamb mode in the context of atmospheric modeling.
    The momentum advection terms in the momentum component of \Cref{eqn:nonlinear_swe} are turned off, making the momentum equation purely linear.
    The thickness equation component of \Cref{eqn:nonlinear_swe} is formally nonlinear, but because the fluid depth is sufficiently large compared to any perturbation in the thickness, this behaves as though it is linear.
    Therefore, we refer to this test case as quasi-linear; using the same notation as \Cref{eqn:nonlinear_swe}, the model equations are given by
    \begin{equation}
    \begin{aligned}
        \diffp{\vec{u}}{t}& = - g\nabla (h + z_b) \\
        \diffp{h}{t}& + \nabla \cdot \left(h\vec{u}\right) = 0 \,.
    \end{aligned}
    \label{eqn:quasilinear_swe}
    \end{equation}
    The fluid velocity is initialized to zero, while the thickness is initialized as a Gaussian bell curve centered at 0\textdegree\ latitude of the form \( e^{-100x^2 - 100y^2} + 500 \).
    The case runs for seven simulated days.

    \item[] \textbf{Geophysical Gravity Wave (GGW)}: The same initial conditions as QLW, but with the full nonlinear SWEs.
    The flow exhibits similar behavior, but the final state, influenced by the nonlinear momentum advection and Coriolis terms, is more complex.
    The case runs for seven simulated days.

    \item[] \textbf{Barotropically Unstable Jet (BUJ):} Originally formulated by \citet{galewsky2004}, this case consists of a strong zonal flow at mid-latitude with a small perturbation that causes the barotropic jet to become unstable. 
    This flow is driven by strongly nonlinear momentum advection tendencies, physical processes that are of particular relevance to weather prediction and climate modeling.
    The case runs for six simulated days. 

    \item[] \textbf{Williamson Test Case 5 (WTC5)}: A zonal flow over an isolated mountain, one of the standard \citet{williamson1992} SWE test cases.
    Here, the flow is initialized to a steady state assuming uniform bottom topography, but the bottom topography is then altered to introduce an isolated mountain at mid-latitude.
    This perturbation alters the geostrophic balance of the flow, triggering the development of advection-driven instabilities.
    The case runs for 50 simulated days.
\end{itemize}


\subsection{Results}
\label{sec:results}

\def\compressionscale{0.95}
\def\diagnosticscale{0.6}
\def\compressionceps{1}
\def\matvecceps{\compressionceps}
\def\matvecsubscale{0.45}

Here we present the results of TT-method performance experiments on the four test cases of increasing complexity described in \Cref{sec:test_cases}.
As discussed previously, we present the results of two primary metrics; first, the TT-compression of both the layer thickness and velocity variables in time, assuming a dynamically chosen \( \varepsilon \) as in \Cref{eqn:dynamic_eps}, and second, the speedup achieved by TT-methods versus an optimized SpMV kernel of performing a representative linear transformation on the model state data (in this case, taking the divergence of the velocity field).
As TT-rounding is an important contributor to the overall performance of a TT-method, a rounding step is accounted for in the timing results.
We additionally give some results not accounting for the rounding step which are clearly marked.

When comparing a TT-obtained solution to the reference solution given by standard methods, we use a normalized Frobenius error, defined as
\begin{equation}
    \label{eqn:norm_frob}
    E_\text{NF} = \frac{\frob{\vec{s} - \vec{t}}}{\frob{\vec{s}}},
\end{equation}
where \( \vec{s} \) is the reference solution, \( \vec{t} \) is the TT-method solution.

Finally, to connect the TT-compression to the complexity of the fluid flow through a metric relevant to geophysical fluids, we also compute the total eddy kinetic energy (EKE) in time on QS0.2.
EKE can be formally defined in different ways, but is generally understood to be the kinetic energy of a perturbation to the mean fluid flow.
Here, we use 
\begin{equation}
    \label{eqn:eke}
    K_\text{EKE} = \frac{1}{2} \left( (\bar{u} - u)^2 + (\bar{v} - v)^2 \right) \,,
\end{equation}
where \( \vec{u} = (u,\, v,\, 0) \) is the fluid velocity and \( \bar{\, \cdot\,} \) denotes an average in time equal to the length of each test case.
We tested different formulations for EKE, but got qualitatively similar answered for the cases considered here.
Then, as we plot the total EKE, we simply integrate \( K_\text{EKE} \) over the spherical spatial domain \( S \) and normalize by the volume of \( S \),
\begin{equation}
    \label{eqn:total_eke}
    T_\text{EKE} = \frac{1}{\vol(S)} \int_S K_\text{EKE} \dif A \,.
\end{equation}
We only plot the EKE on the QS0.2 because the EKE is qualitatively similar on each mesh, and this is the highest resolution mesh by all test cases. 
The only difference between resolutions is the order of magnitude of the EKE, which increases with resolution.

The purpose of our calculation of the total EKE is to provide a diagnostic quantity common to GFD applications that we can compare to the TT-rank over the course of each simulation.
Below, we observe that this is a potentially useful metric that may suggest whether certain flows can be well compressed with TT; in our results, high total EKE corresponds to high TT-rank.
However, we note that it is not true in general that high EKE implies high TT-rank, or vice-versa. 
One can easily contrive examples where a given state exhibits high EKE and low TT-rank, as well as examples with low EKE and high TT-rank.
For example, the initial state of the BUJ case has low TT-rank; if we were to run this case without the thickness perturbation so the barotropic jet was stable for all time, but we suddenly forced the jet to the south during the final hours of the simulation, the resulting flow would have the same low TT-rank but high EKE.  
As such, we include EKE results and plots not to make a general claim connecting high EKE and high TT-rank, but instead to suggest that it may provide a valuable diagnostic for a certain class of flows.

\subsubsection{Quasi-Linear Wave}
\label{sec:qlw}

The first and simplest of our four test cases.
The momentum advection and Coriolis terms are turned off, leaving us with a quasi-linear system; the model state in QLW is entirely characterized by a small number of superimposed gravity waves.
The case starts with a single gravity wave, but more are generated when the original collides with the domain boundaries at the north and south poles, resulting in a partial reflection back towards the original starting point.

\begin{figure}
    \centering
    \includegraphics[width=\compressionscale\linewidth]{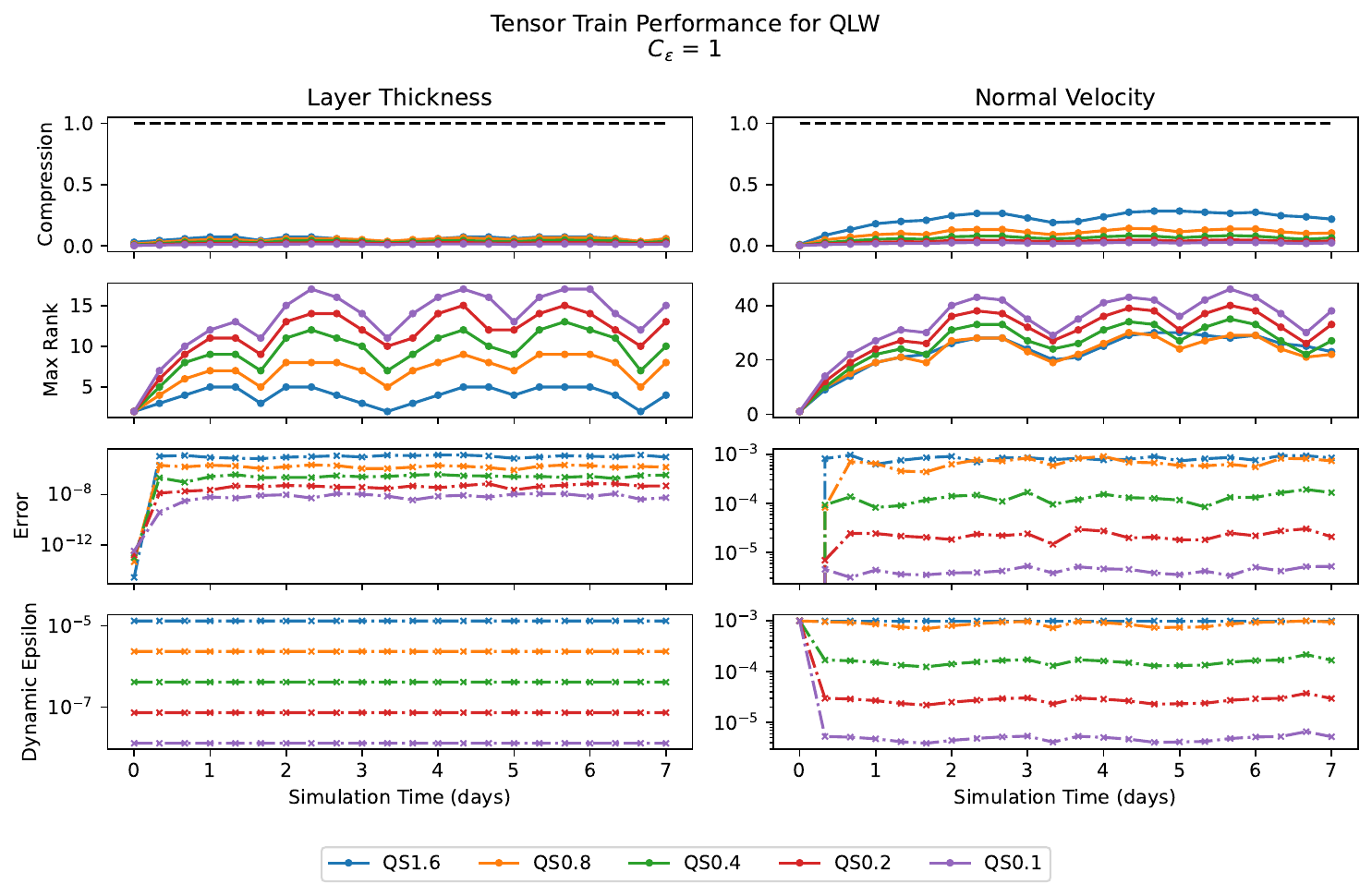}
    \caption{
        TT-compression results for QLW with \( C_\epsilon = \compressionceps \).
    }
    \label{fig:compression_qlw_Ceps\compressionceps}
\end{figure}

\Cref{fig:compression_qlw_Ceps\compressionceps} shows that both the TT of the layer thickness and normal velocity data remain well compressed for the entirety of simulation.
Further, as expected, the compression rate improves as the mesh resolution increases.
Of course, the overall rank of the TT approximations to these data both increase with resolution, but this is simply explained by the overall problem size increasing by a factor of four between each of our meshes.
As can be seen in both the compression curves and rank curves, there is a clear structure where the curves rise and dip over the course of the QLW simulation.
The four troughs occur with when the gravity wave first contacts the north/south pole boarder, when the unreflected portion of the wave coalesces with itself on the opposite side of the globe, when it contacts the boarder again, and finally when the wave returns to its starting position. 

Again as expected, the TT-error (as defined in \Cref{eqn:norm_frob}) improves with mesh resolution as the dynamic \( \epsilon \) is restricted to account for theoretically smaller spatial truncation errors.
We note that for the normal velocity data on QS1.6 and QS0.8, the dynamically chosen \( \epsilon \) defaults to around \( 10^{-3} \) as defined in \Cref{eqn:dynamic_eps}.
This is explained by the fact that the grid resolution in both cases is simply large enough that the maximum allowed \( \epsilon \) is used.
Compare this to \Cref{fig:compression_qlw_Ceps0.1} where we take \( C_\epsilon = 0.1 \) and see that the \( \epsilon \) chosen for the two meshes is in fact different.

\begin{figure}
    \centering
    \includegraphics[width=\diagnosticscale\linewidth]{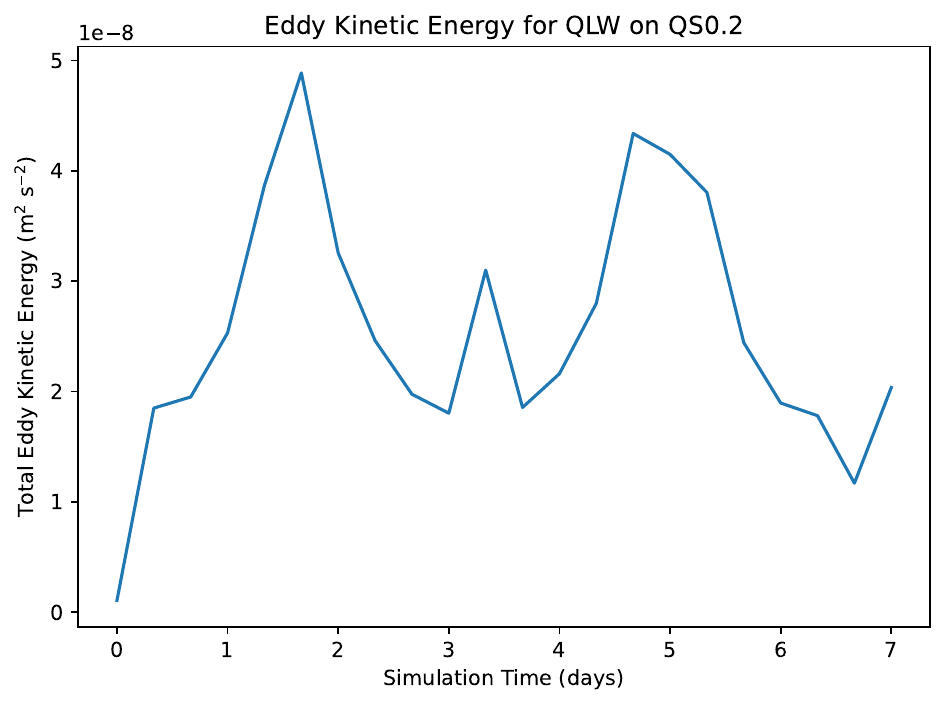}
    \caption{
        Total eddy kinetic energy for QLW on QS0.2.
    }
    \label{fig:diagnostic_qlw_qs0.2}
\end{figure}

\Cref{fig:diagnostic_qlw_qs0.2} shows the total EKE over the course of the QLW case.
We see that the energy of the perturbation on the mean flow remains very small for the entirety of the simulation (note that the scale of the vertical axis is \( 10^{-7} \)).
In particular, compare this to similar plots for BUJ and WTC5 in \Cref{fig:diagnostic_buj_qs0.2,fig:diagnostic_wtc5_qs0.2} respectively.

\begin{figure}
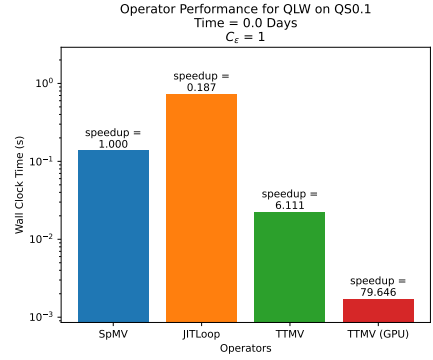
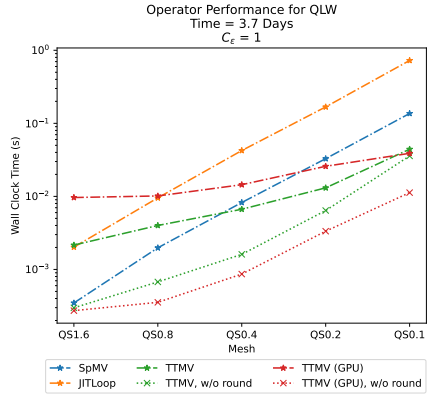
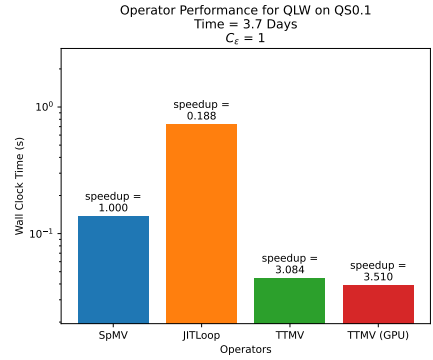
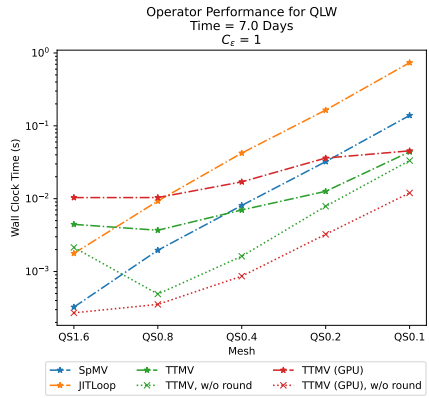
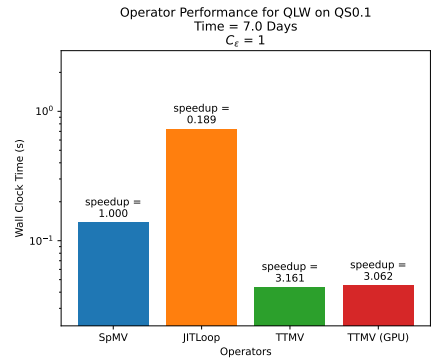

    \centering
    \begin{subfigure}{\matvecsubscale\linewidth}
        \includegraphics[width=\linewidth]{matvec/matvec_qlw_timeind0_Ceps\matvecceps.pdf}
        \caption{~}
        \label{fig:matvec_qlw_timeind0_Ceps\matvecceps}
    \end{subfigure}
    \hfill
    \begin{subfigure}{\matvecsubscale\linewidth}
        \includegraphics[width=\linewidth]{matvec/matvec_bar_qlw_qs0.1_timeind0_Ceps\matvecceps.pdf}
        \caption{~}
        \label{fig:matvec_bar_qlw_qs0.1_timeind0_Ceps\matvecceps}
    \end{subfigure}

    \medskip
    
    \begin{subfigure}{\matvecsubscale\linewidth}
        \includegraphics[width=\linewidth]{matvec/matvec_qlw_timeind11_Ceps\matvecceps.pdf}
        \caption{~}
        \label{fig:matvec_qlw_timeind11_Ceps\matvecceps}
    \end{subfigure}
    \hfill
    \begin{subfigure}{\matvecsubscale\linewidth}
        \includegraphics[width=\linewidth]{matvec/matvec_bar_qlw_qs0.1_timeind11_Ceps\matvecceps.pdf}
        \caption{~}
        \label{fig:matvec_bar_qlw_qs0.1_timeind11_Ceps\matvecceps}
    \end{subfigure}

    \medskip
    
    \begin{subfigure}{\matvecsubscale\linewidth}
        \includegraphics[width=\linewidth]{matvec/matvec_qlw_timeind21_Ceps\matvecceps.pdf}
        \caption{~}
        \label{fig:matvec_qlw_timeind21_Ceps\matvecceps}
    \end{subfigure}
    \hfill
    \begin{subfigure}{\matvecsubscale\linewidth}
        \includegraphics[width=\linewidth]{matvec/matvec_bar_qlw_qs0.1_timeind21_Ceps\matvecceps.pdf}
        \caption{~}
        \label{fig:matvec_bar_qlw_qs0.1_timeind21_Ceps\matvecceps}
    \end{subfigure}
    \caption{
        TT-format operation performance in the QLW test case.
    }
    \label{fig:matvec_qlw_\matvecceps}
\end{figure}

\begin{figure}
    \centering
    \includegraphics[width=\diagnosticscale\linewidth]{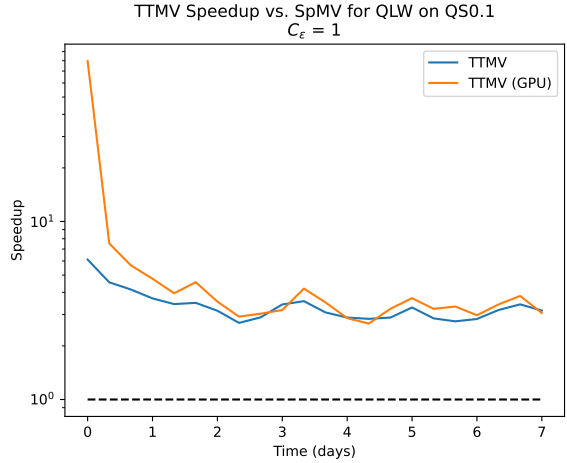}
    \caption{
        TTMV speedup versus SpMV in time for the QLW test case on QS0.1.
    }
    \label{fig:matvec_intime_qlw_\matvecceps}
\end{figure}

Next, \Cref{fig:matvec_qlw_\matvecceps} shows the speedup for our TT-methods at three different points in the simulation; at the beginning, approximately half-way through, and at the end.
This is done to show the speedup achieved at different levels of TT-compression/rank.
An overview of the speedup over the course of the entire simulation on QS0.1 is given in \Cref{fig:matvec_intime_qlw_\matvecceps}.
QS0.1 was chosen as a representative mesh because it is by far the largest problem, where the TT-methods perform best.

Throughout the simulation, TTMV outperforms SpMV by a factor of approximately three, except at the very beginning of the simulation, where the fluid velocity is initialized to zero so the TT-approximation to the velocity data is trivial (\Cref{fig:matvec_qlw_timeind0_Ceps\matvecceps,fig:matvec_bar_qlw_qs0.1_timeind0_Ceps\matvecceps}).
Both TTMV and SpMV outperform JITLoop, which serves as a worst-case point of comparison.
The speedup obtained by TTMV as compared to JITLoop is approximately \( \nicefrac{3}{0.18} = 16.\overline{66} \).
This serves to demonstrate that the implementation to which we compare TT-methods is of the utmost importance when considering possible speedups.

\FloatBarrier

\subsubsection{Geophysical Gravity Wave}
\label{sec:ggw}

This second test case is initialized the same as QLW, except uses the full nonlinear SWEs.
This serves to show that simply adding nonlinear dynamics to the model does not necessarily cause issues with TT-performance.
The case behaves similarly to QLW, but the gravity waves spawned from the initial Gaussian perturbation to the layer thickness and the subsequent collision with the mesh boundaries are subject to nonlinear effects due to momentum advection and the rotation of the sphere.

\begin{figure}
    \centering
    \includegraphics[width=\compressionscale\linewidth]{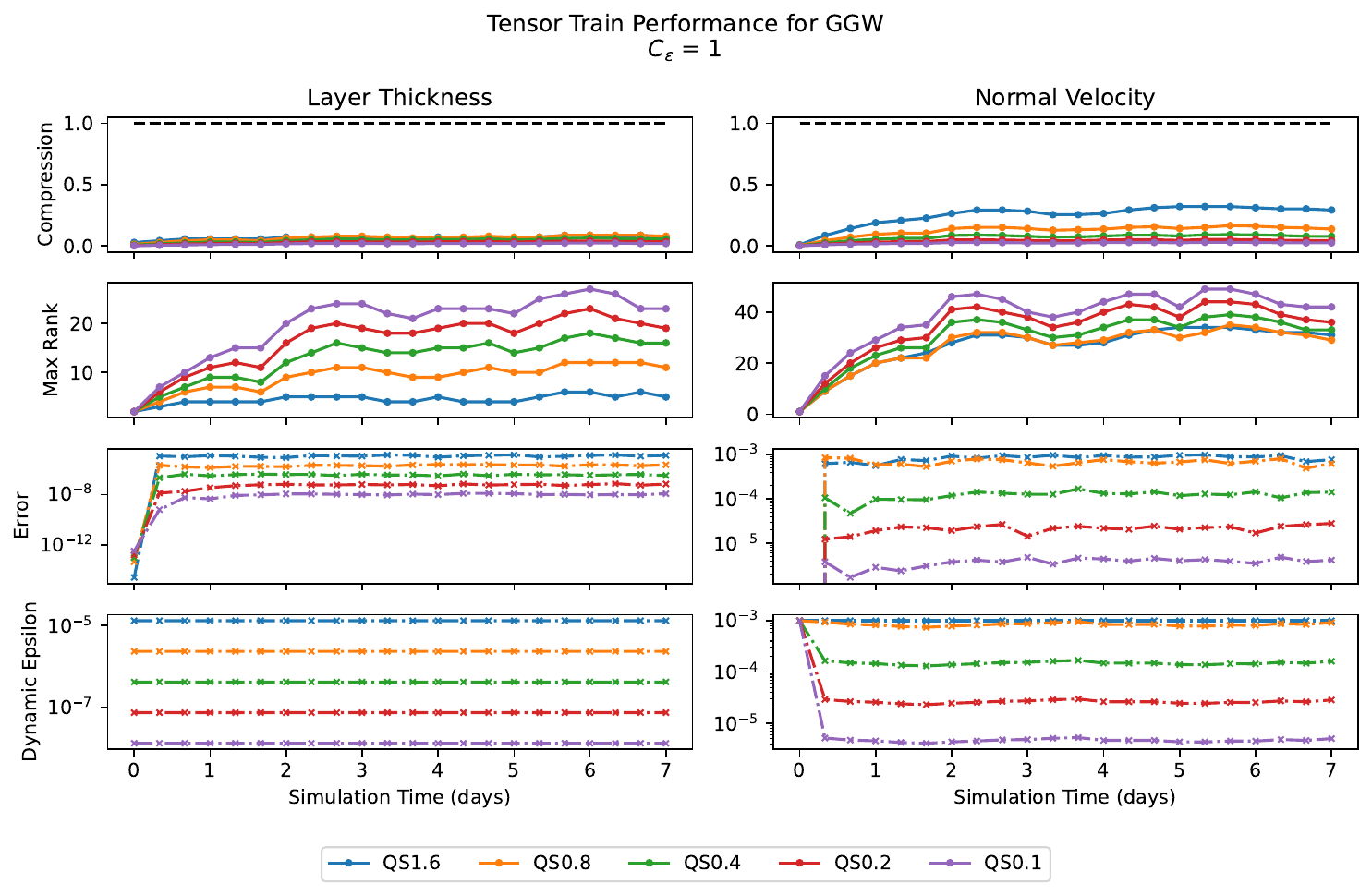}
    \caption{
        TT-compression results for GGW with \( C_\epsilon = \compressionceps \).
    }
    \label{fig:compression_ggw_Ceps\compressionceps}
\end{figure}

As in \Cref{fig:compression_qlw_Ceps\compressionceps} for QLW, \Cref{fig:compression_ggw_Ceps\compressionceps} shows the TT-compression for layer thickness and normal velocity during the GGW test case.
Similar to the previous case, both prognostic quantities remain well compressed throughout the simulation on sufficiently large problems.
The compression is slightly less than than in QLW however, as can be seen from the max rank plots in \Cref{fig:compression_ggw_Ceps\compressionceps} as compared to those in \Cref{fig:compression_qlw_Ceps\compressionceps}; we can see that on QS0.1, the TT-ranks for both thickness and velocity are slightly higher, clearly due to the nonlinear dynamics.
However, because these nonlinear dynamics are not dominant, the effect on the compression is relatively small.

\begin{figure}
    \centering
    \includegraphics[width=\diagnosticscale\linewidth]{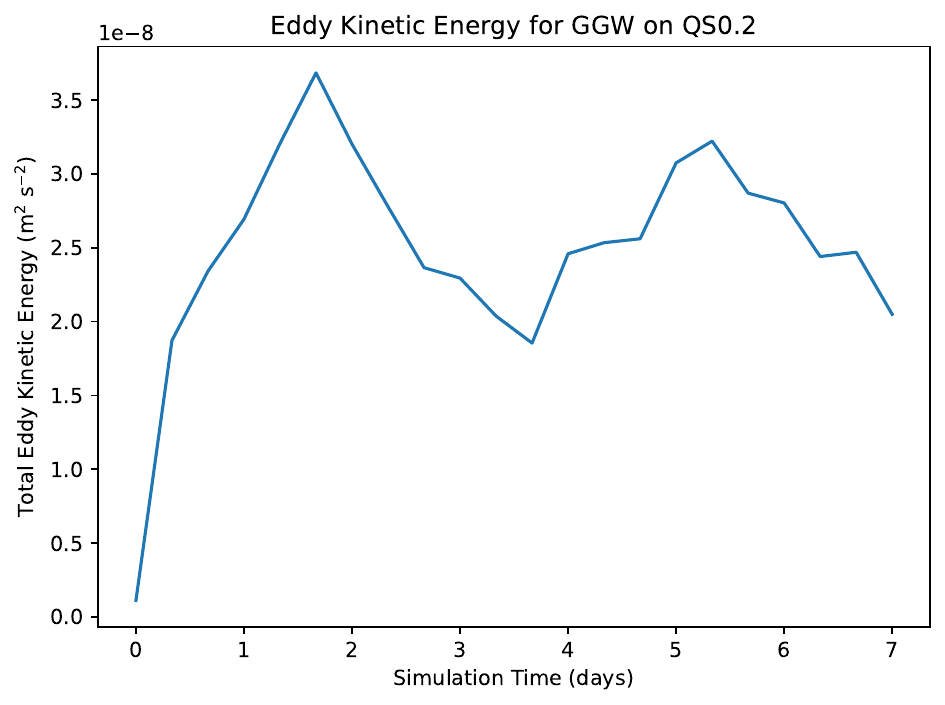}
    \caption{
        Total eddy kinetic energy for GGW on QS0.2.
    }
    \label{fig:diagnostic_ggw_qs0.2}
\end{figure}

As evidence that the nonlinear terms are not dominant in GGW, we can compare the total EKE plots between the present case (\Cref{fig:diagnostic_ggw_qs0.2}) and QLW (\Cref{fig:diagnostic_qlw_qs0.2}).
Both show that the total EKE is on the order of \( 10^{-7} \) and show a similar evolution in time.
Again, this shows that the presence of nonlinear dynamics in the problem does not necessarily lead to a loss of TT-compression. 

\begin{figure}
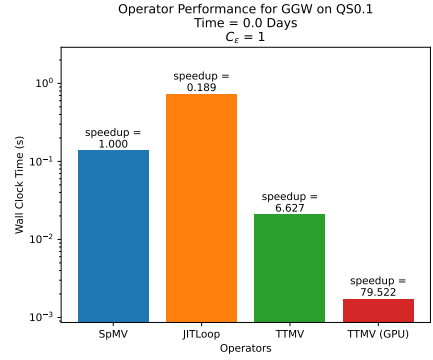
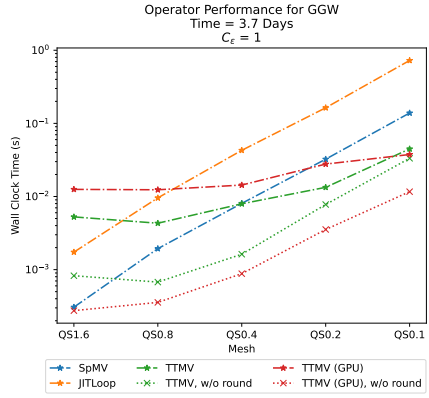
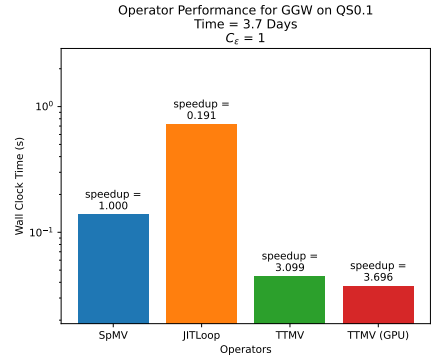
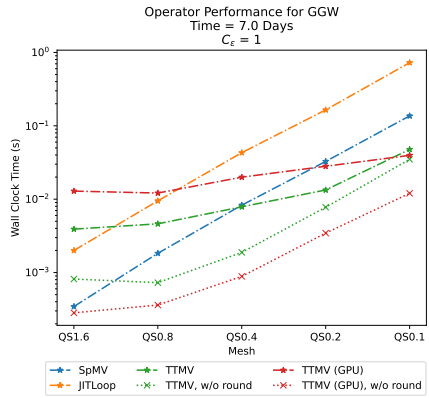
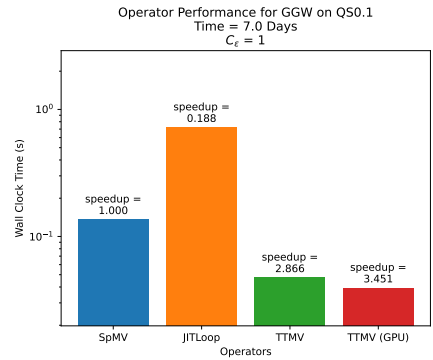

    \centering
    \begin{subfigure}{\matvecsubscale\linewidth}
        \includegraphics[width=\linewidth]{matvec/matvec_ggw_timeind0_Ceps\matvecceps.pdf}
        \caption{~}
        \label{fig:matvec_ggw_timeind0_Ceps\matvecceps}
    \end{subfigure}
    \hfill
    \begin{subfigure}{\matvecsubscale\linewidth}
        \includegraphics[width=\linewidth]{matvec/matvec_bar_ggw_qs0.1_timeind0_Ceps\matvecceps.pdf}
        \caption{~}
        \label{fig:matvec_bar_ggw_qs0.1_timeind0_Ceps\matvecceps}
    \end{subfigure}

    \medskip
    
    \begin{subfigure}{\matvecsubscale\linewidth}
        \includegraphics[width=\linewidth]{matvec/matvec_ggw_timeind11_Ceps\matvecceps.pdf}
        \caption{~}
        \label{fig:matvec_ggw_timeind11_Ceps\matvecceps}
    \end{subfigure}
    \hfill
    \begin{subfigure}{\matvecsubscale\linewidth}
        \includegraphics[width=\linewidth]{matvec/matvec_bar_ggw_qs0.1_timeind11_Ceps\matvecceps.pdf}
        \caption{~}
        \label{fig:matvec_bar_ggw_qs0.1_timeind11_Ceps\matvecceps}
    \end{subfigure}

    \medskip
    
    \begin{subfigure}{\matvecsubscale\linewidth}
        \includegraphics[width=\linewidth]{matvec/matvec_ggw_timeind21_Ceps\matvecceps.pdf}
        \caption{~}
        \label{fig:matvec_ggw_timeind21_Ceps\matvecceps}
    \end{subfigure}
    \hfill
    \begin{subfigure}{\matvecsubscale\linewidth}
        \includegraphics[width=\linewidth]{matvec/matvec_bar_ggw_qs0.1_timeind21_Ceps\matvecceps.pdf}
        \caption{~}
        \label{fig:matvec_bar_ggw_qs0.1_timeind21_Ceps\matvecceps}
    \end{subfigure}
    \caption{
        TT-format operation performance in the GGW test case.
    }
    \label{fig:matvec_ggw_\matvecceps}
\end{figure}

\Cref{fig:matvec_ggw_\matvecceps} shows the speedup for our TT-methods at the same points in time in \Cref{fig:matvec_qlw_\matvecceps} for QLW. 
The results here are very similar to those from before, due to the fact that the TT-compression rates in both cases are also similar.

\begin{figure}
    \centering
    \includegraphics[width=\diagnosticscale\linewidth]{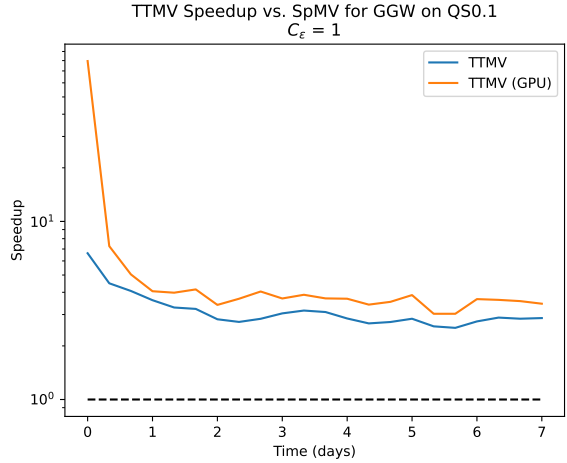}
    \caption{
        TTMV speedup versus SpMV in time for the GGW test case on QS0.1.
    }
    \label{fig:matvec_intime_ggw_\matvecceps}
\end{figure}

\FloatBarrier

\subsubsection{Barotropically Unstable Jet}
\label{sec:buj}

Our third test case is a canonical case for the SWEs that begins with a very simple initial condition and evolves into a flow with structures at multiple scales relevant to realistic applications.
As such, this is perhaps the most illustrative test case for our purposes.
We expect that TT to effectively compress the initial states for both thickness and velocity, then will see how the TT-compression evolves in time with the flow.

\begin{figure}
    \centering
    \includegraphics[width=\compressionscale\linewidth]{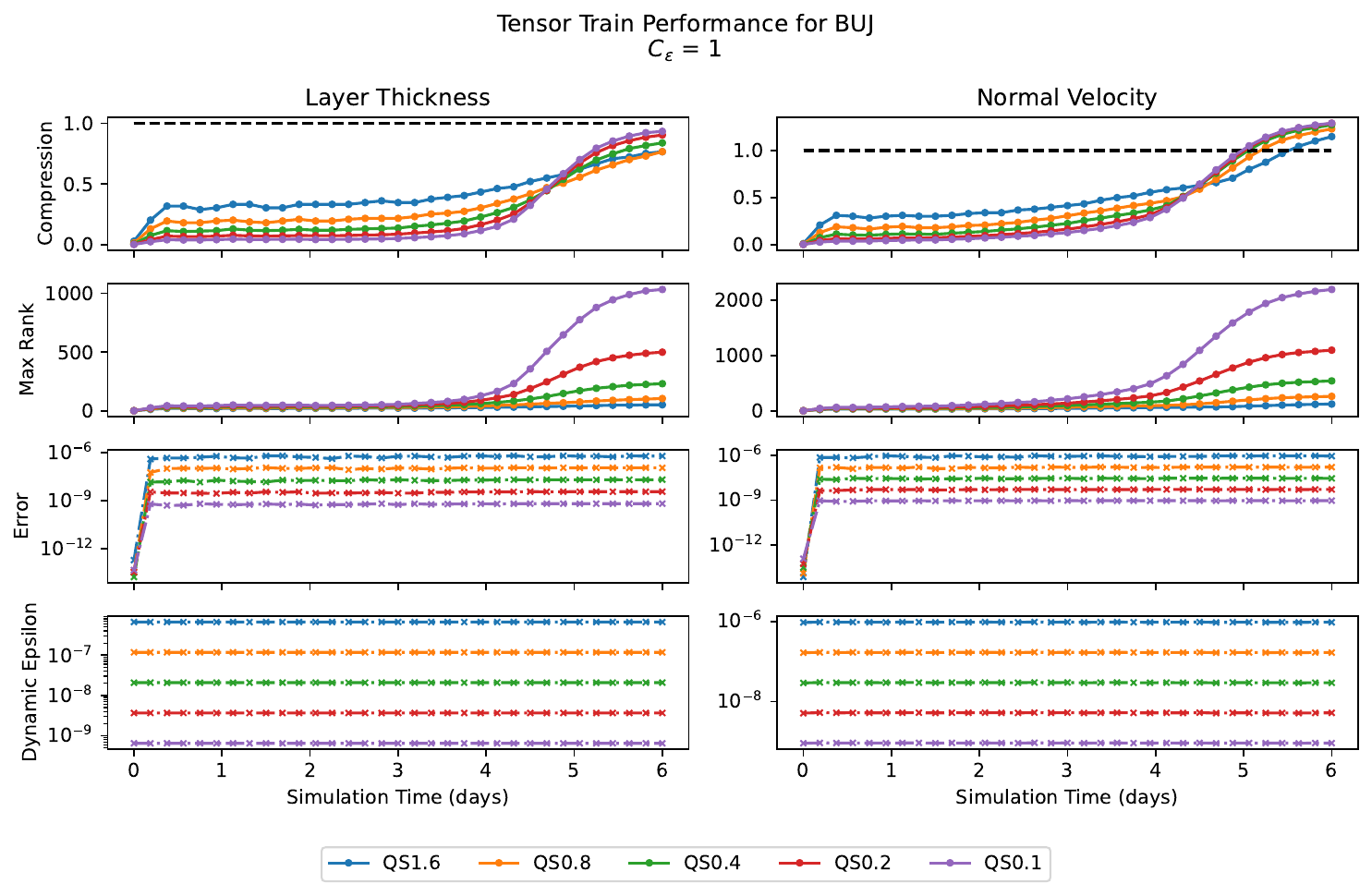}
    \caption{
        TT-compression results for BUJ with \( C_\epsilon = \compressionceps \).
    }
    \label{fig:compression_buj_Ceps\compressionceps}
\end{figure}

As we can see in \Cref{fig:compression_buj_Ceps\compressionceps}, the model state does indeed begin as well compressed, and remains so through approximately the first half of the simulation where the barotropic jet begins to roll up due to the perturbation in the initial condition.
After the half-way mark however, the compression, especially for the velocity data, begins to worsen.
By the end of the simulation, compression is lost entirely.
The TT-compression for the velocity even surpasses a ratio of 1, meaning that the TT of the data is larger than the original uncompressed data, exhibiting anti-compression.
Notably, during the final two days of the simulation, the compression on the larger meshes (in particular QS0.1) is worse than the smaller meshes.
This is likely because the high resolution meshes resolve smaller scale structures in the flow, requiring the TT-rank to grow more quickly to maintain fidelity -- this can be seen in the plots of the maximum rank for both the thickness and velocity.

\begin{figure}
    \centering
    \includegraphics[width=\diagnosticscale\linewidth]{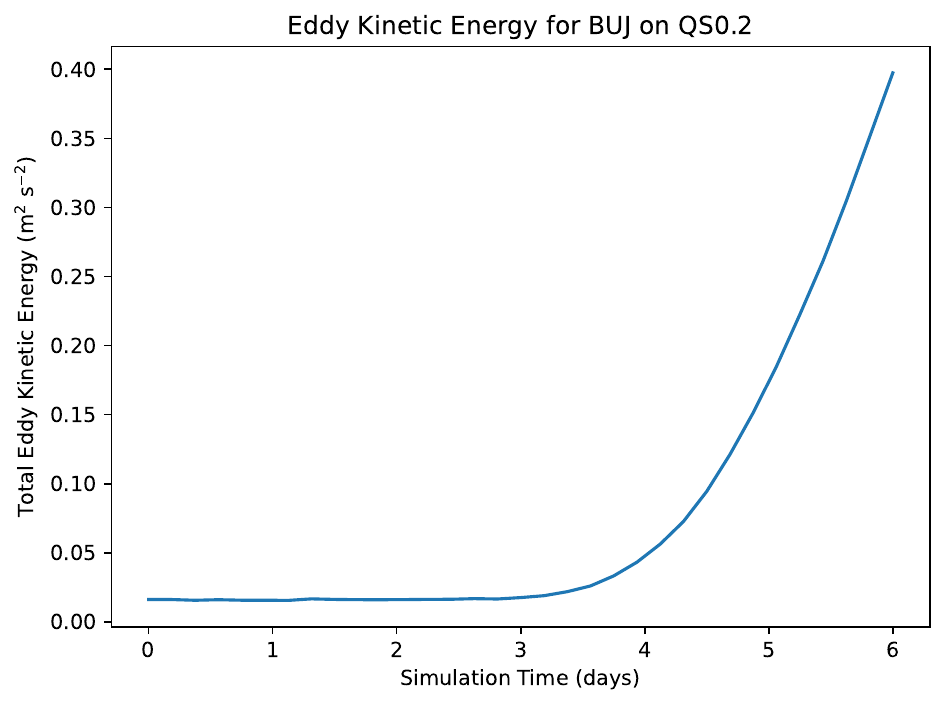}
    \caption{
        Total eddy kinetic energy for BUJ on QS0.2.
    }
    \label{fig:diagnostic_buj_qs0.2}
\end{figure}

It is also notable that the total EKE (\Cref{fig:diagnostic_buj_qs0.2}) and the compression are positively correlated.
The total EKE starts low as the flow remains close to its simple initial condition, but begins to quickly grow between days three and four as the jet begins its roll-up.
This coincides with the loss of compression for both prognostic quantities. 

\begin{figure}
    \centering
    \begin{subfigure}{\matvecsubscale\linewidth}
        \includegraphics[width=\linewidth]{matvec/matvec_buj_timeind0_Ceps\matvecceps.pdf}
        \caption{~}
        \label{fig:matvec_buj_timeind0_Ceps\matvecceps}
    \end{subfigure}
    \hfill
    \begin{subfigure}{\matvecsubscale\linewidth}
        \includegraphics[width=\linewidth]{matvec/matvec_bar_buj_qs0.1_timeind0_Ceps\matvecceps.pdf}
        \caption{~}
        \label{fig:matvec_bar_buj_qs0.1_timeind0_Ceps\matvecceps}
    \end{subfigure}

    \medskip
    
    \begin{subfigure}{\matvecsubscale\linewidth}
        \includegraphics[width=\linewidth]{matvec/matvec_buj_timeind16_Ceps\matvecceps.pdf}
        \caption{~}
        \label{fig:matvec_buj_timeind16_Ceps\matvecceps}
    \end{subfigure}
    \hfill
    \begin{subfigure}{\matvecsubscale\linewidth}
        \includegraphics[width=\linewidth]{matvec/matvec_bar_buj_qs0.1_timeind16_Ceps\matvecceps.pdf}
        \caption{~}
        \label{fig:matvec_bar_buj_qs0.1_timeind16_Ceps\matvecceps}
    \end{subfigure}

    \medskip
    
    \begin{subfigure}{\matvecsubscale\linewidth}
        \includegraphics[width=\linewidth]{matvec/matvec_buj_timeind32_Ceps\matvecceps.pdf}
        \caption{~}
        \label{fig:matvec_buj_timeind32_Ceps\matvecceps}
    \end{subfigure}
    \hfill
    \begin{subfigure}{\matvecsubscale\linewidth}
        \includegraphics[width=\linewidth]{matvec/matvec_bar_buj_qs0.1_timeind32_Ceps\matvecceps.pdf}
        \caption{~}
        \label{fig:matvec_bar_buj_qs0.1_timeind32_Ceps\matvecceps}
    \end{subfigure}
    \caption{
        TT-format operation performance in the BUJ test case.
    }
    \label{fig:matvec_buj_\matvecceps}
\end{figure}

The speedup achieved by TT in the BUJ case suffers similarly; we see nontrivial speedup during the first few days of the simulation, but the loss of compression results in TTMV being slower than SpMV for the majority of the run (\Cref{fig:matvec_intime_buj_\matvecceps}), with the slowdown being particularly problematic at the end of the simulation.
TTMV remains faster than JITLoop for the majority of BUJ (\Cref{fig:matvec_buj_\matvecceps}), but falls well behind near the end. Regardless, the comparison to the optimized SpMV gives a much more realistic approximation to a high performance code.

\begin{figure}
    \centering
    \includegraphics[width=\diagnosticscale\linewidth]{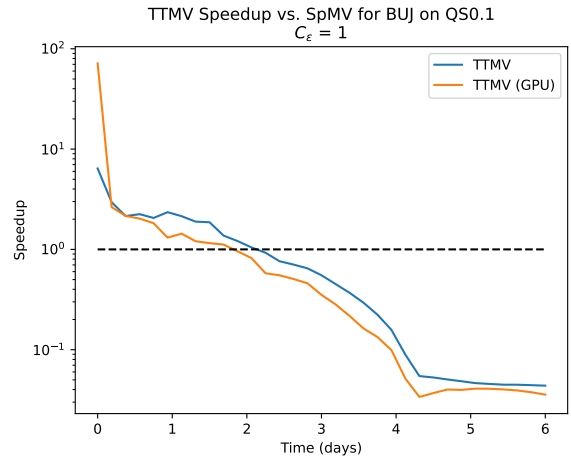}
    \caption{
        TTMV speedup versus SpMV in time for the BUJ test case on QS0.1.
    }
    \label{fig:matvec_intime_buj_\matvecceps}
\end{figure}

\FloatBarrier

\subsubsection{Williamson Test Case 5}
\label{sec:wtc5}

Our final test case exhibits the most complicated dynamics, and is run for a full 50 simulated days.
Note that we have omitted results for WTC5 on QS0.1 because of the computational expense of running swe-python on this high resolution mesh.

\begin{figure}
    \centering
    \includegraphics[width=\compressionscale\linewidth]{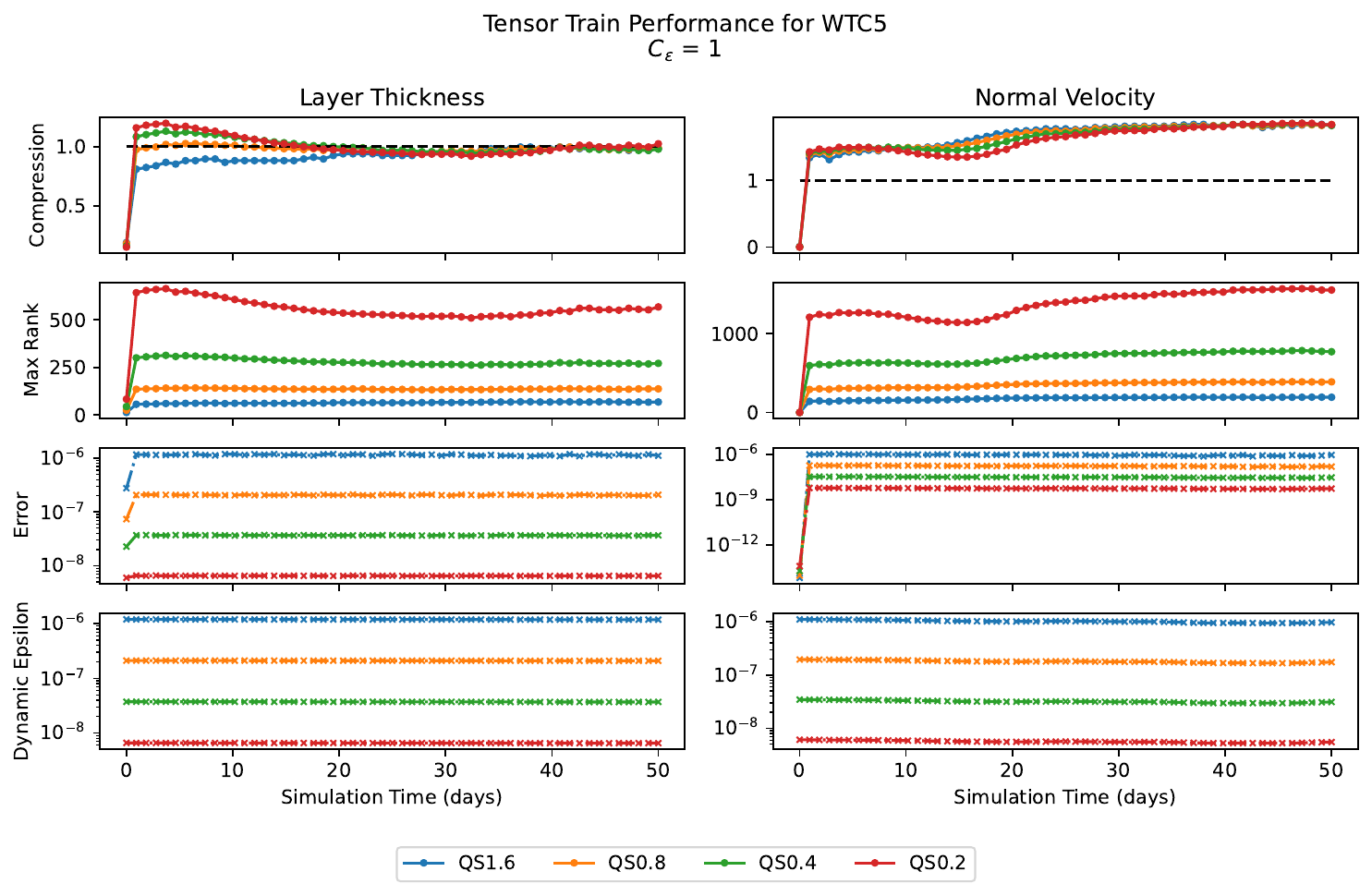}
    \caption{
        TT-compression results for WTC5 with \( C_\epsilon = \compressionceps \).
    }
    \label{fig:compression_wtc5_Ceps\compressionceps}
\end{figure}

Looking at \Cref{fig:compression_wtc5_Ceps\compressionceps}, we can see that compression is lost very soon after the beginning of the simulation.
As a reminder, the loss of compression is not instantaneous (i.e. after a single time-step), because we are only plotting the compression (and related quantities) at a subset of time-levels.
Nonetheless, the loss of compression is very rapid, and does turn into anti-compression, especially for the TT of the normal velocity.

\begin{figure}
    \centering
    \includegraphics[width=\diagnosticscale\linewidth]{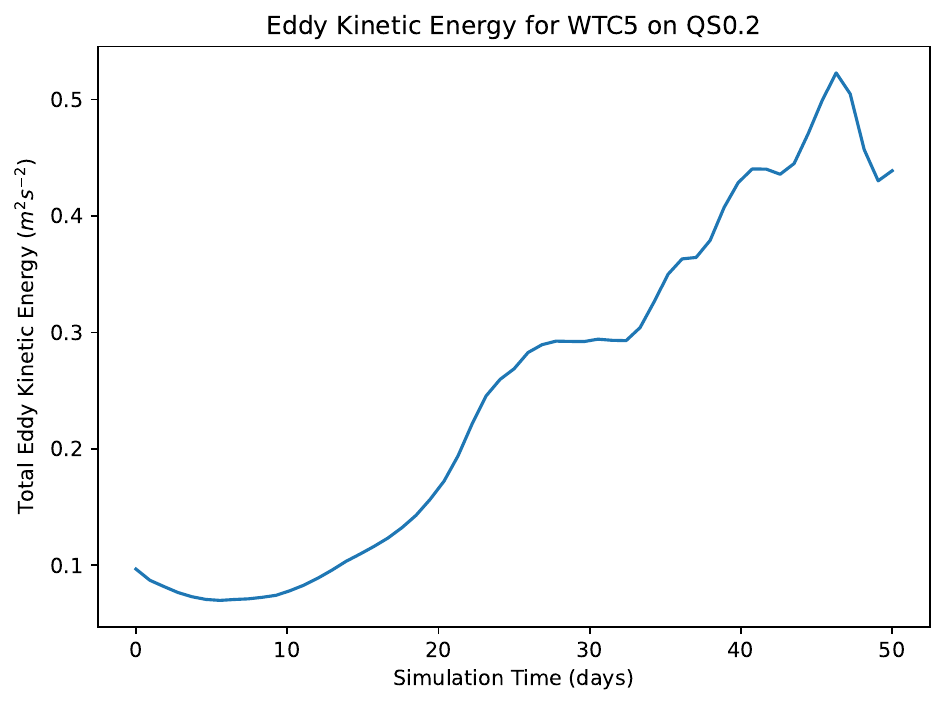}
    \caption{
        Total eddy kinetic energy for WTC5 on QS0.2.
    }
    \label{fig:diagnostic_wtc5_qs0.2}
\end{figure}

Compare \Cref{fig:diagnostic_buj_qs0.2,fig:diagnostic_wtc5_qs0.2}, which show the total EKE for BUJ and WTC5 respectively.
The EKE in the WTC5 case starts near 0.1, while the EKE in the BUJ is close to zero for much of the simulation, only reaching 0.1 around four and a half days.
This four and a half day mark is also when TT-compression in the BUJ case begins to quickly degrade.
That is, the total EKE for WTC5 starts at a level not reached by BUJ until it begins to lose TT-compression.
Taken together with the results from the QLW and GGW test cases (which have EKE near zero), this points to a possible connection between EKE (i.e. the difference from the mean velocity state) and TT-compression in certain flows.


\begin{figure}
    \centering
    \begin{subfigure}{\matvecsubscale\linewidth}
        \includegraphics[width=\linewidth]{matvec/matvec_wtc5_timeind0_Ceps\matvecceps.pdf}
        \caption{~}
        \label{fig:matvec_wtc5_timeind0_Ceps\matvecceps}
    \end{subfigure}
    \hfill
    \begin{subfigure}{\matvecsubscale\linewidth}
        \includegraphics[width=\linewidth]{matvec/matvec_bar_wtc5_qs0.2_timeind0_Ceps\matvecceps.pdf}
        \caption{~}
        \label{fig:matvec_bar_wtc5_qs0.2_timeind0_Ceps\matvecceps}
    \end{subfigure}

    \medskip
    
    \begin{subfigure}{\matvecsubscale\linewidth}
        \includegraphics[width=\linewidth]{matvec/matvec_wtc5_timeind27_Ceps\matvecceps.pdf}
        \caption{~}
        \label{fig:matvec_wtc5_timeind27_Ceps\matvecceps}
    \end{subfigure}
    \hfill
    \begin{subfigure}{\matvecsubscale\linewidth}
        \includegraphics[width=\linewidth]{matvec/matvec_bar_wtc5_qs0.2_timeind27_Ceps\matvecceps.pdf}
        \caption{~}
        \label{fig:matvec_bar_wtc5_qs0.2_timeind27_Ceps\matvecceps}
    \end{subfigure}

    \medskip
    
    \begin{subfigure}{\matvecsubscale\linewidth}
        \includegraphics[width=\linewidth]{matvec/matvec_wtc5_timeind54_Ceps\matvecceps.pdf}
        \caption{~}
        \label{fig:matvec_wtc5_timeind54_Ceps\matvecceps}
    \end{subfigure}
    \hfill
    \begin{subfigure}{\matvecsubscale\linewidth}
        \includegraphics[width=\linewidth]{matvec/matvec_bar_wtc5_qs0.2_timeind54_Ceps\matvecceps.pdf}
        \caption{~}
        \label{fig:matvec_bar_wtc5_qs0.2_timeind54_Ceps\matvecceps}
    \end{subfigure}
    \caption{
        TT-format operation performance in the WTC5 test case.
    }
    \label{fig:matvec_wtc5_\matvecceps}
\end{figure}

\begin{figure}
    \centering
    \includegraphics[width=\diagnosticscale\linewidth]{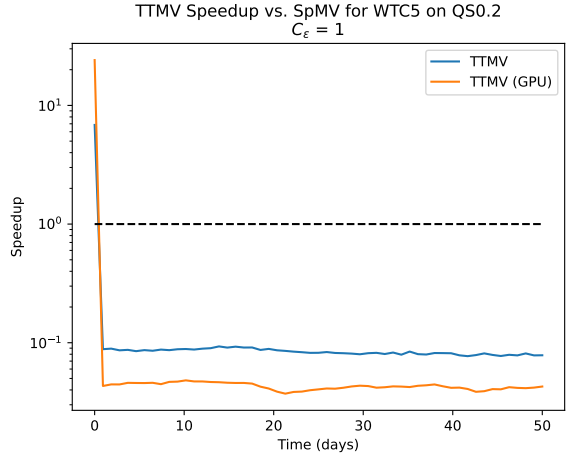}
    \caption{
        TTMV speedup versus SpMV in time for the WTC5 test case on QS0.2.
    }
    \label{fig:matvec_intime_wtc5_\matvecceps}
\end{figure}

Because of the immediate loss of compression, the speedups achieved by TTMV are poor as compared to SpMV.
Of course, we should note that the mesh on which \Cref{fig:matvec_intime_wtc5_\matvecceps} is generated is four times smaller than corresponding plots for the other three test cases; the shown speedup is likely worse than it would be on QS0.1.
Nonetheless, the qualitative behavior of the speedup curves would likely be consistent across cases.

\FloatBarrier


\subsection{Discussion}
\label{sec:discussion}

\begin{figure}
    \centering
    \begin{subfigure}{0.45\linewidth}
        \includegraphics[width=\linewidth]{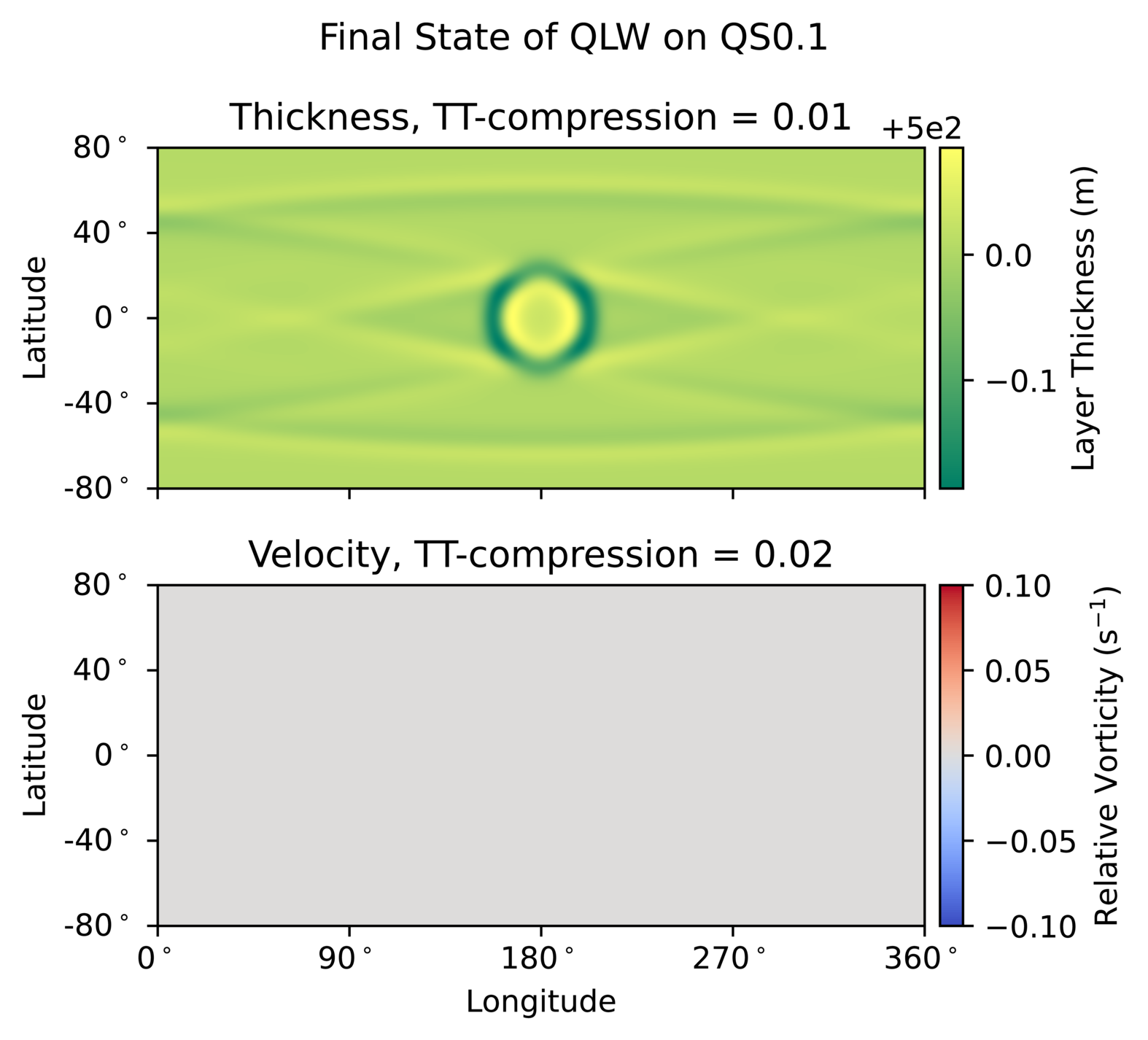}
        \caption{~}
        \label{fig:qlw_compression_summary}
    \end{subfigure}
    \hfill
    \begin{subfigure}{0.45\linewidth}
        \includegraphics[width=\linewidth]{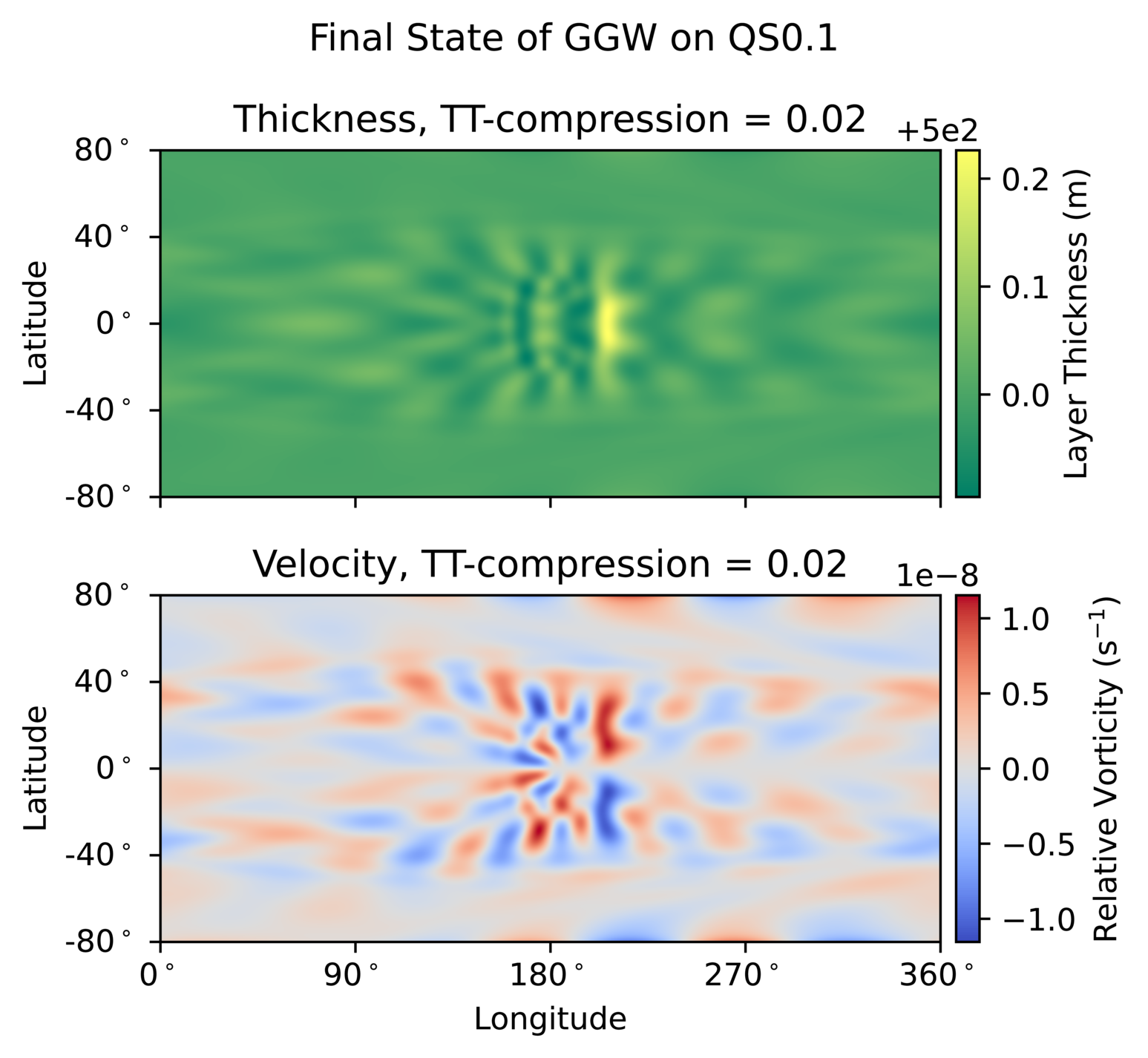}
        \caption{~}
        \label{fig:ggw_compression_summary}
    \end{subfigure}

    \begin{subfigure}{0.45\linewidth}
        \includegraphics[width=\linewidth]{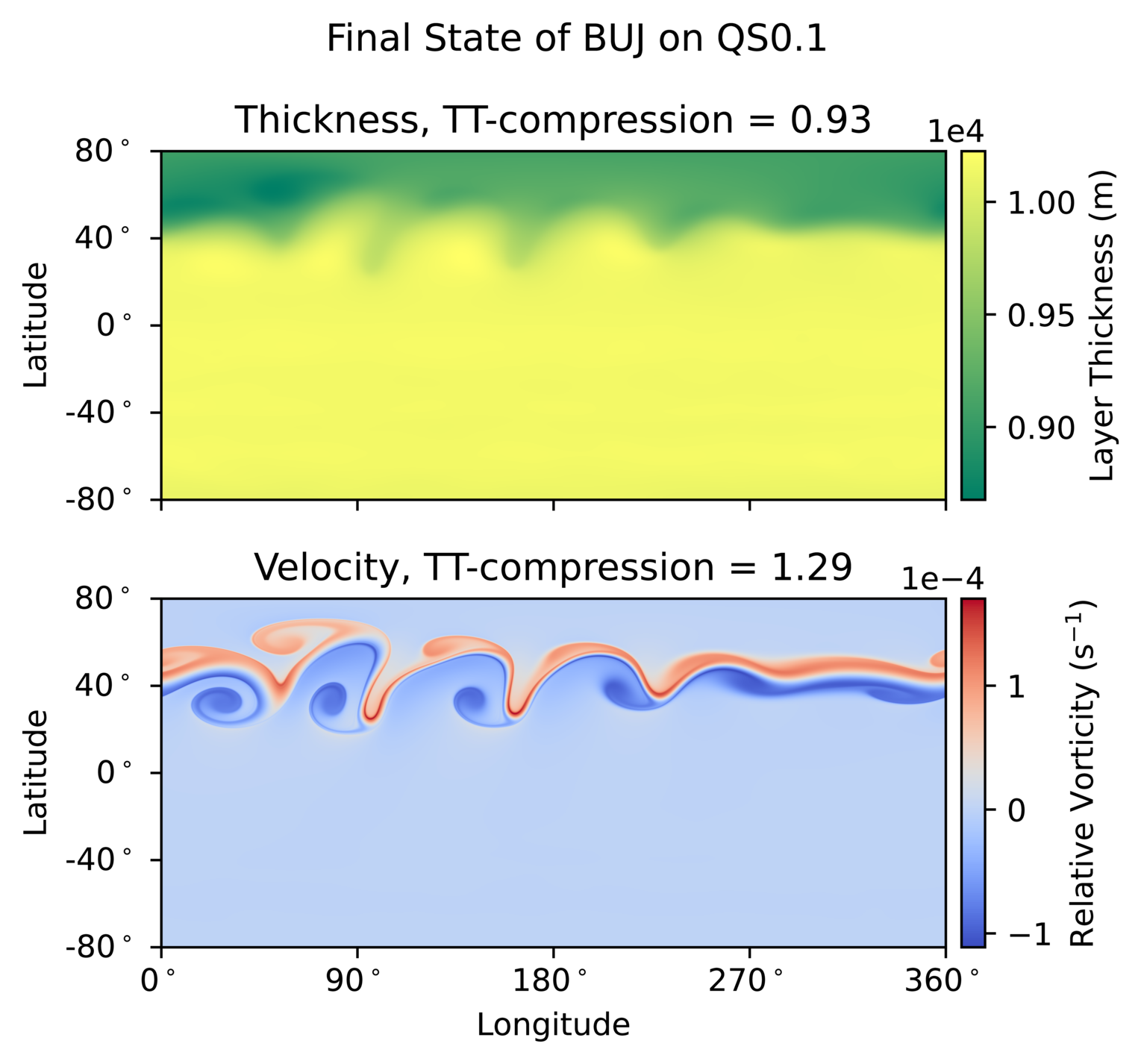}
        \caption{~}
        \label{fig:buj_compression_summary}
    \end{subfigure}
    \hfill
    \begin{subfigure}{0.45\linewidth}
        \includegraphics[width=\linewidth]{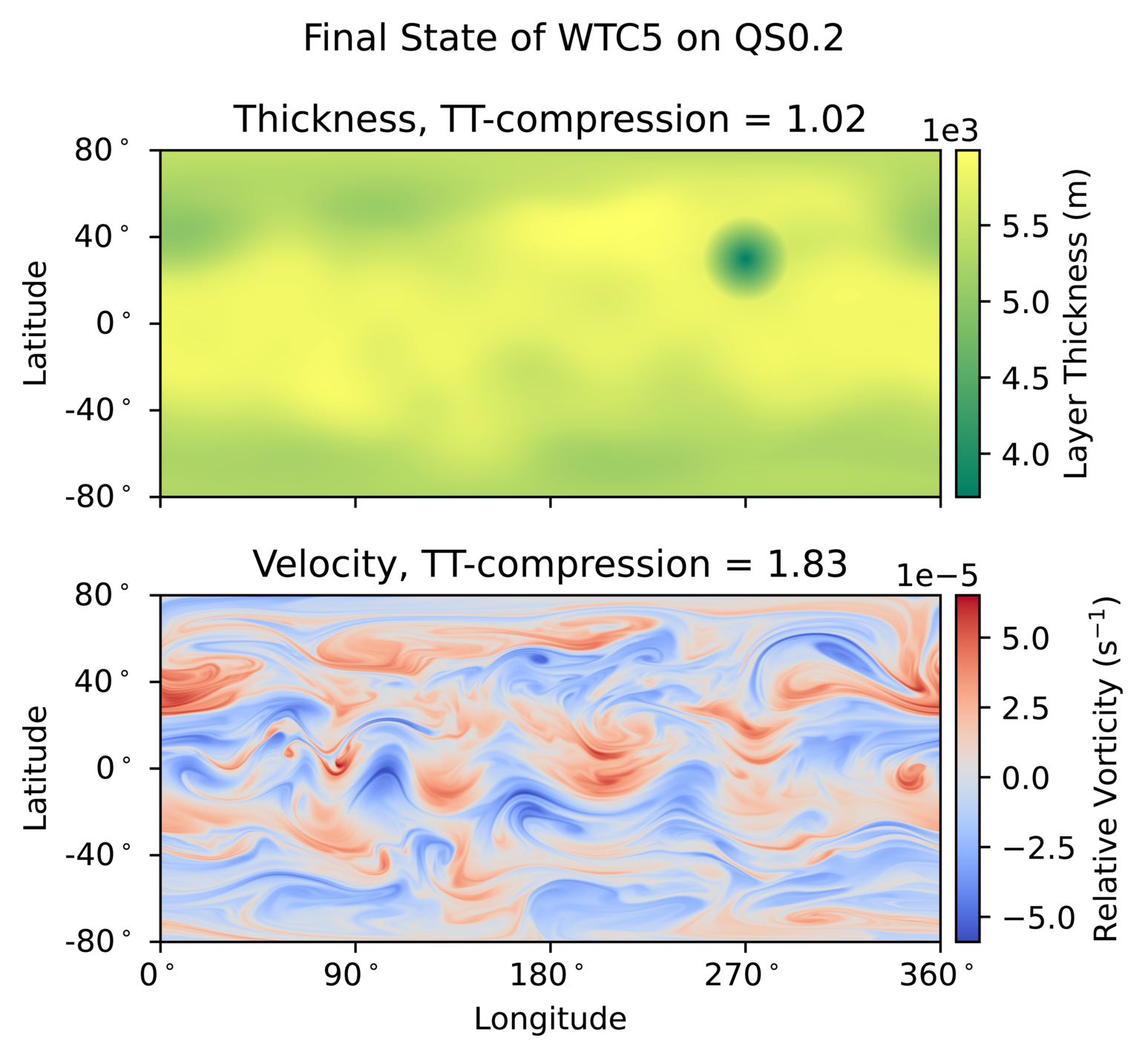}
        \caption{~}
        \label{fig:wtc5_compression_summary}
    \end{subfigure}
    
    \caption{
        Snapshots of the final states for each of our four test cases, along with the TT-compression rate achieved with \( C_\epsilon = 1 \) (recall that compression greater than one is anti-compression).
        Note that the velocity plots do not show the velocity itself, but rather the relative vorticity \( \nabla \times \vec{u} \), which provides a better visualization of the relevant data.
        The relative vorticity plot for the QLW case shows nothing because the case does not exhibit nonzero vorticity.
    }
    \label{fig:compression_summary}
\end{figure}

Taking the results of all four test cases together, we see a clear progression in the effectiveness of our TT-methods.
In the QLW case where the model dynamics are relatively simple for the entirety of the simulation, we see strong TT-compression and speedup compared to the optimized SpMV kernel.
Adding nonlinear dynamics to the system in the GGW case, we see a slight increase in the TT-rank for both the layer thickness and normal velocity data, but the overall compression remains strong, leading to good speedup similar to that in the QLW case.
The BUJ test case starts with a relatively simple state that shows itself to be compressible, but develops to exhibit complex structures at multiple scales which require high TT-rank to accurately represent, resulting in poor compression and eventually anti-compression. 
Finally, while WTC5's initial condition is compressible, its state quickly develops so as to require very high TT-rank, leading to strong anti-compression.
\Cref{fig:compression_summary} serves as a summary and visualization of our main results.

Of course, the speedup obtained by TT-methods versus traditional methods is directly tied to the TT-compression of the state being evolved; as demonstrated above, the TT-compression is effectively representative of the obtainable speedup.
An important takeaway from this fact is that the specific speedup results shown here need not be definitive for the overall results to hold.
The actual speedups obtained depend on a number of highly variable factors including algorithm implementation, programming language, compilers, and computing hardware. 
For example, it is likely that our the reported TT speedups would improve over time as implementations mature.
However, the lack of TT-compression in physically relevant cases like BUJ and WTC5 points to a more fundamental challenge in applying TT-methods to geophysical fluids applications.

The high TT-rank in BUJ and WTC5 are also problematic for the expense that it imposes on the TT-rounding procedure.
Taking the final time \( t = 50 \) days for WTC5 (\Cref{fig:matvec_wtc5_timeind54_Ceps\matvecceps,fig:matvec_bar_wtc5_qs0.2_timeind54_Ceps\matvecceps}) as an example, observe the cost incurred by the TT-rounding procedure for TTMV on both CPU and GPU.
On the GPU, the cost of the operation without the TT-round is almost as fast as SpMV despite the high anti-compression.
Similarly, on the CPU, the TTMV without the TT-round is faster than JITLoop.
Accounting for the TT-round, both cases are significancy slower because of the high rank of the result and the \( \ord{dnr^6} \) cost of rounding \citep{oseledets2011}.
Of course, one does not necessarily need to round after each and every TT operation, but the longer this is delayed, the more the rank will grow, and the more the \( \ord{dnr^6} \) cost will effect performance -- especially in high-rank cases like BUJ and WTC5.

Finally, we note that because the TT-compression depends on the dimension of the data tensor (\Cref{eqn:compression_order}), the compression in all given test cases would likely improve in a higher-dimensional domain such as the full 3-dimensional ocean.
Additionally, in a real-world application, cell-centered quantities like the thickness and additional tracers could be bundled into a single higher denominational tensor, potentially furthering the effectiveness of the TT method.
This warrants further investigation.
However, the poor TT-compression in the velocity field would likely remain problematic, especially as 3-dimensional dynamics further exacerbated the TT-rank growth we see in BUJ and WTC5.
Because of the additional computational cost of TT operations, including TT-rounding, the TT-compression must be strong to result in computational speedup (e.g. \Cref{fig:compression_buj_Ceps1,fig:matvec_buj_1}); simply having a compression rate less than one is not enough.



\section{Conclusion}
\label{sec:conclusion}

The results presented in this work show that TT methods can be very effective for simple problems (\Cref{fig:compression_qlw_Ceps1,fig:compression_ggw_Ceps1}), but struggle to efficiently represent more realistic flows (\Cref{fig:compression_buj_Ceps1,fig:compression_wtc5_Ceps1}).
\Cref{fig:diagnostic_buj_qs0.2,fig:diagnostic_wtc5_qs0.2} suggest a correlation between the eddy kinetic energy (i.e. the difference between the velocity state and a smooth mean state) and the TT-compression, with high EKE cases showing poor TT-compression.
As such, it is not likely GFD models can achieve meaningful speedups via implementations of full TT-format numerics.
As noted in the introduction and in \Cref{sec:approximating_with_tt}, the reason that TT is so effective in the quantum materials space is that it can be shown analytically that all realistic quantum states can be efficiently represented by a TT \citep{orus2014}.
Here, we have shown that this is not the case for GFD applications; some states can be efficiently represented with TT, but there exists realistic states that cannot be.

Of course, this does not mean that TT methods, or tensor decomposition methods more generally, cannot be of use of in GFD applications.
Since TT is very memory efficient and fast for simpler problems, there is a opportunity to apply TT methods to appropriate subproblems or subsystems within existing GFD codes.
For example, one could potentially apply TT only to the gravity wave subsystem within a larger ocean or atmosphere model as this is relatively simple, but computationally expensive because of the rapid motions it entails.
Further, it may be that other tensor decomposition methods more general than tensor train could be more effective in this application, particularly for dealing with unstructured grids.

Future work on this application of tensor decomposition methods could include an investigation of the effectiveness of different decomposition formats on different meshes.
Additionally, a particularly useful result would be an analytic proof connecting some diagnostic quantity of a given flow to high TT-rank (e.g. if it could be shown that in certain cases, high EKE implies that the flow will have high TT-rank).
This would provide a result analogous to that in \citet{orus2014} that showed what subset of flow states can be efficiently represented with TT, allowing researchers across applications to know if TT can be applied to their particular problem in advance.



\clearpage
\section*{Acknowledgments}

This work was supported by the U.S. Department of Energy through Los Alamos National Laboratory.
Los Alamos National Laboratory is operated by Triad National Security, LLC, for the National Nuclear Security Administration of U.S. Department of Energy (Contract No. 89233218CNA000001).
The research presented in this article was supported by the Laboratory Directed Research and Development program of Los Alamos National Laboratory under project numbers 20240782ER (LDRD Seedling) and 20258172CT-IST (ISTI Rapid Response), as well as the Center for Nonlinear Studies under project number 20250614CR-NLS.
Computing resources were supplied by the Center for Nonlinear Studies.


\section*{Data Statement}

No AI was used in the production of this work, including the writing of this document and the software used to run simulations and generate figures.

All software and data used in this work is open source and freely available; documentation is generally sparse, but the corresponding author is happy to respond to questions.
\begin{itemize}
    \item[] sparsett \citep{sparsett}:
    \begin{itemize}
        \item[] GitLab: \url{https://gitlab.com/jeremy-lilly/sparsett}
    \end{itemize}
    
    \item[] swe-python \citep{swe-python}:
    \begin{itemize}
        \item[] GitHub: \url{https://github.com/jeremy-lilly/swe-python/tree/tt-dev}
    \end{itemize}
\end{itemize}


\appendix
\renewcommand{\thesection}{\Alph{section}}
\crefalias{section}{appendix}



\section{Tensor Train on Unstructured Grids}
\label{apx:unstructured_tt}

\begin{figure}
    \centering
    \begin{subfigure}{0.49\linewidth}
        \includegraphics[width=\linewidth]{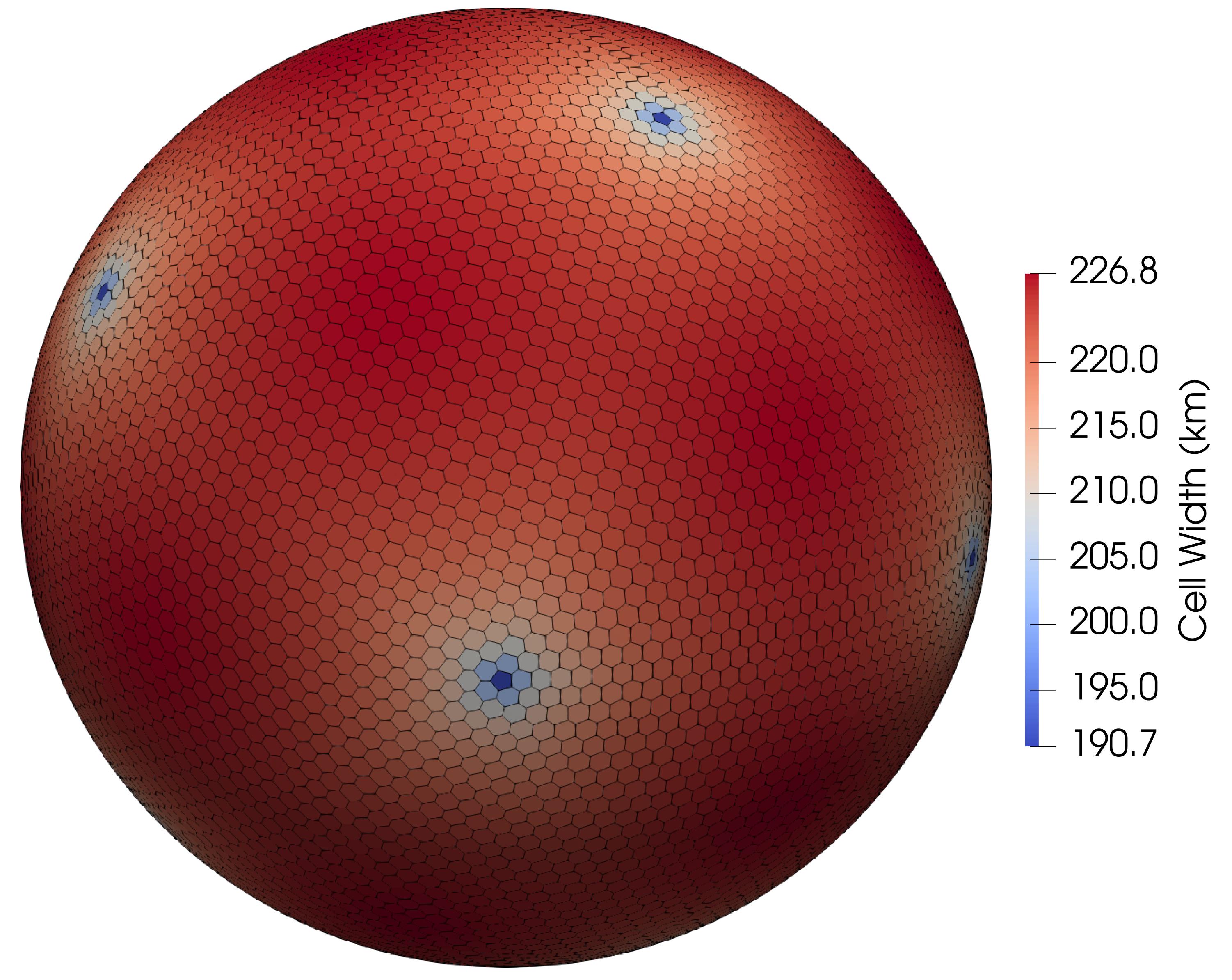}
        \caption{~}
        \label{fig:icos_mesh}
    \end{subfigure}
    \hfill
    \begin{subfigure}{0.49\linewidth}
        \def\dotscale{0.6}
\def\mydot{\scalebox{\dotscale}{\( \mathcolor{black}{\bullet} \)}}
\def\mysquare{\scalebox{\dotscale}{{\tiny \( \mathcolor{black}{\blacksquare} \)}}}
\def\mytriangle{\scalebox{\dotscale}{\raisebox{1px}{\small\( \mathcolor{black}{\blacktriangle} \)}}}

\tikzstyle{tn}=[font=\tiny, align=center]
\tikzstyle{ar}=[->, dotted, very thick]
\tikzstyle{background grid}=[draw, black!50, step=1cm]

\def\scale{1.25}
\begin{tikzpicture}[black,
                    scale=\scale,
                    every node/.style={scale=\scale}]
    \coordinate (primal_center) at (0,0);
    \coordinate (dual_center) at (30:1);
    \coordinate (edge_center) at (60:0.866);
    \coordinate (mid_norm) at (60:0.5);

    \node[tn] at ($(edge_center)+(105:0.25cm)$) {\( \mathbf{x}_e \)};

    \node[tn, below left=-1pt of primal_center] {\( \mathbf{x}_i \)};
    \foreach \x in {30,90,...,330}{
        \draw[thick] (\x:1) -- (\x+60:1);
    }
    \draw[thick] (dual_center) -- ( $(dual_center)+(dual_center)$ );

    \node[tn, below right=1pt of dual_center] {\( \mathbf{x}_v \)};
    \foreach \x in {-150,-30,90}{
        \draw[thick, dashed] ($(\x:1cm)+(dual_center)$) -- ($(\x+120:1cm)+(dual_center)$);
        \node at ($(\x:1cm)+(dual_center)$) {\mydot};
    }


    \foreach \x in {30,90,...,330}{
        \node at (\x+60:1)  {\mytriangle};
        \node[rotate=\x-150] at (\x+30:0.866)  {\mysquare};
    }
    \node[rotate=30] at ($(30:0.5cm)+(dual_center)$)  {\mysquare};



    \node[tn, align=left, text width=3.5cm, anchor=west] (legend) at (20:2) {
        \( \mathbf{x}_i = \) thickness points \newline
        \( \mathbf{x}_e = \) normal velocity points \newline
        \( \mathbf{x}_v = \) potential vorticity points
    };
\end{tikzpicture}
        \caption{~}
        \label{fig:trisk_hex}
    \end{subfigure}
    \caption{
        Example TRiSK grid from a Voronoi tessellation, where the primal cells are hexagons (and some pentagons) and the dual cells are triangles centered at primal cell vertices.
    }
    \label{fig:trisk_grid}
\end{figure}
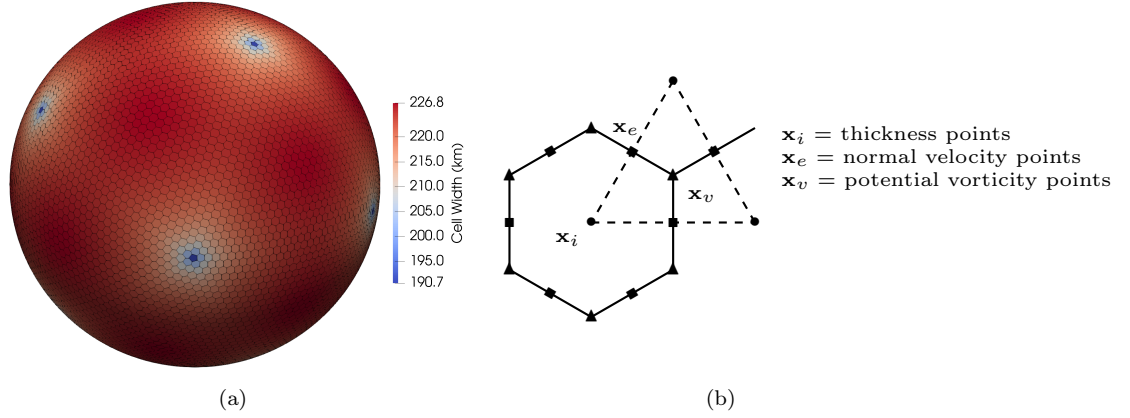

The goal of this appendix section is to outline and discuss the challenges in adapting TT methods to unstructured, variable resolution meshes.
Our discussion will focus on the types of meshes used by the ocean and atmosphere components of E3SM, which are Voronoi grids \citep{ju2011, okabe2017} consisting primarily of hexagons as the primal mesh, with a dual mesh consisting of triangles.
The spatial discretization used is a C-grid-type discretization \citep{arakawa1977} wherein the mass variable is computed on cell centers and the normal component of velocity is computed on cell edges (\Cref{fig:trisk_grid}).

\subsection{Tensorizing an Unstructured Mesh}
\label{apx:tensorizing}

\begin{figure}
    \centering
    \includegraphics[width=0.67\linewidth]{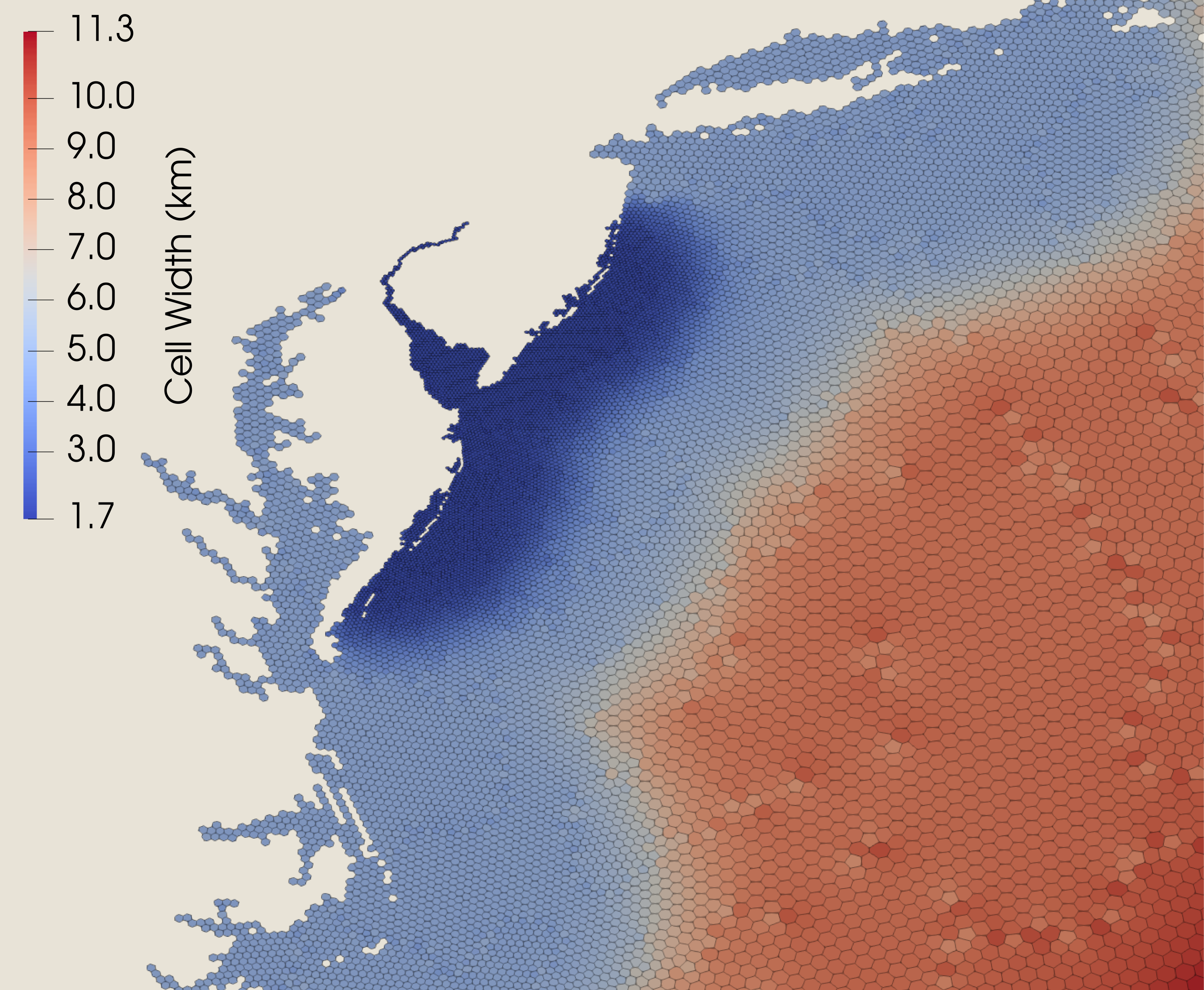}
    \caption{
        A realistic, variable resolution, E3SM ocean mesh zoomed in to Delaware Bay on the east coast of the United States.
    }
    \label{fig:del_bay}
\end{figure}

In a word, the difficulty is finding a map from a general unstructured mesh to a tensor space \( \Rtens{n} \) that allows TT to effectively compress the model state.
In the case of the structured meshes as used in the main body of this work (\Cref{fig:qs1.6}), there is a simple map between the mesh and a tensor space that preserves the spatial locality of data in such a way that the separation of variables performed by taking the TT of the data can exploit that locality to (potentially) find a low-rank representation of the data.
That is, the structured 2-dimensional mesh is configured in a way that it can be mapped naturally to a 2-dimensional tensor space.
\Cref{fig:icos_mesh} shows the simplest incarnation of an E3SM-like unstructured Voronoi mesh; there is no obvious mapping from this mesh to a tensor space that is appropriate for decomposition via TT.
One can come up with any number of mappings between such a mesh and a tensor space that allows the model data to be given as tensors, but not necessarily one that preserves the data locality, adjacency, and directionality that allows for a low-rank TT representation.
This problem only becomes more difficult when we move way from simple meshes like that in \Cref{fig:icos_mesh}, to more realistic meshes as in \Cref{fig:del_bay}.

For now, we will focus on the simpler case from \Cref{fig:icos_mesh}, which we will call ICOS (icosahedral centroidal Voronoi tessellation).
In essence, the mapping between such an ICOS mesh and a tensor space can be broken up into two steps; first, ordering the mesh data into a single vector with length equal to the number of data points in the mesh \( N \), and second, folding this vector into a \( d \)-dimensional tensor for \( d \geq 2 \).
To ease the discussion, we will refer to these two steps together as the \emph{tensorization} of the model data.

As a simple example, one possible tensorization of an ICOS mesh could be given by first selecting a starting cell (or edge), then visiting each cell (or edge) in a spiral pattern radiating out from the chosen starting element.
The vectors of model data could then be sorted according to the order that each element was visited during this spiral traversal.
This ordering somewhat preserves data locality, but not data adjacency.
The ordered vectors could then be folded according to the prime factorization of \( N \), where a vector of length \( N = 2^{\alpha_1}\, 3^{\alpha_2}\, 5^{\alpha_3}\, 7^{\alpha_4}\, \cdots \) is folded into a tensor with shape
\begin{equation}
    \label{eqn:qttish_shape}
    [\underbrace{2,\, \cdots,\, 2}_{\alpha_1 \text{ times}},\,
     \underbrace{3,\, \cdots,\, 3}_{\alpha_2 \text{ times}},\,  
     \underbrace{5,\, \cdots,\, 5}_{\alpha_3 \text{ times}},\,
     \underbrace{7,\, \cdots,\, 7}_{\alpha_4 \text{ times}},\,
     \cdots] \,.
\end{equation}
The choice of folding is essentially arbitrary, though is inspired by so-called quantized tensor train (QTT) methods \citep{oseledets2010,khoromskij2011}, which fold a vector of length \( N = 2^L \) into shape
\begin{equation}
\label{eqn:qtt_shape}
    [\underbrace{2,\, 2,\, \cdots,\, 2}_{L \text{ times}}] \,.
\end{equation}

\begin{figure}
    \centering
    \includegraphics[width=0.5\linewidth]{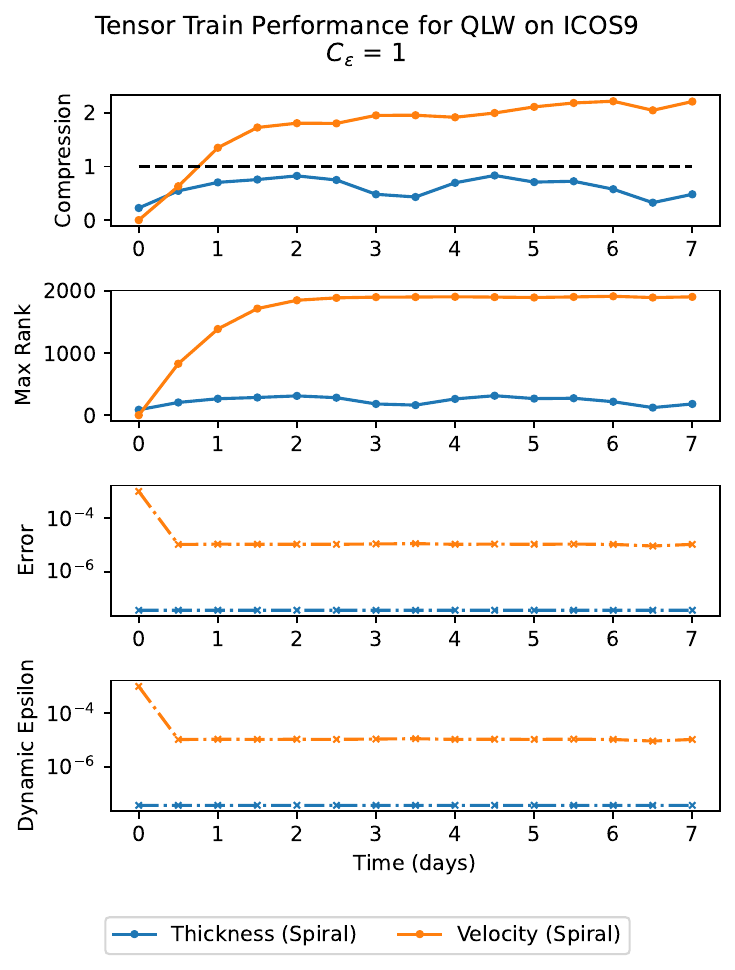}
    \caption{
        TT performance for QLW on a 9th level ICOS mesh (ICOS9).
        Model data is tensorized using the spiral ordering and QTT-like folding described in \Cref{apx:tensorizing}.
    }
    \label{fig:compression_icos_qttish}
\end{figure}

The reader might suspect that this is not likely to result in favorable TT performance for a number of reasons, and they would be correct.
\Cref{fig:compression_icos_qttish} shows the TT compression and error results obtained using these two tensorizations for QLW on an ICOS mesh consisting of 2,621,442 cells (comparable to QS0.2).
The layer thickness shows poor compression, and the normal velocity shows very high anti-compression.

The authors have experimented with many other tensorization schemes, but none proved to be effective enough to warrant further testing.
Further, we note again that an ICOS mesh gives a greatly simplified tensorization problem; more realistic meshes such as that in \Cref{fig:del_bay} can only greatly increase the difficulty of the problem.
One could re-grid the model data, interpolating it onto a more TT-friendly grid, but this would introduce additional complications related to the fact that our unstructured grids have a nonuniform number of data points along each given spatial axis (i.e. there are more cells at 0\textdegree\, latitude than at 80\textdegree\, latitude).
Then, one would need to make decisions about how the re-gridded data was to be loaded into a tensor -- if a TT decomposition similar to that used for the QS meshes in the body of this work was wanted, one would need to pad the data tensor with some set of placeholder values so that the tensor was of a uniform size along each axis.
In the case of very highly variable resolution meshes, this could lead to a tensor full of mostly placeholder values that could still effect the rank of its TT.

Another avenue that was briefly explored was the possible use of tensor decomposition methods more general than TT.
Also called tensor networks, there are tensor decomposition formats that can be viewed as generalizations of the TT format.
For example, the tensor ring (TR) format \citep{wu2023} is similar to TT, but the boundary ranks \( r_0 \) and \( r_d \) do not have to be equal to one -- we only require that \( r_0 = r_d \).
The resulting tensor network can be viewed as a ``ring'' of tensors where there are no boundary cores. 
Another format, generally refereed to as Projected Entangled Pair States (PEPS) \citep{orus2014} generalizes TT to higher dimensions; instead of a one-dimensional array of core tensors, one could have a ``matrix'' of cores, or even higher dimensional structures.
However, with any tensor network, the fundamental problem of finding a map between a mesh and a tensor space to be decomposed remains, and is not necessarily made easier by different network topology.


\subsection{Space Filling Curves}
\label{apx:sfc}

\begin{figure}
    \centering
    \begin{subfigure}{0.49\linewidth}
        \includegraphics[width=\linewidth]{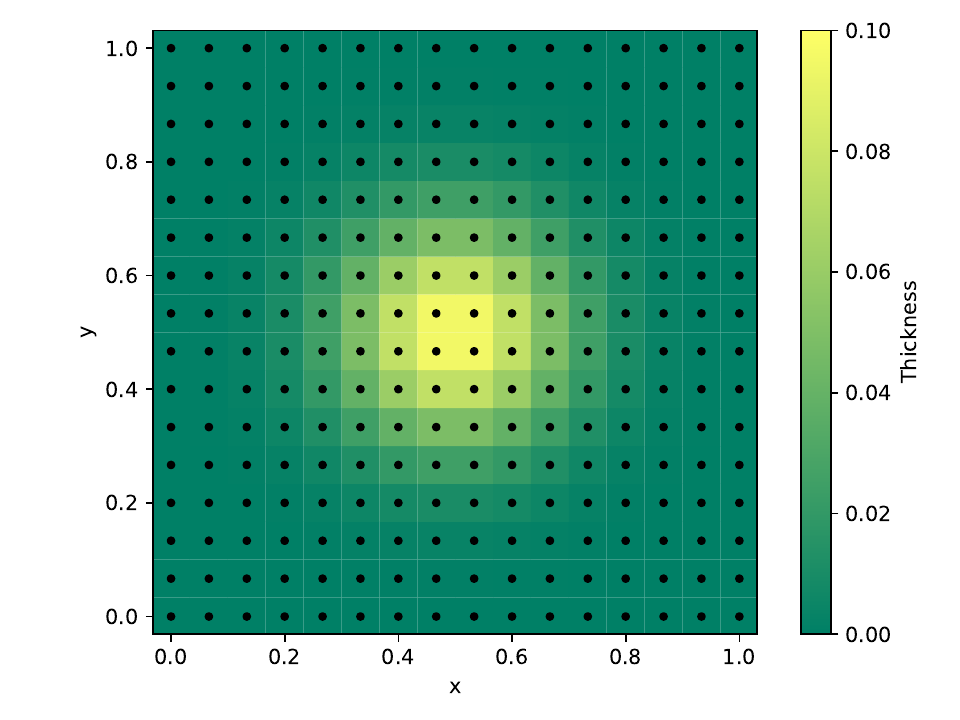}
        \caption{~}
        \label{fig:gbump}
    \end{subfigure}
    \hfill
    \begin{subfigure}{0.49\linewidth}
        \includegraphics[width=\linewidth]{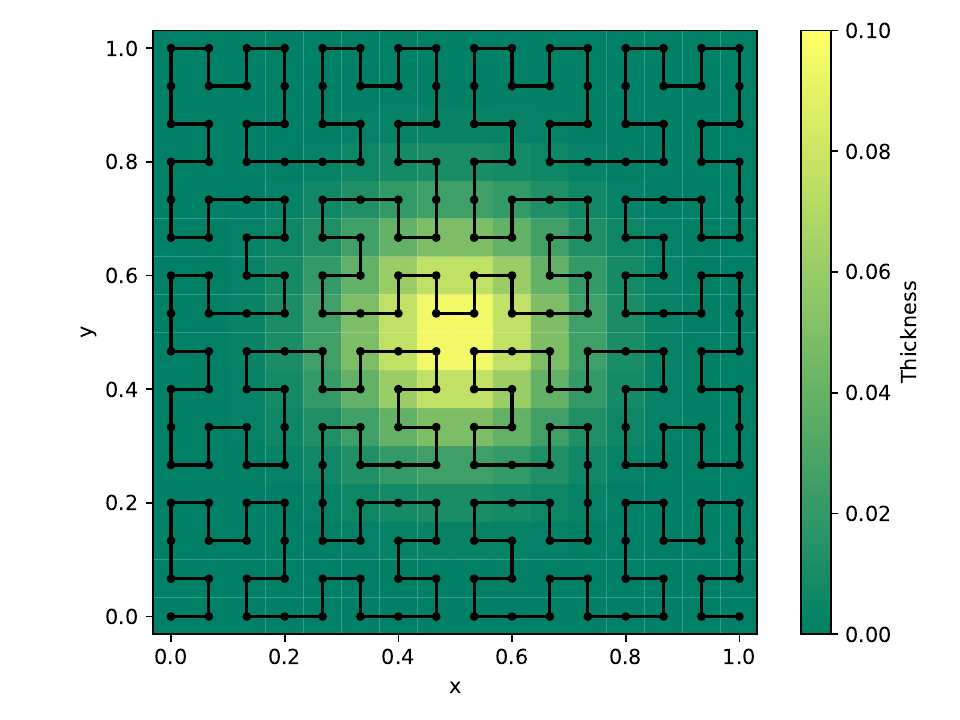}
        \caption{~}
        \label{fig:gbump_hilbert}
    \end{subfigure}
    \caption{
        A Gaussian bump on a simple \( 2^\ell \) by \( 2^\ell \) grid discretizing \( [0,\, 1] \times [0,\, 1] \) (a) and an ordering of the grid data given by a Hilbert curve (b).
    }
    \label{fig:qtt_gbump}
\end{figure}

Another possible avenue for tensorizing a general unstructured grid like that in \Cref{fig:del_bay} could involve space filling curves (SFCs).
Given a set of points in \( \mathbb{R}^d \) for \( d \geq 1 \), it is possible to build a curve that visits each point in turn.
The simplest example of such a curve is the Hilbert curve (\Cref{fig:gbump_hilbert}).

In the present context, a SFC could be used to provide the needed 1-dimensional ordering our \( d \)-dimensional model data -- this type of strategy was suggested to the authors in multiple instances.
Since this seemed to be a popular and possibly promising idea, we have attempted this strategy on a simplified problem and record the results here.
Let \( \ell \) be a positive integer, and consider a uniform \( 2^\ell \) by \( 2^\ell \) grid discretizing \( [0,\, 1] \times [0,\, 1] \) (the black dots in \Cref{fig:gbump} shows such a grid for \( \ell = 4 \)).
Let \( L = 2\ell \), then there are a total of \( 2^L \) grid points which need to be ordered.
Then, we assign each of these points a quantity according to
\begin{equation}
    \label{eqn:gbump}
    g(x,\, y) = \frac{1}{10}e^{-25\left[(x - \nicefrac{1}{2})^2 + (y - \nicefrac{1}{2})^2\right]} \,;
\end{equation}
this is essentially the initial condition for the thickness variable in the QLW test case.
The goal here is to see how well this state can be compressed with TT when ordered via a Hilbert curve (\Cref{fig:gbump_hilbert}) and folded into a \( L \)-dimensional tensor according to \Cref{eqn:qtt_shape}.
The standard method to order the data is the same used in the body of this work, with the rows of the resulting matrix then concatenated to form a 1-dimensional vector to be folded as in \Cref{eqn:qtt_shape} again.
Both methods result in tensors of the same dimension and shape, but with a different data layout.
The idea is that while the standard method is not easily generalized to unstructured meshes, the Hilbert curve method is.

\begin{table}
    \centering
    \begin{tabular}{lcc}
        \null & Compression & Max Rank \\
        \hline
        \multicolumn{3}{c}{\( \ell = 4 \); 256 grid points} \\
        \hline
        Standard & 0.313 & 4 \\
        Hilbert curve & 2.656 & 16 \\
        \hline
        \multicolumn{3}{c}{\( \ell = 8 \); 65,536 grid points} \\
        \hline
        Standard & 0.012 & 8 \\
        Hilbert curve & 0.419 & 68 \\
        \hline
        \multicolumn{3}{c}{\( \ell = 11 \); 4,194,304 grid points} \\
        \hline
        Standard & 0.0003 & 8 \\
        Hilbert curve & 0.0081 & 68 \\
        \hline
    \end{tabular}
    \caption{
        TT-compression of a simple Gaussian bump for different data orderings using \( \epsilon = 10^{-10} \).
    }
    \label{tbl:hc_results}
\end{table}

\Cref{tbl:hc_results} shows that the TT-compression achieved with the Hilbert curve method is, in each case, an order of magnitude worse than that achieved by the standard method.
For large problem sizes the Hilbert curve method compression is still strong, but recall that the state being compressed is very simple.
The fact that the maximum TT-rank needed to compress this simple state is high compared to the standard method's maximum rank suggests that this will not extend well to more complicated states.
Of course, the use of SFCs for this application is still a largely unexplored avenue; how the above strategy performs when generalized to realistic meshes is unknown and may be the topic of future work.
One could apply a general SFC algorithm, such as that given by \citet{sasidharan2015}, to an unstructured mesh and perform a study similar to the one done in this work.

The tests in this appendix show that even in the best case scenarios of quad and quasi-uniform meshes (i.e. ICOS), unstructured data ordering prevents TT methods from effective compression.
In the more challenging cases of variable-resolution meshes TT would likely perform even worse.
This small investigation led to the use of two-dimensional structured grids for the thrust of the study, in order to present the best possible range of results for TT in a spherical shallow water setting.



\section{Additional Results}
\label{apx:additional_results}

\def\compressionscale{0.95}
\def\diagnosticscale{0.6}
\def\matvecsubscale{0.45}

Here, we provide a collection of additional results for the four test cases from \Cref{sec:experiments}.
TT-compression results are given for \( C_\epsilon = 0.1,\, 10,\, 100  \) for each test case, and additional speedup results are given for BUJ using \( C_\epsilon = 10,\, 100 \).
In general, we observe that changing \( C_\epsilon \) does not meaningfully effect the qualitative performance of our TT methods; it does change the TT-compression in each case, but the behavior of the TT-compression curves in each case remains largely the same.

In particular, we draw the readers attention to \Cref{fig:matvec_intime_buj_10,fig:matvec_intime_buj_100}, which show the speedup achieved by TTMV versus SpMV for the BUJ case.
Compare these to \Cref{fig:matvec_intime_buj_1} and observe that increasing \( C_\epsilon \) doesn't change the overall behavior of the speedup curves.
Rather, this only shifts them up on the vertical axis slightly, with the effect that the TTMV methods remain faster than SpMV slightly longer into the simulation before SpMV overtakes them again.
This suggests that finding speedup for complex flows like BUJ and WTC5 is not the simple matter of a well-chosen \( C_\epsilon \).
If we were to increase \( C_\epsilon \) enough that TTMV was faster than SpMV for the whole of the simulation, the introduced error would simply be too large for the results to be valuable.


\subsection{Quasi-Linear Wave}
\label{apx:qlw}

\def\compressionceps{0.1}
\begin{figure}
    \centering
    \includegraphics[width=\compressionscale\linewidth]{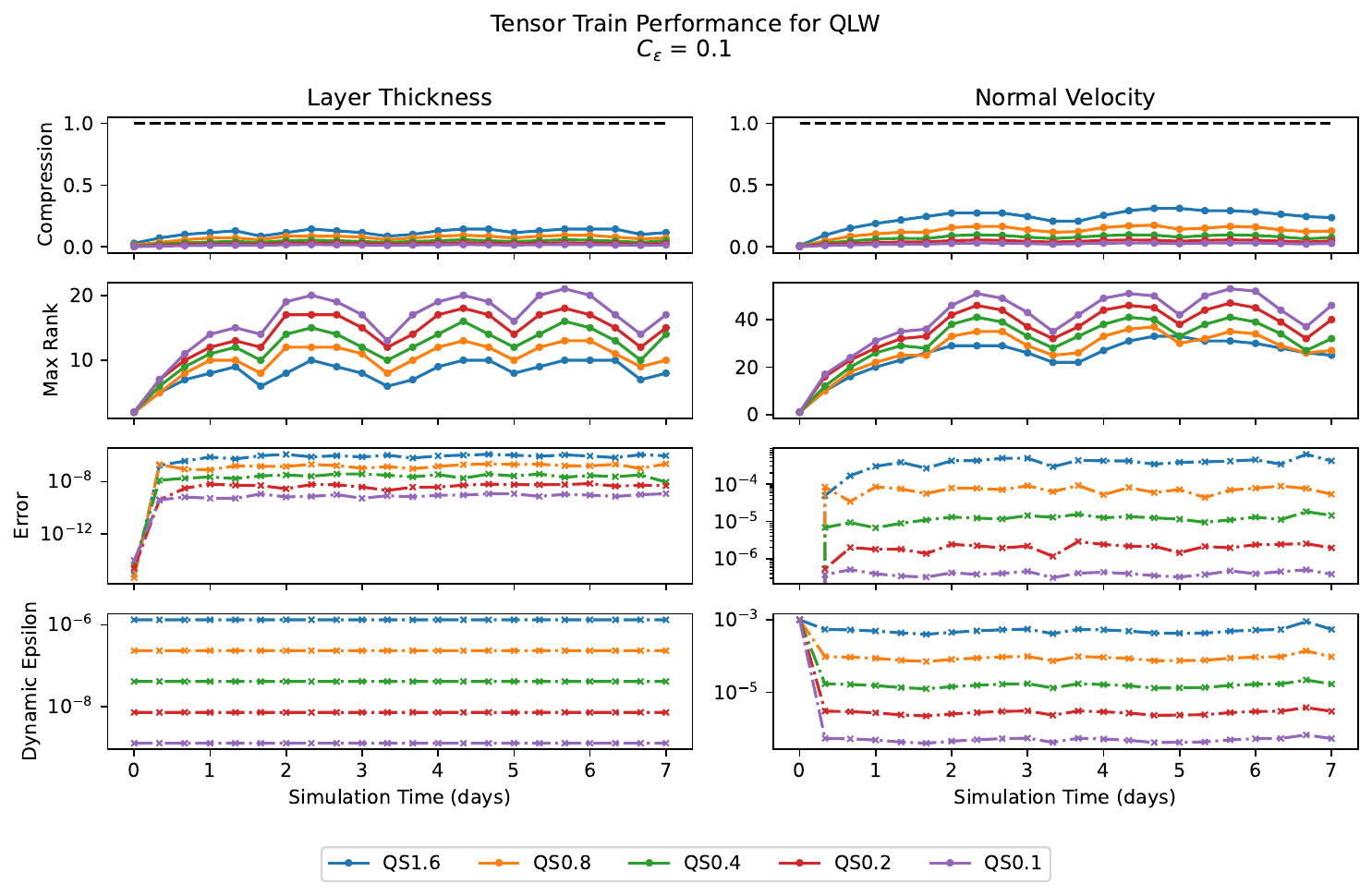}
    \caption{
        TT-compression results for QLW with \( C_\epsilon = \compressionceps \).
    }
    \label{fig:compression_qlw_Ceps\compressionceps}
\end{figure}

\def\compressionceps{10}
\begin{figure}
    \centering
    \includegraphics[width=\compressionscale\linewidth]{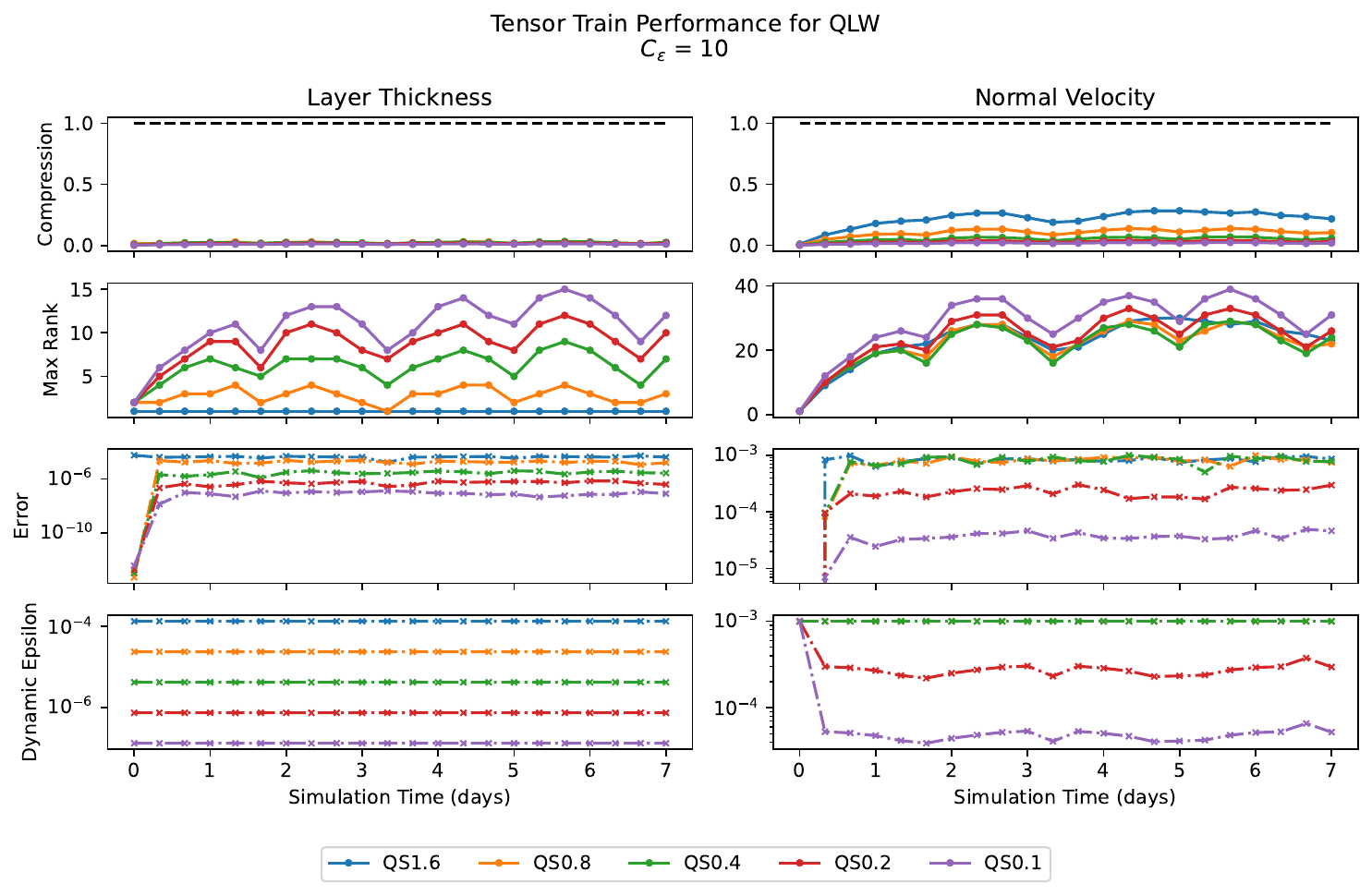}
    \caption{
        TT-compression results for QLW with \( C_\epsilon = \compressionceps \).
    }
    \label{fig:compression_qlw_Ceps\compressionceps}
\end{figure}

\def\compressionceps{100}
\begin{figure}
    \centering
    \includegraphics[width=\compressionscale\linewidth]{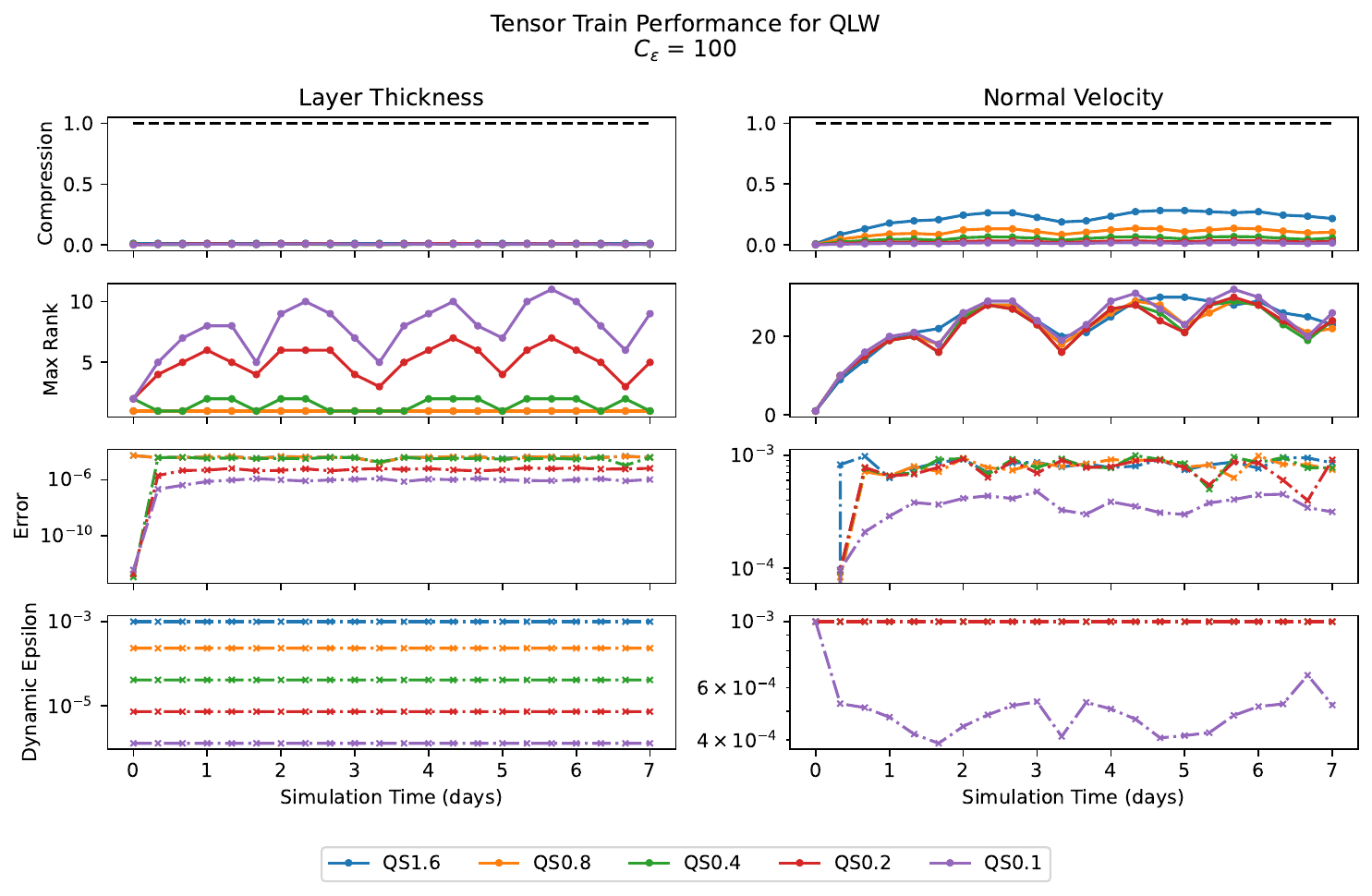}
    \caption{
        TT-compression results for QLW with \( C_\epsilon = \compressionceps \).
    }
    \label{fig:compression_qlw_Ceps\compressionceps}
\end{figure}

\FloatBarrier


\subsection{Geophysical Gravity Wave}
\label{apx:ggw}

\def\compressionceps{0.1}
\begin{figure}
    \centering
    \includegraphics[width=\compressionscale\linewidth]{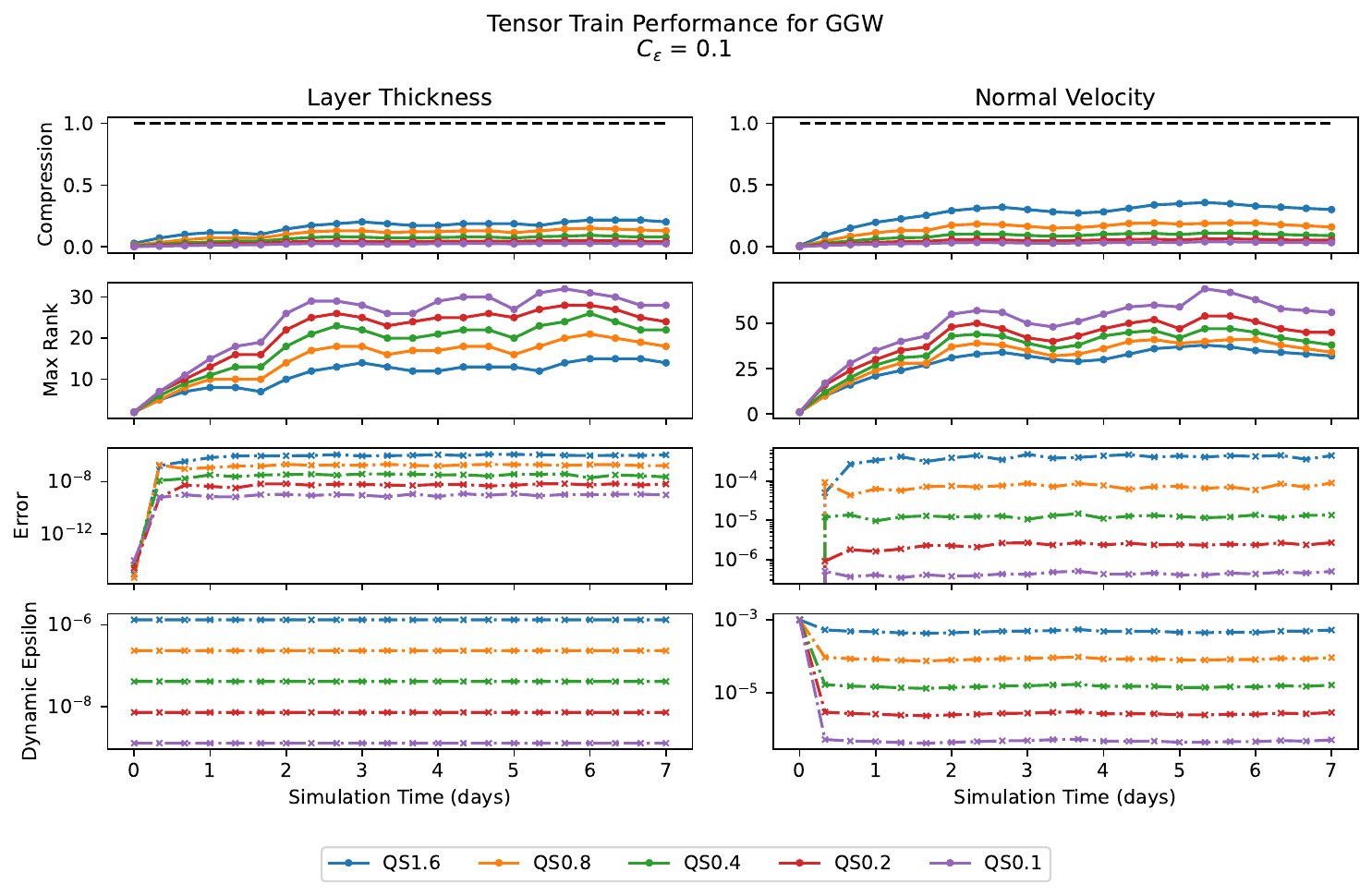}
    \caption{
        TT-compression results for GGW with \( C_\epsilon = \compressionceps \).
    }
    \label{fig:compression_GGW_Ceps\compressionceps}
\end{figure}

\def\compressionceps{10}
\begin{figure}
    \centering
    \includegraphics[width=\compressionscale\linewidth]{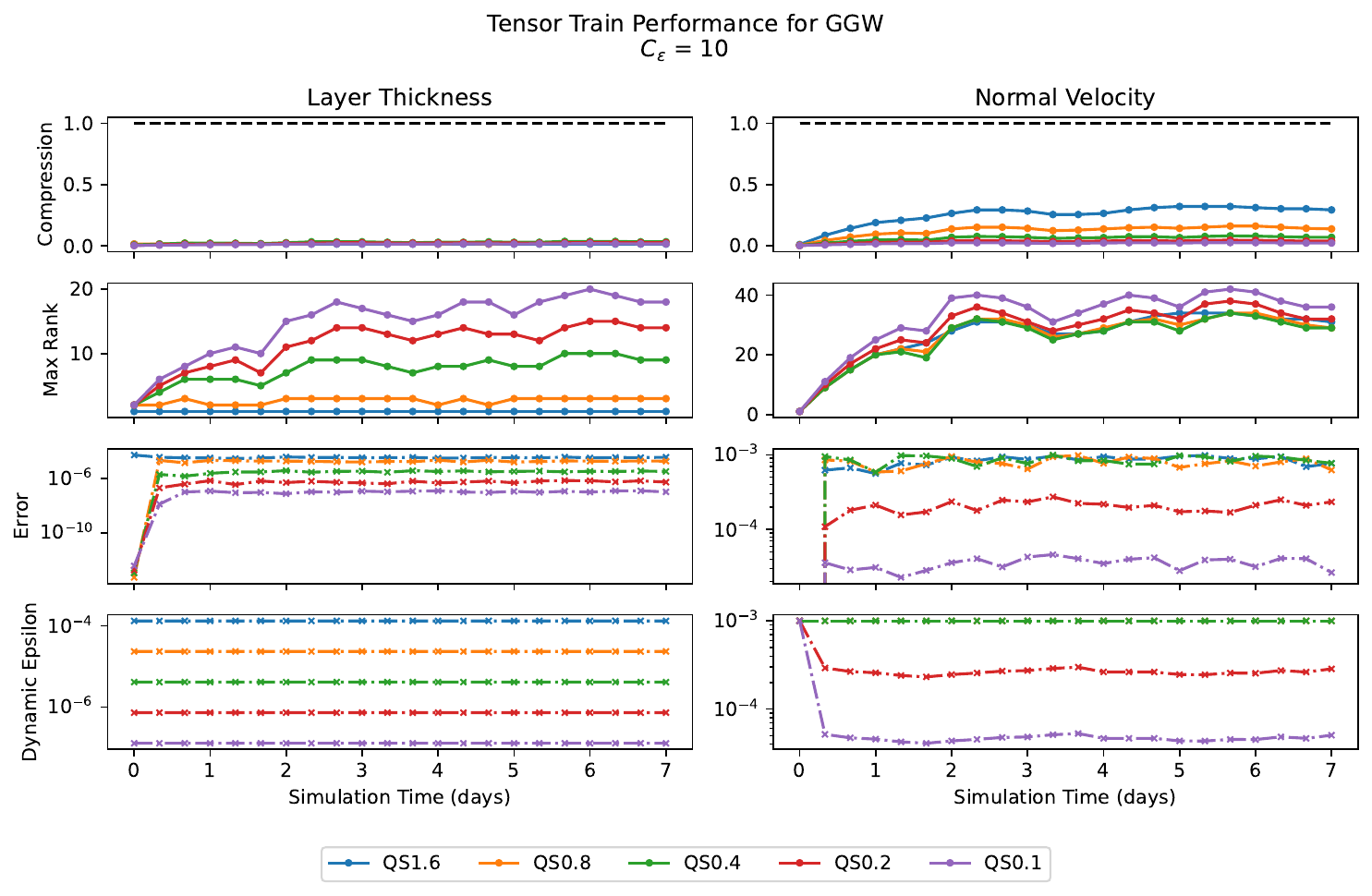}
    \caption{
        TT-compression results for GGW with \( C_\epsilon = \compressionceps \).
    }
    \label{fig:compression_GGW_Ceps\compressionceps}
\end{figure}

\def\compressionceps{100}
\begin{figure}
    \centering
    \includegraphics[width=\compressionscale\linewidth]{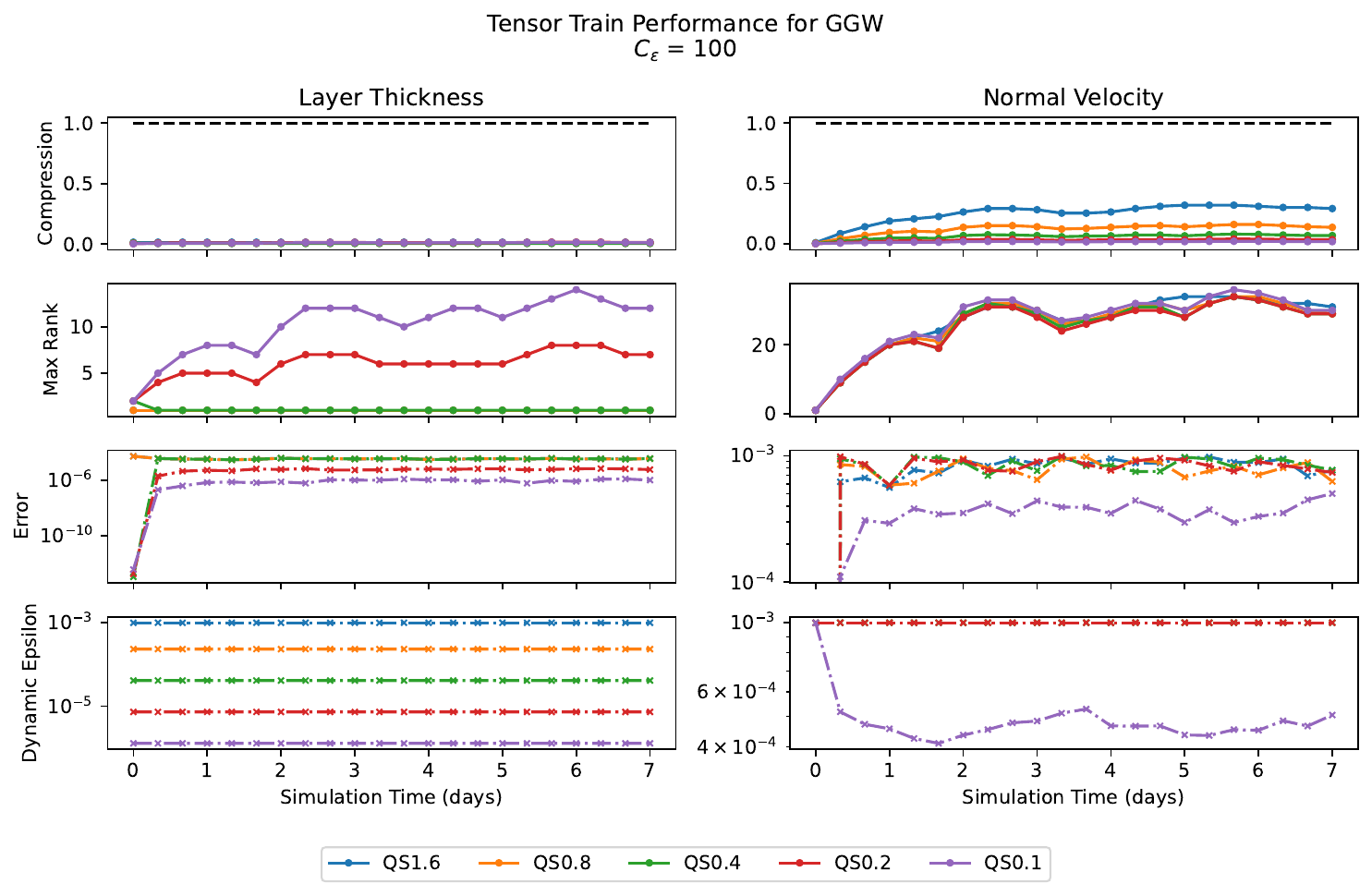}
    \caption{
        TT-compression results for GGW with \( C_\epsilon = \compressionceps \).
    }
    \label{fig:compression_GGW_Ceps\compressionceps}
\end{figure}

\FloatBarrier


\subsection{Barotropically Unstable Jet}
\label{apx:buj}

\def\compressionceps{0.1}
\begin{figure}
    \centering
    \includegraphics[width=\compressionscale\linewidth]{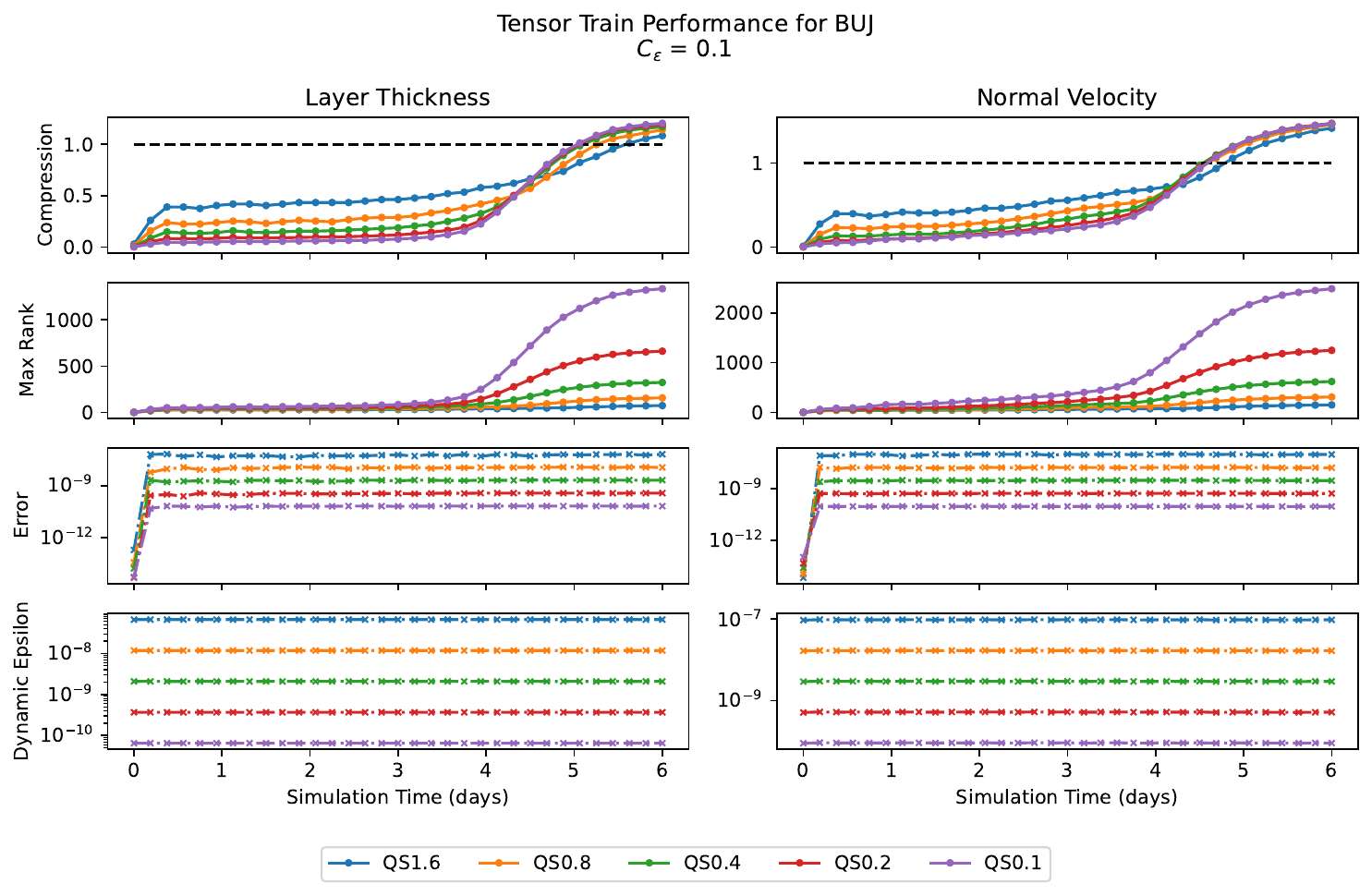}
    \caption{
        TT-compression results for BUJ with \( C_\epsilon = \compressionceps \).
    }
    \label{fig:compression_buj_Ceps\compressionceps}
\end{figure}

\def\compressionceps{10}
\begin{figure}
    \centering
    \includegraphics[width=\compressionscale\linewidth]{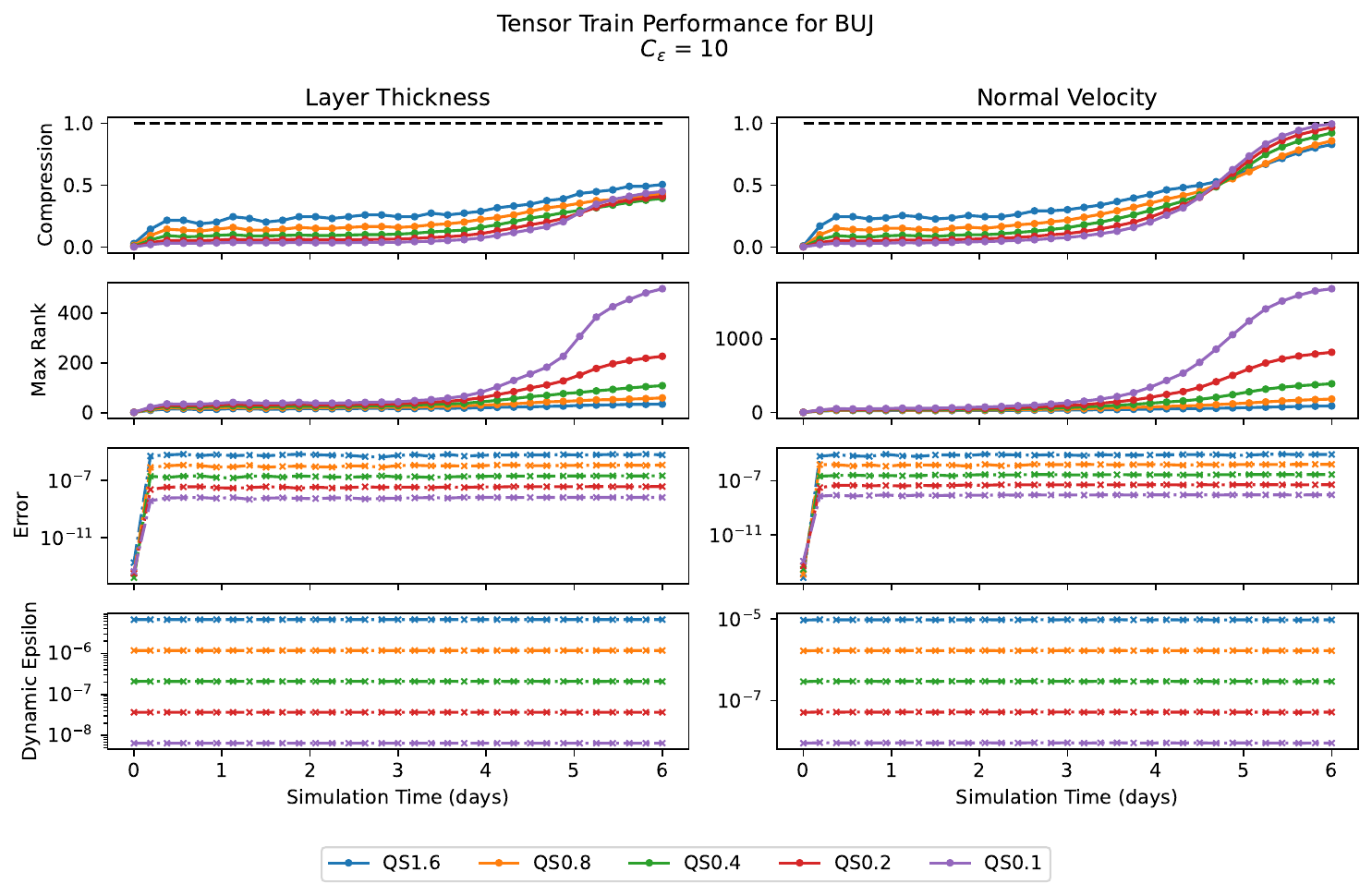}
    \caption{
        TT-compression results for BUJ with \( C_\epsilon = \compressionceps \).
    }
    \label{fig:compression_buj_Ceps\compressionceps}
\end{figure}

\def\compressionceps{100}
\begin{figure}
    \centering
    \includegraphics[width=\compressionscale\linewidth]{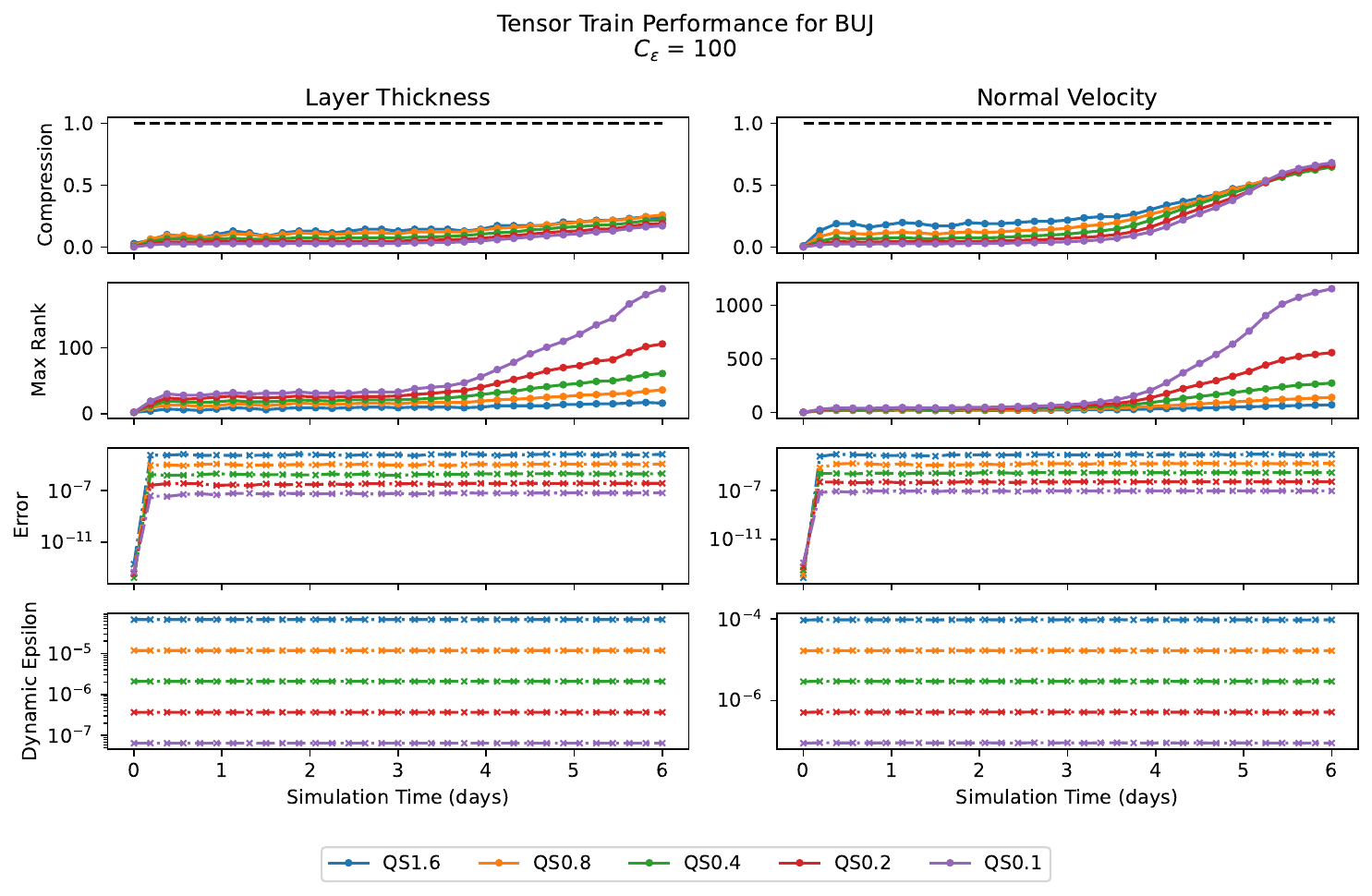}
    \caption{
        TT-compression results for BUJ with \( C_\epsilon = \compressionceps \).
    }
    \label{fig:compression_buj_Ceps\compressionceps}
\end{figure}

\def\matvecceps{10}
\begin{figure}
    \centering
    \begin{subfigure}{\matvecsubscale\linewidth}
        \includegraphics[width=\linewidth]{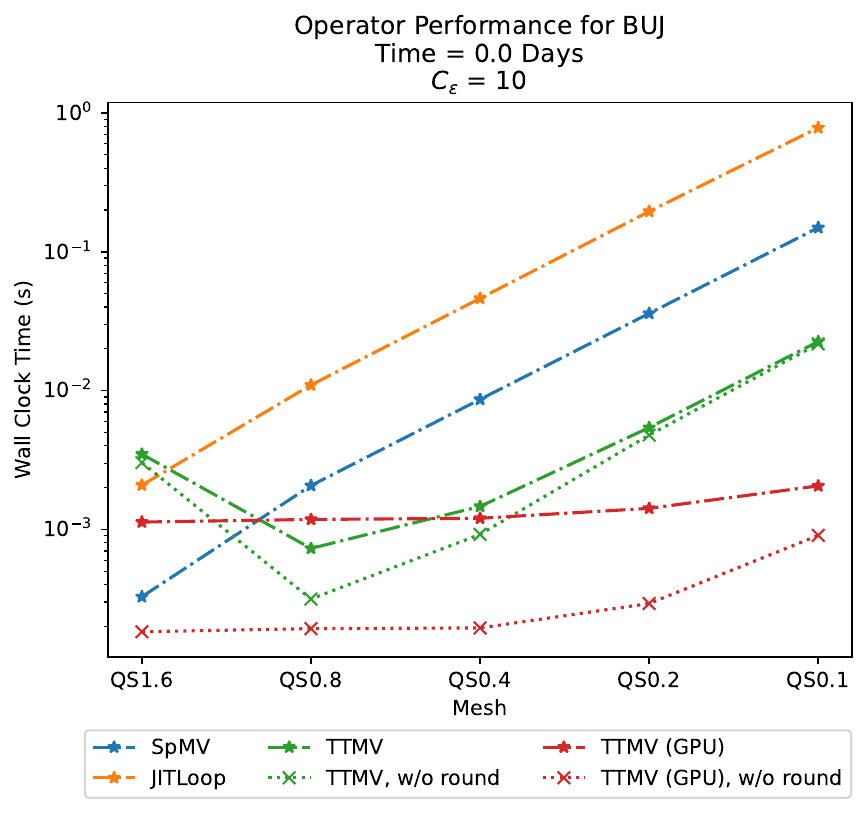}
        \caption{~}
        \label{fig:matvec_buj_timeind0_Ceps\matvecceps}
    \end{subfigure}
    \hfill
    \begin{subfigure}{\matvecsubscale\linewidth}
        \includegraphics[width=\linewidth]{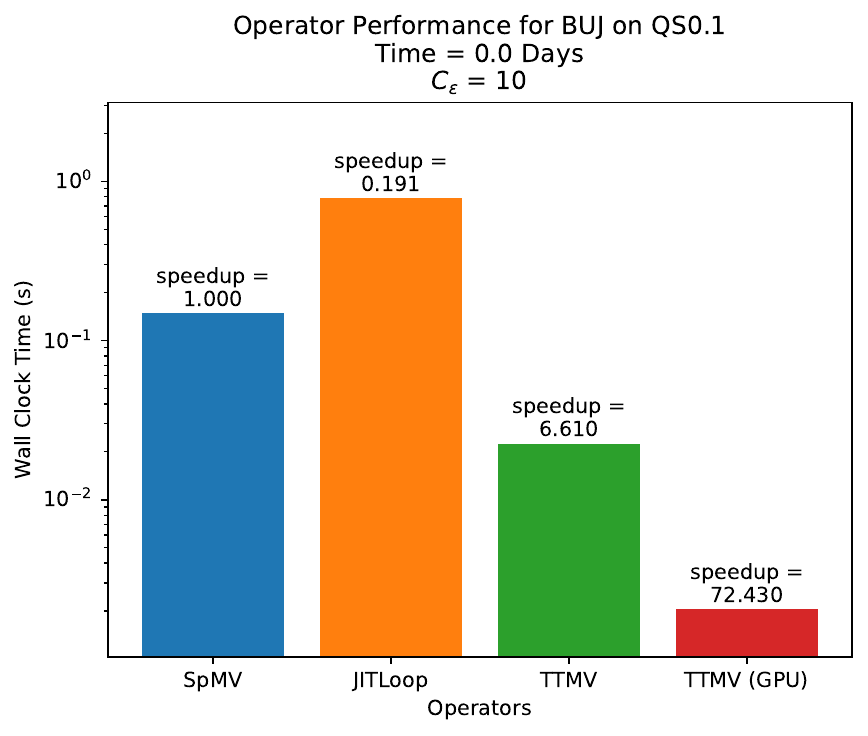}
        \caption{~}
        \label{fig:matvec_bar_buj_qs0.1_timeind0_Ceps\matvecceps}
    \end{subfigure}

    \medskip
    
    \begin{subfigure}{\matvecsubscale\linewidth}
        \includegraphics[width=\linewidth]{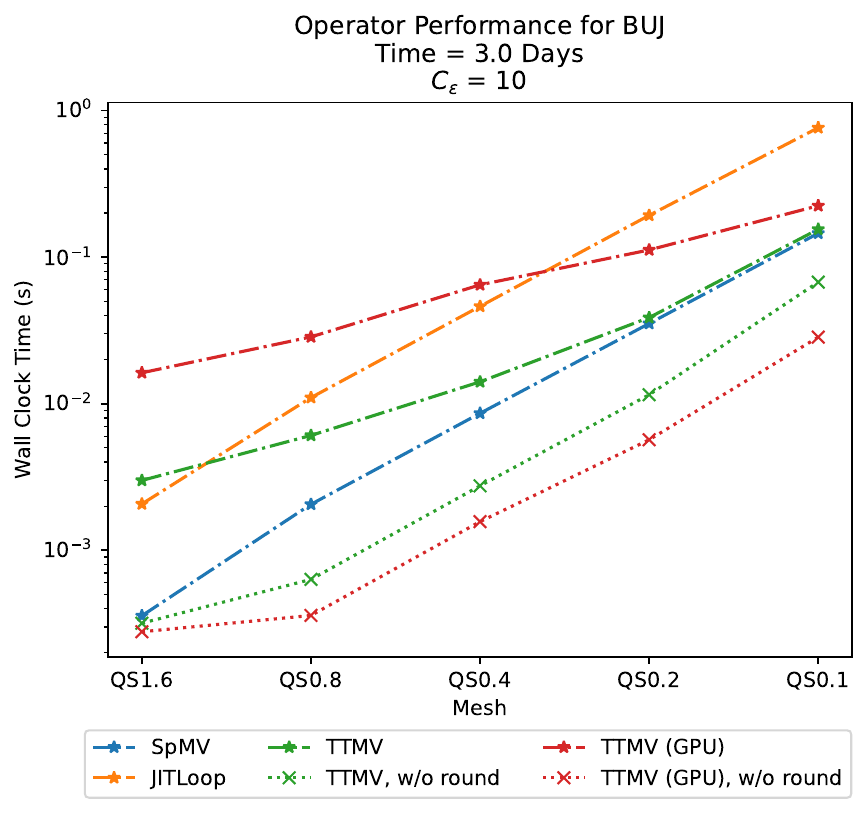}
        \caption{~}
        \label{fig:matvec_buj_timeind16_Ceps\matvecceps}
    \end{subfigure}
    \hfill
    \begin{subfigure}{\matvecsubscale\linewidth}
        \includegraphics[width=\linewidth]{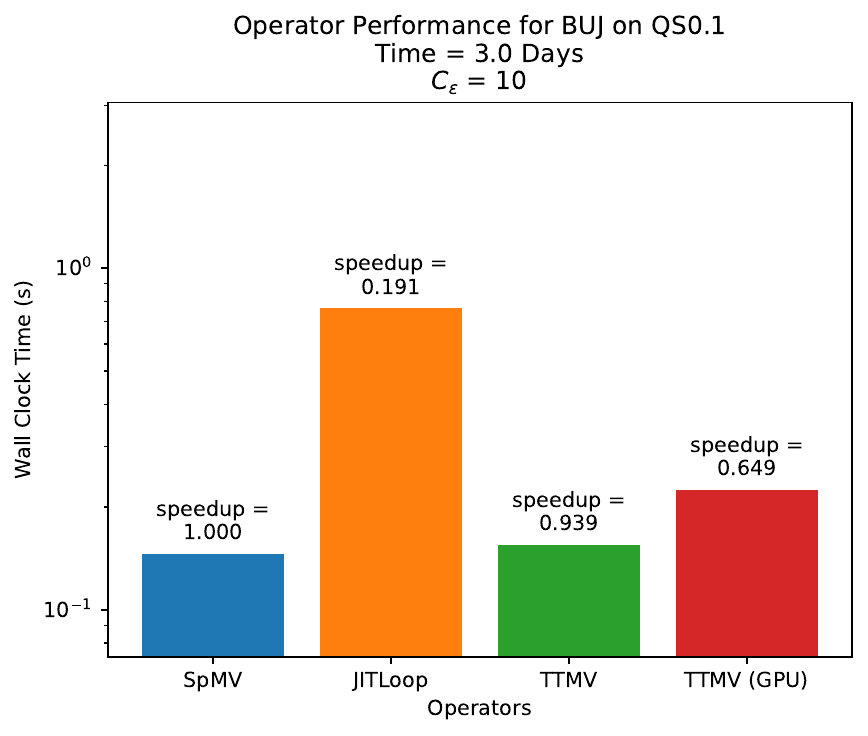}
        \caption{~}
        \label{fig:matvec_bar_buj_qs0.1_timeind16_Ceps\matvecceps}
    \end{subfigure}

    \medskip
    
    \begin{subfigure}{\matvecsubscale\linewidth}
        \includegraphics[width=\linewidth]{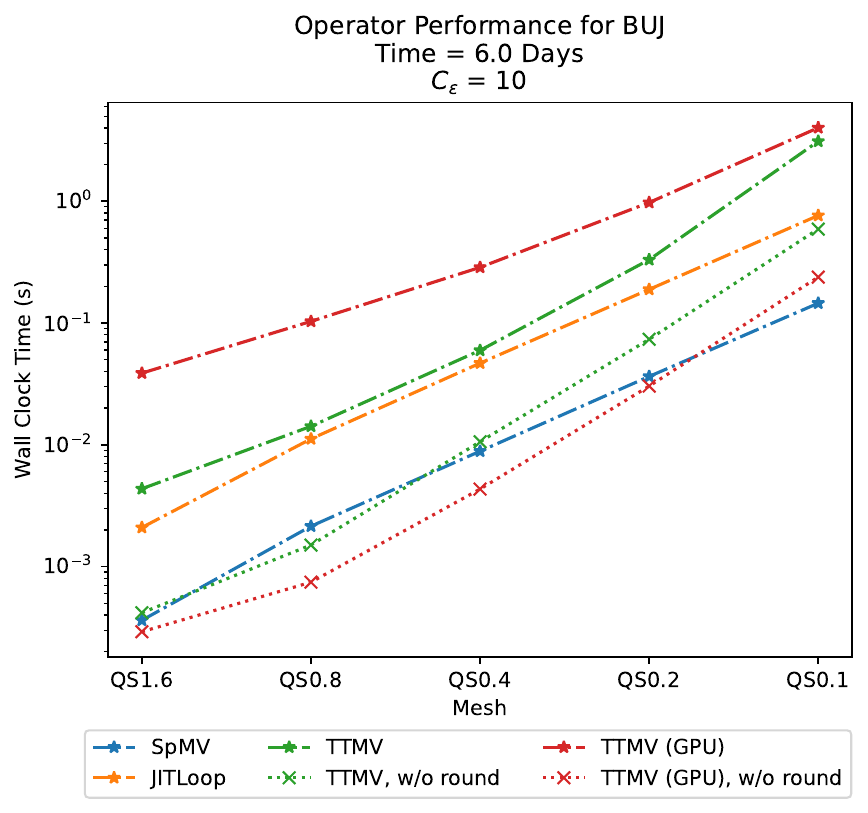}
        \caption{~}
        \label{fig:matvec_buj_timeind32_Ceps\matvecceps}
    \end{subfigure}
    \hfill
    \begin{subfigure}{\matvecsubscale\linewidth}
        \includegraphics[width=\linewidth]{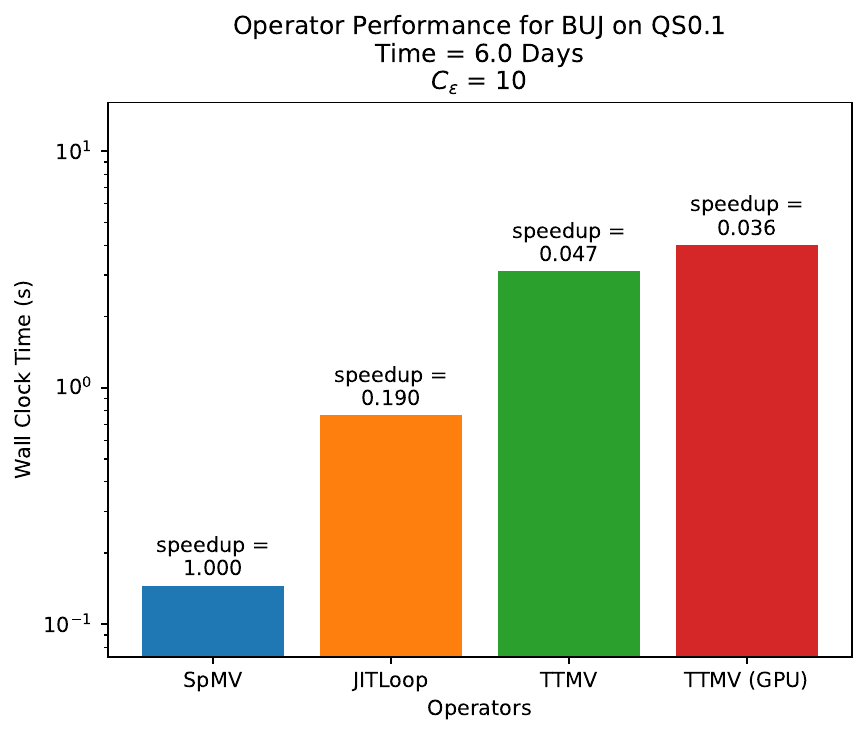}
        \caption{~}
        \label{fig:matvec_bar_buj_qs0.1_timeind32_Ceps\matvecceps}
    \end{subfigure}
    \caption{
        TT-format operation performance in the BUJ test case.
    }
    \label{fig:matvec_buj_\matvecceps}
\end{figure}

\begin{figure}
    \centering
    \includegraphics[width=\diagnosticscale\linewidth]{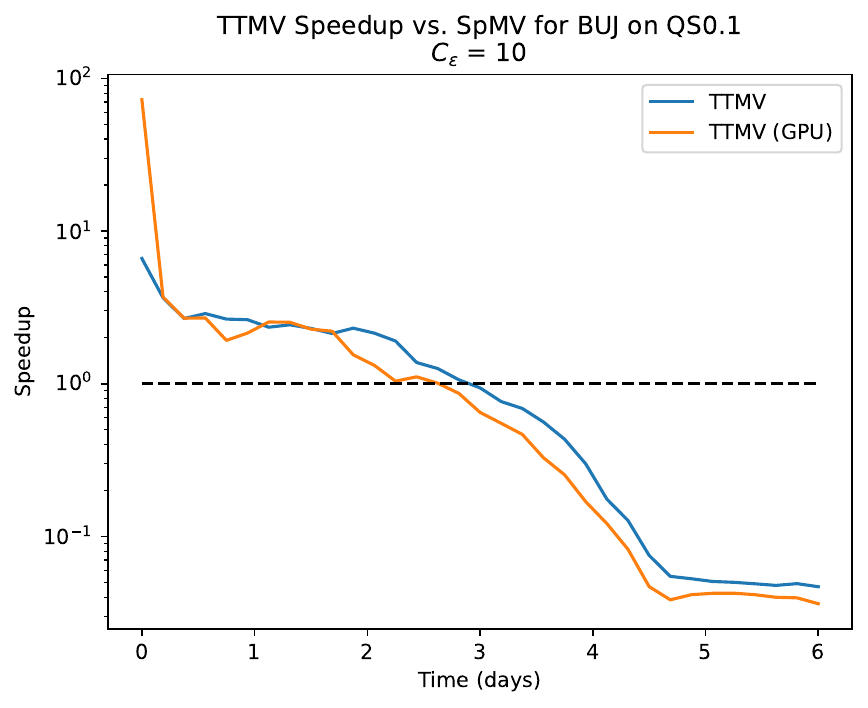}
    \caption{
        TTMV speedup versus SpMV in time for the BUJ test case.
    }
    \label{fig:matvec_intime_buj_\matvecceps}
\end{figure}

\def\matvecceps{100}
\begin{figure}
    \centering
    \begin{subfigure}{\matvecsubscale\linewidth}
        \includegraphics[width=\linewidth]{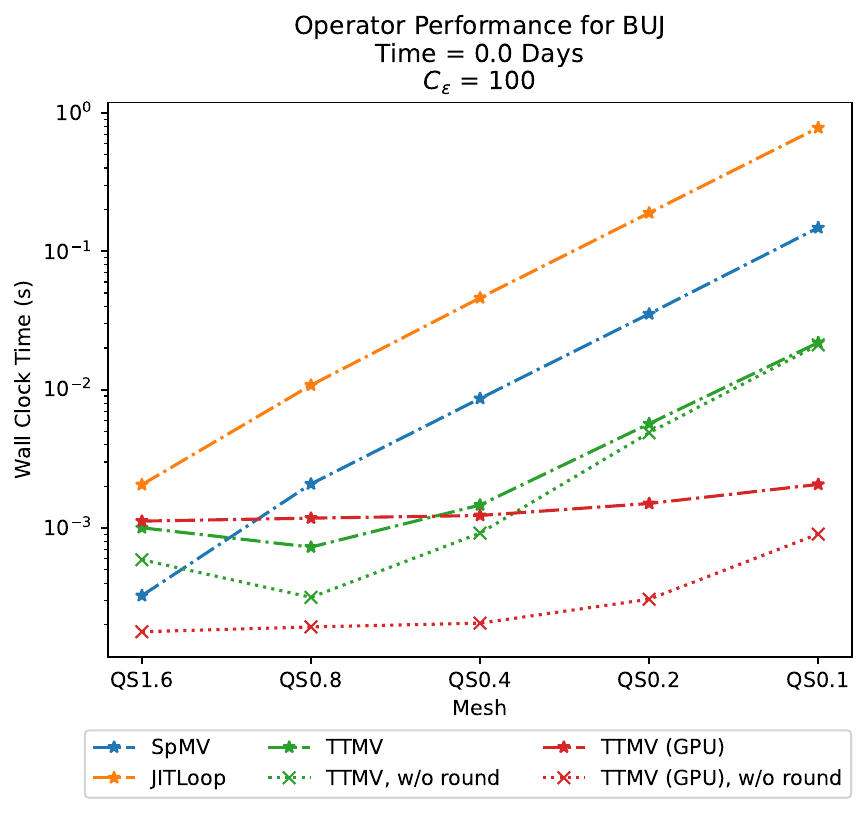}
        \caption{~}
        \label{fig:matvec_buj_timeind0_Ceps\matvecceps}
    \end{subfigure}
    \hfill
    \begin{subfigure}{\matvecsubscale\linewidth}
        \includegraphics[width=\linewidth]{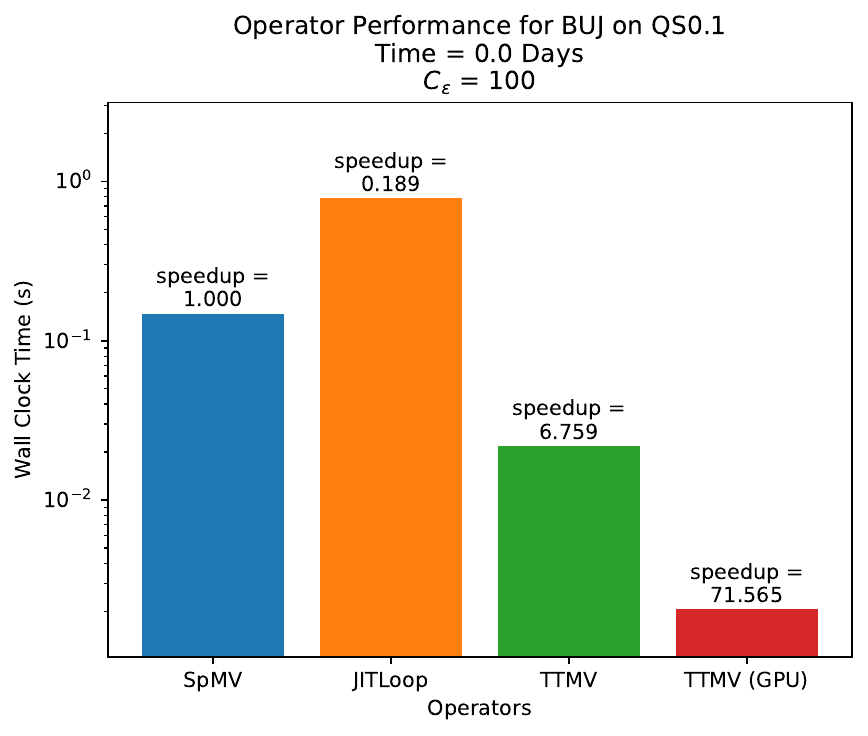}
        \caption{~}
        \label{fig:matvec_bar_buj_qs0.1_timeind0_Ceps\matvecceps}
    \end{subfigure}

    \medskip
    
    \begin{subfigure}{\matvecsubscale\linewidth}
        \includegraphics[width=\linewidth]{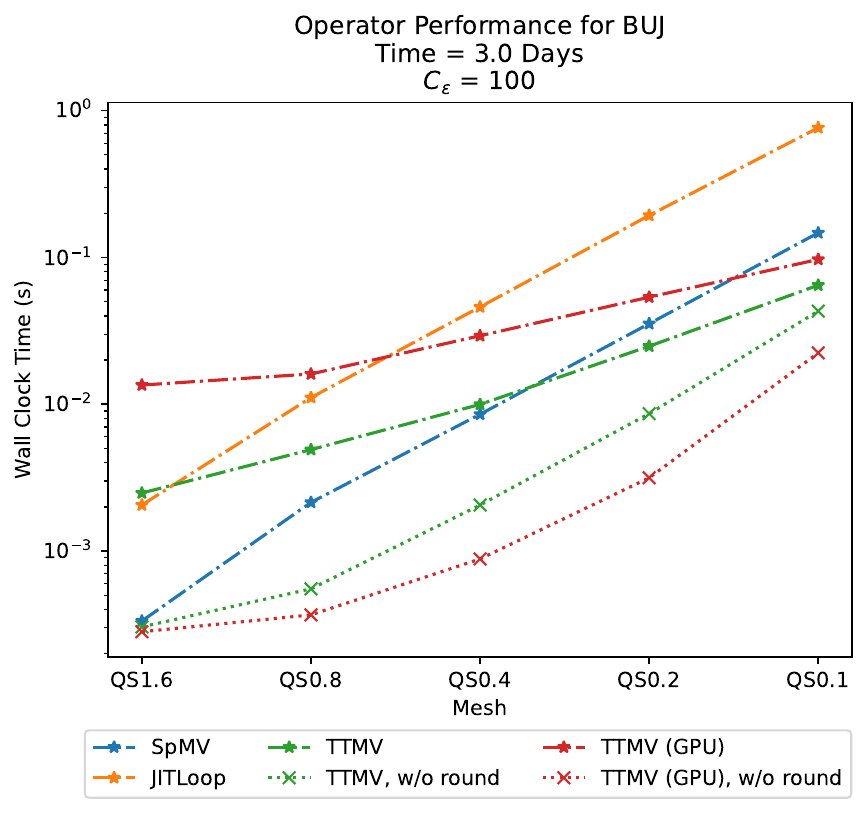}
        \caption{~}
        \label{fig:matvec_buj_timeind16_Ceps\matvecceps}
    \end{subfigure}
    \hfill
    \begin{subfigure}{\matvecsubscale\linewidth}
        \includegraphics[width=\linewidth]{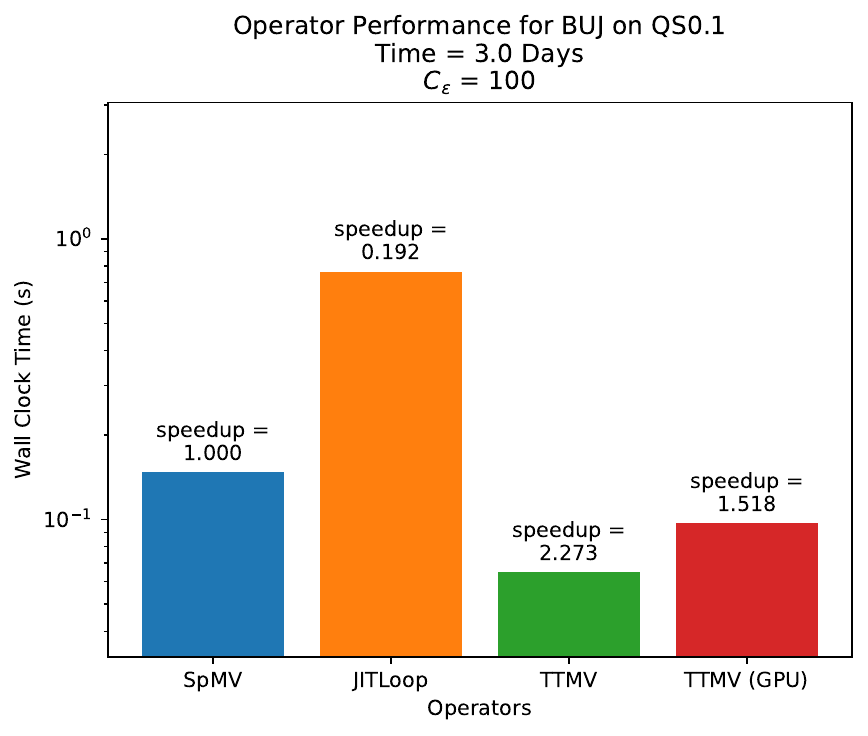}
        \caption{~}
        \label{fig:matvec_bar_buj_qs0.1_timeind16_Ceps\matvecceps}
    \end{subfigure}

    \medskip
    
    \begin{subfigure}{\matvecsubscale\linewidth}
        \includegraphics[width=\linewidth]{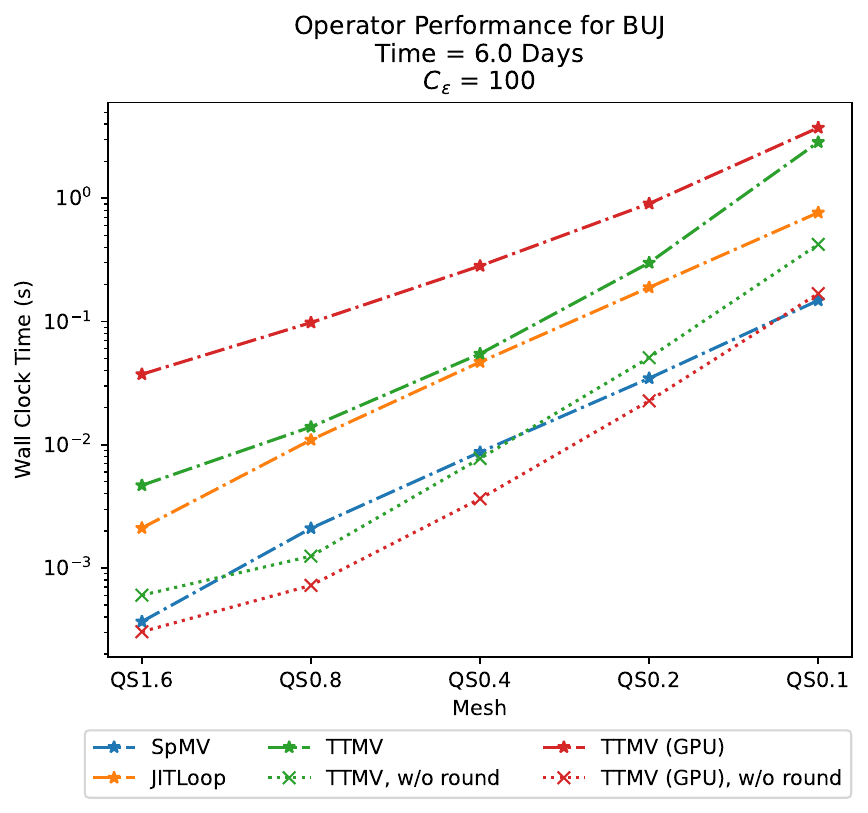}
        \caption{~}
        \label{fig:matvec_buj_timeind32_Ceps\matvecceps}
    \end{subfigure}
    \hfill
    \begin{subfigure}{\matvecsubscale\linewidth}
        \includegraphics[width=\linewidth]{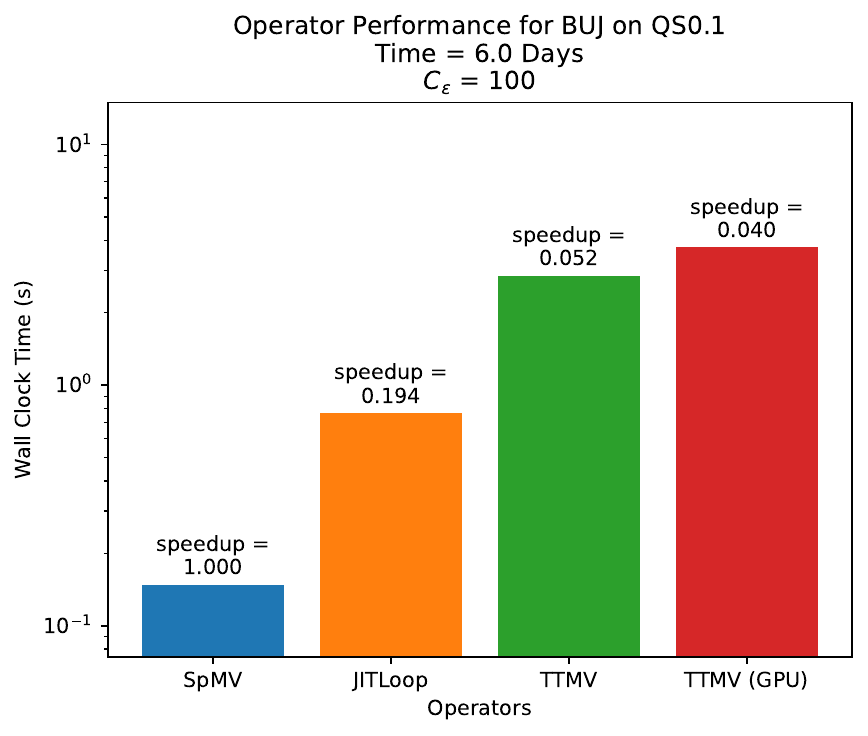}
        \caption{~}
        \label{fig:matvec_bar_buj_qs0.1_timeind32_Ceps\matvecceps}
    \end{subfigure}
    \caption{
        TT-format operation performance in the BUJ test case.
    }
    \label{fig:matvec_buj_\matvecceps}
\end{figure}

\begin{figure}
    \centering
    \includegraphics[width=\diagnosticscale\linewidth]{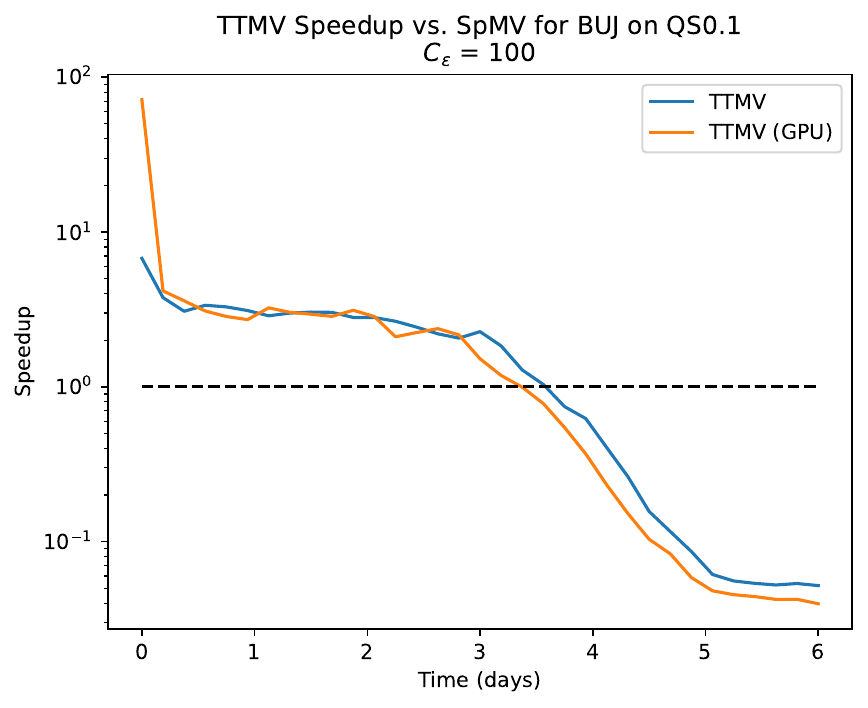}
    \caption{
        TTMV speedup versus SpMV in time for the BUJ test case.
    }
    \label{fig:matvec_intime_buj_\matvecceps}
\end{figure}

\FloatBarrier


\subsection{Williamson Test Case 5}
\label{apx:wtc5}

\def\compressionceps{0.1}
\begin{figure}
    \centering
    \includegraphics[width=\compressionscale\linewidth]{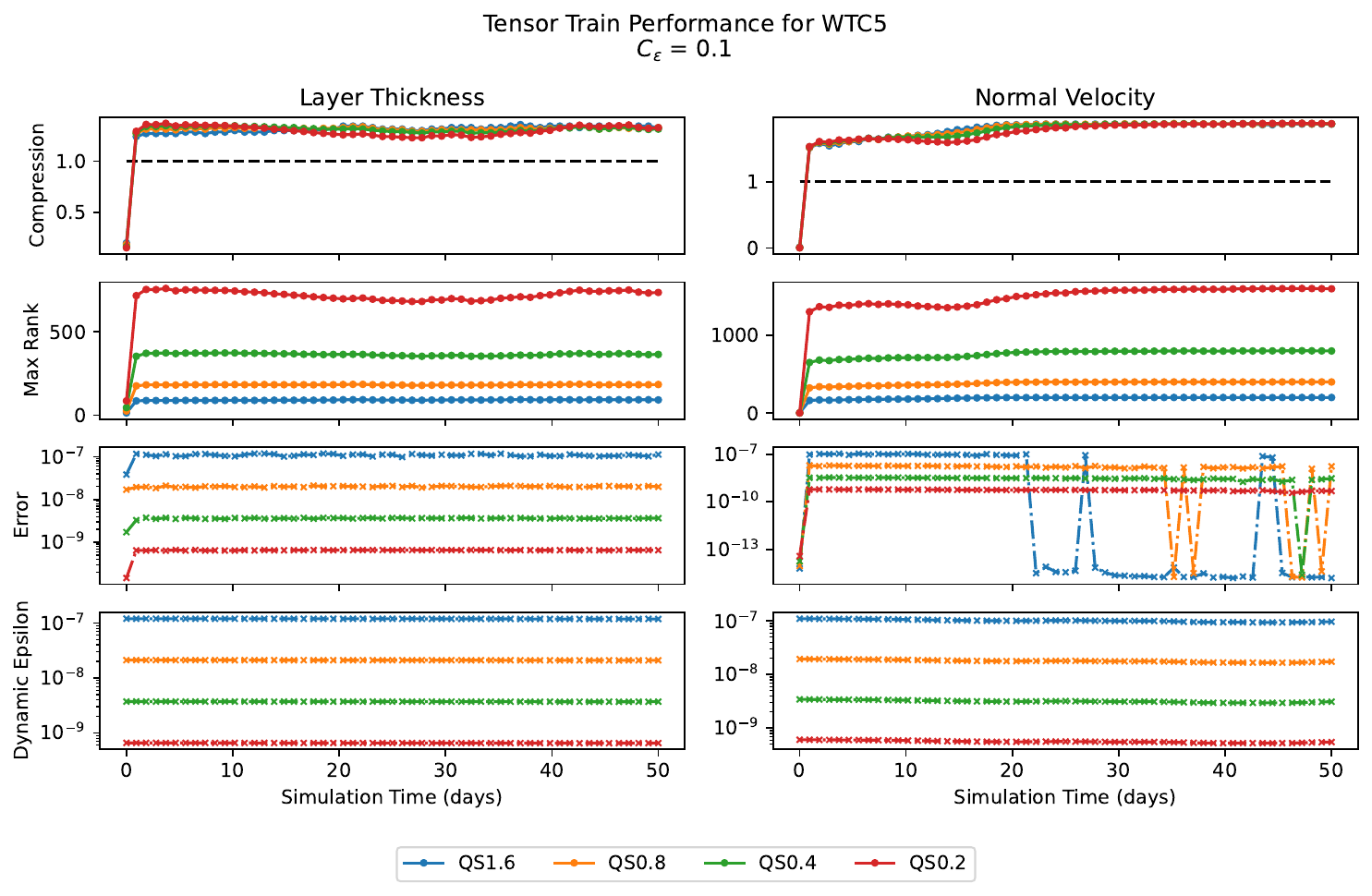}
    \caption{
        TT-compression results for WTC5 with \( C_\epsilon = \compressionceps \).
    }
    \label{fig:compression_wtc5_Ceps\compressionceps}
\end{figure}

\def\compressionceps{10}
\begin{figure}
    \centering
    \includegraphics[width=\compressionscale\linewidth]{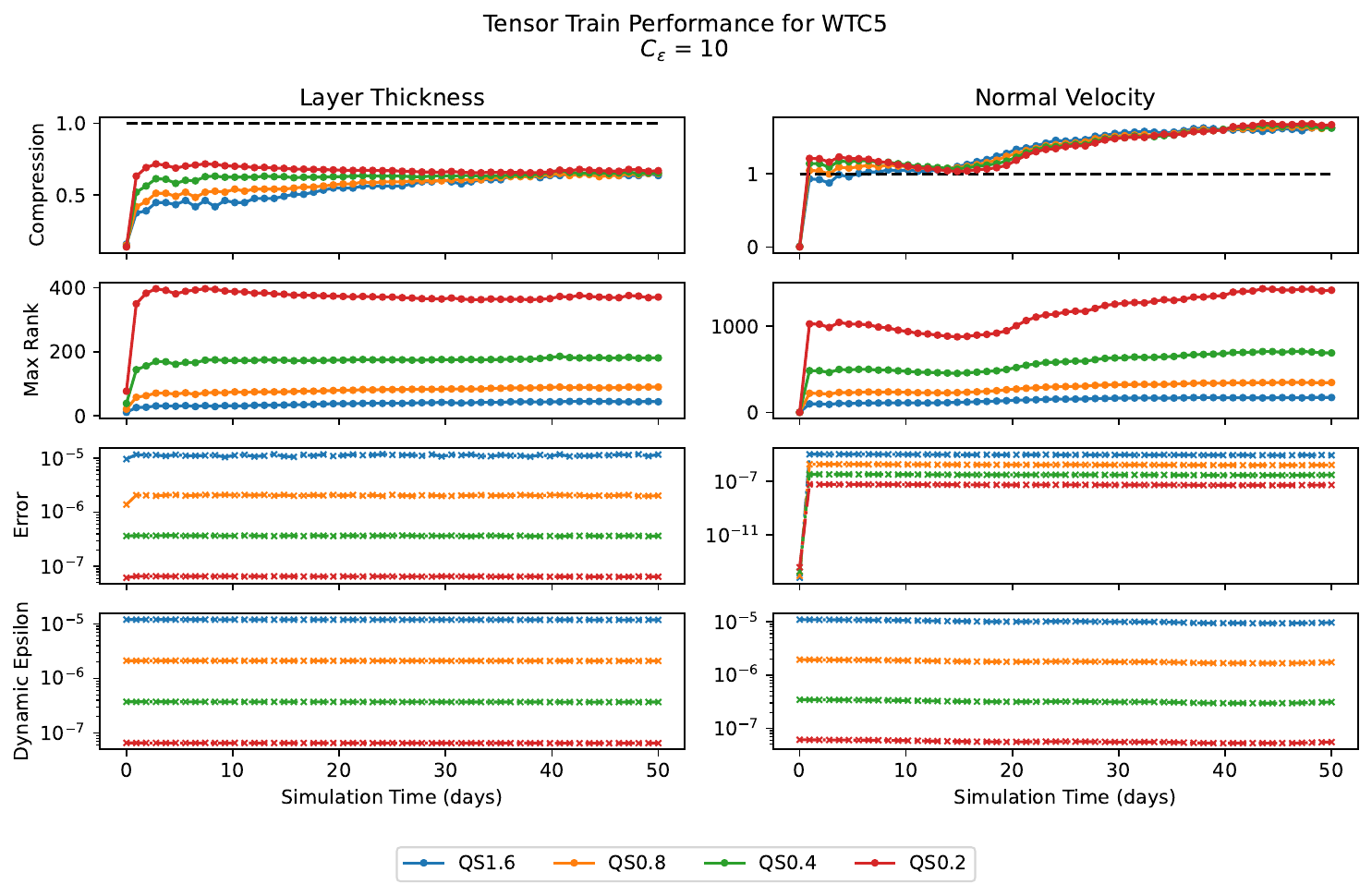}
    \caption{
        TT-compression results for WTC5 with \( C_\epsilon = \compressionceps \).
    }
    \label{fig:compression_wtc5_Ceps\compressionceps}
\end{figure}

\def\compressionceps{100}
\begin{figure}
    \centering
    \includegraphics[width=\compressionscale\linewidth]{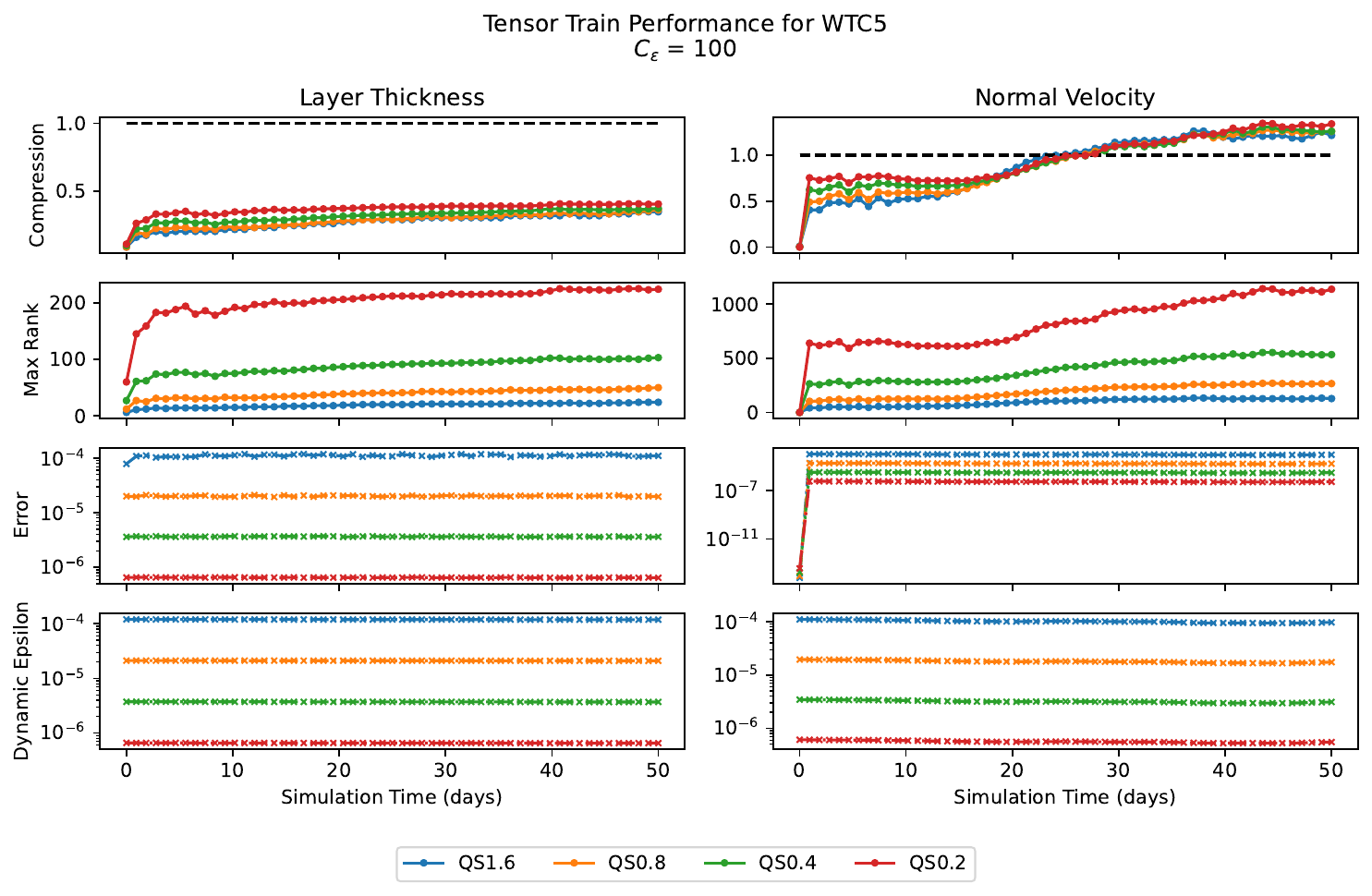}
    \caption{
        TT-compression results for WTC5 with \( C_\epsilon = \compressionceps \).
    }
    \label{fig:compression_wtc5_Ceps\compressionceps}
\end{figure}

\FloatBarrier



\bibliographystyle{elsarticle-harv}
\bibliography{references}


\end{document}